\documentclass[12pt,preprint]{aastex}

\begin{document}
\title{A 2\,mm spectral line survey of the starburst galaxy NGC\,253}
\shorttitle{2\,mm line survey of NGC\,253}

\author{S. Mart\'{\i}n}
\affil{Instituto de Radioastronom\'{\i}a Milim\'etrica (IRAM), Avda. Divina Pastora 7, Local 20, E-18012 Granada, Spain}
\email{martin@iram.es}
\author{R. Mauersberger}
\affil{Instituto de Radioastronom\'{\i}a Milim\'etrica (IRAM), Avda. Divina Pastora 7, Local 20, E-18012 Granada, Spain}
\author{J. Mart\'{\i}n-Pintado}
\affil{Departamento de Astrofis\'{\i}ca Molecular e Infrarroja, Instituto de Estructura de la Materia, CSIC, Serrano 121, E-28006 Madrid, Spain}
\author{C. Henkel}
\affil{Max-Planck-Institut f\"ur Radioastronomie, Auf dem H\"ugel 69, D-53121 Bonn, Germany}
\and
\author{S. Garc\'{\i}a-Burillo}
\affil{Observatorio de Madrid, Alfonso XII, 3, 28014 Madrid, Spain}

\shortauthors{Mart\'{\i}n et al.}

\begin{abstract}
We present the first unbiased molecular line survey towards an extragalactic source,
namely the nuclear region of the starburst galaxy NGC\,253.
The scan covers the frequency band from 129.1 to 175.2\,GHz, i.e. most of the 2\,mm atmospheric window.
We identify 111 spectral features as transitions from 25 different molecular species. 
Eight of which (three tentatively) are detected for the first time in the extragalactic interstellar medium.
Among these newly detected species, we detected the rare isotopomers $^{34}$SO and HC$^{18}$O$^+$.
Tentative detections of two deuterated species, DNC and N$_2$D$^+$,
are reported for the first time from 
a target beyond the Magellanic Clouds.
Additionally, three hydrogen recombination lines are identified,
while no organic molecules larger than methanol are detected. 
Column densities and rotation temperatures are calculated for all the species, including 
an upper limit to the ethanol abundance.
A comparison of the chemical composition of the nuclear environment of NGC\,253 with those of selected nearby galaxies 
demonstrates the chemical resemblance of IC\,342 and NGC\,4945 to that of NGC\,253.
On the other hand, the chemistries characterizing NGC\,253 and M\,82 are clearly different.
We also present a comparison of the chemical composition of NGC\,253 with those observed in Galactic prototypical sources.
The chemistry of NGC\,253 shows a striking similarity with the chemistry observed toward the Galactic center molecular 
clouds, which are thought to be dominated by low-velocity shocks.
This resemblance strongly suggests that the heating in the nuclear environment of NGC\,253 is dominated by the same mechanism 
as that in the central region of the Milky Way.

\end{abstract}
\keywords{ surveys ---  galaxies: abundances ---  galaxies: individual (NGC\,253) ---  galaxies: ISM ---  galaxies: nuclei ---  galaxies: starburst ---  radio lines: galaxies}
%
\section{Introduction}
Our knowledge of the chemical composition of the interstellar medium (ISM) in the nuclei of external galaxies has so far been 
restricted to only $\sim 30$ molecular species
(see Sect.~\ref{sect.census}),
which is still far from the 129 molecules detected in the interstellar and circumstellar medium within our Galaxy \citep{Lovas04}.
The initial detection of most of these 129 molecules has been the result of unbiased frequency scans towards specific Galactic
sources, such as massive star-forming regions (e.g. Sgr B2(M), \citealt{Cummins}; OriMC-1, \citealt{Lee01}), 
cold molecular clouds \citep[e.g. TMC-1,][]{Kaifu04}, and evolved stars \citep[e.g. IRC+10216,][]{Cerni00}. 
Spectral line scans towards Galactic sources focused mainly on the hot cores associated with massive star-forming regions such 
as Sgr\,B2 and Orion-KL due to their complex chemistry as well as the brightness of their lines. 
Extragalactic molecular spectroscopy has been so far limited to selecting the strongest features seen in the Galactic center and
disk sources and observing them towards well selected extragalactic targets.
The most ambitious census of the molecular content of an external galaxy is the compilation of data toward the starburst galaxy
NGC\,4945 by \citet{Wang04}.
Their multi-line study combines observations of 19 molecular species previously known in other extragalactic sources.

Up to now unbiased molecular line surveys of extragalactic sources do not exist in spite of providing
a powerful tool for determining the physical parameters of the molecular ISM.
Observations of a large number of molecular lines with similar angular resolution allow us to describe the chemical 
complexity of the source which provides fundamental information on the physical processes heating the medium in heavily obscured
regions.
Additionally, frequency scans lead to 
serendipitous detections of new species.

The nuclear starburst 
of the galaxy NGC\,253, the target of this survey, is one of the brightest and most prolific (in terms of molecular line detections)
extragalactic molecular line sources.
The Sculptor galaxy NGC\,253 is an almost edge-on barred spiral \citep[$i=72^\circ-78^\circ$,][]{Puche91,Pence81} classified as type SAB(s)c \citep{Vaucou91} or
Sc(s) \citep{Sandage87}.
At a distance of $\sim 3$\,Mpc \citep[e.g.][]{Mouhcine05}, NGC\,253 is one of the nearest archetypes of nuclear starburst galaxies \citep{Rieke80}.
Its nuclear region contains one of the brightest extragalactic IRAS sources with a 100$\mu$m flux of 1860\,Jy \citep{Soifer89}.
In fact, most of the overall IR luminosity
of this galaxy
\citep[$L_{\rm IR}=2.1\,10^{10}\,{\rm  L}_\odot, SFR_{\rm IR}=3.6\, \rm M_\odot yr^{-1}$,][]{Strickland04} stems from the 
regions of intense massive star formation within its central few hundred parsecs.
Violent massive star formation is also revealed by the high supernova rate of $0.05-0.3$\,yr$^{-1}$ in the nuclear region of this 
galaxy \citep{Mattila01,Ulvestad97}.
 
The high nuclear star formation activity is driven by huge amounts of molecular gas in the central few hundred parsecs of NGC\,253 
\citep[$\rm 1.3-2.6\,10^{9}M_\odot$,][]{Canzian88,Mauers96}, likely powered by the bar structure first detected at near-infrared
wavelengths \citep{Scoville85, Forbes92}. 
Position-velocity diagrams from high-resolution observations of the nuclear region of NGC\,253 show that the molecular material
seems to follow orbital motions around the dynamical center which are interpreted as 
$x_1$ and $x_2$ orbits within the context of a bar-like potential
(\citealp[OH,][]{Turner85}; \citealp[CO,][]{Canzian88,Das01,Paglione04};
\citealp[HCN,][]{Paglione95}; \citealp[CS,][]{Peng96}; \citealp[SiO, H$^{13}$CO$^+$,][]{Burillo00}).

Here we present the first unbiased molecular line survey carried out towards a source outside the Milky Way.
Technically, this survey was driven by the availability of two new wideband (1\,GHz) receivers at 2\,mm and new 1\,GHz 
wide filter bank spectrometers, together with the large collecting area of the IRAM 30\,m telescope.
Species such as NO, NS, SO$_2$, H$_2$S and H$_2$CS, detected for the first time towards a starburst environment, 
were previously reported by \citet{Martin03, Martin05}.
Additional new molecules together with a full analysis of the data are presented.

\section{Observations and results}
\clearpage
\begin{figure}[!t]
\centering
	\includegraphics[angle=-90,width=\linewidth]{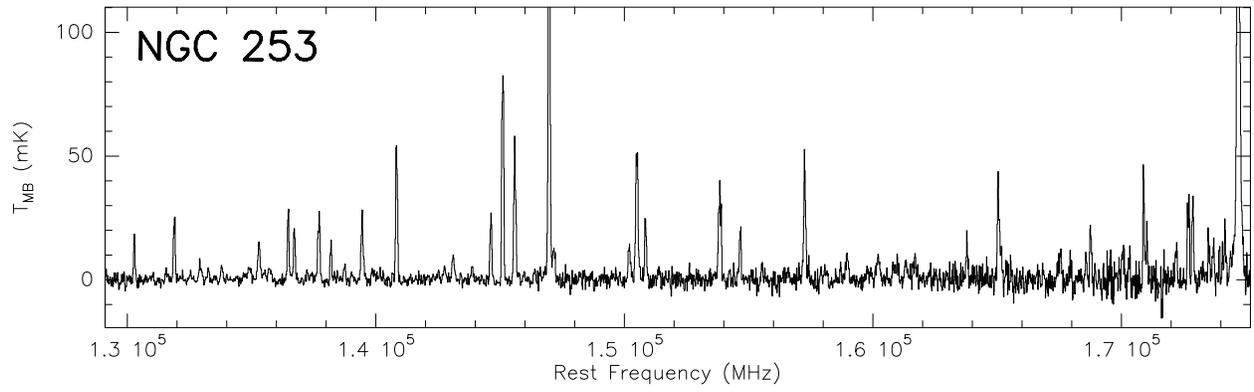}
	\caption{Complete 2\,mm spectral frequency scan toward the nuclear region of NGC\,253.}
	\label{fig.SurveyTot}
\end{figure}
\begin{figure}[!t]
\centering
	\includegraphics[angle=-90,width=0.9\linewidth]{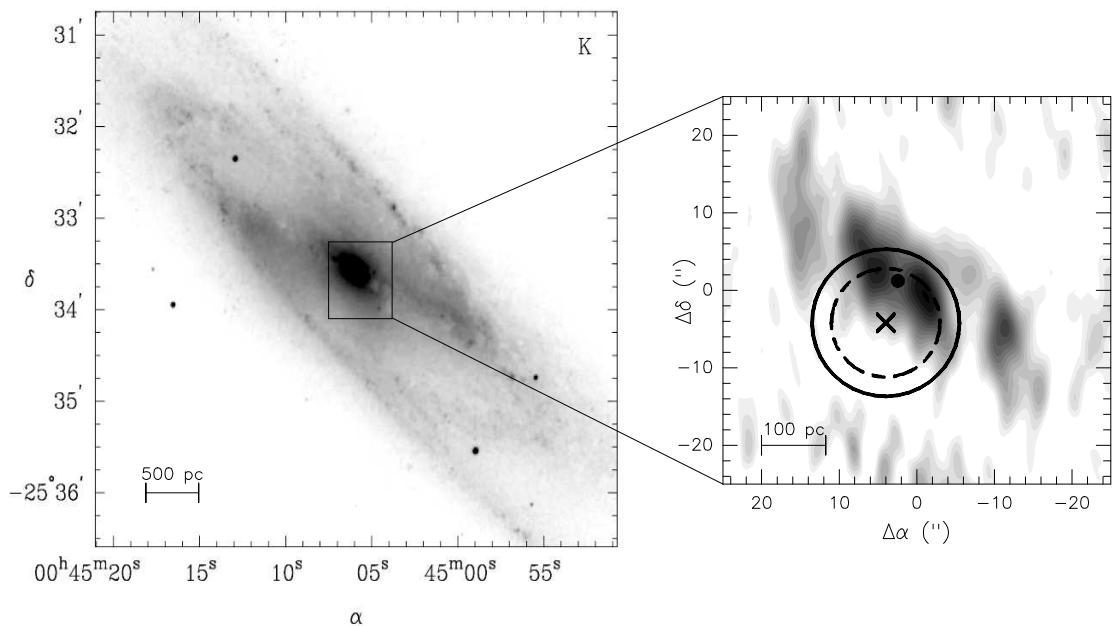}
	\caption{
	Grey-scale, $K-$band image of NGC\,253 \citep[left panel,][]{Engel98} and high-resolution SiO emission
	\citep[right panel,][]{Burillo00}.
	Beam sizes of the IRAM 30\,m telescope for the extreme frequencies observed in the survey are shown as circles on top of 
	the SiO map.
	The continuous line shows the 19$''$ beam at 129\,GHz and the dashed line the 14$''$ beam at 175\,GHz.
	A cross indicates the nominal position of the 2\,mm scan in this work and a dot the dynamical center as derived
	by \citet{Burillo00}.}
	\label{fig.Beamsize}
\end{figure}

\clearpage
The first molecular frequency scan of an external galaxy was carried out at 2\,mm with the 30\,m IRAM telescope at Pico
Veleta, Spain, between 2001 and 2004.
It covers $\sim$86\% of the observable 2\,mm atmospheric window, from 129.1 to 175.2\,GHz (Fig.~\ref{fig.SurveyTot}).
At these frequencies, the telescope beamwidth ranges from $19''$ to $14''$.
Fig.~\ref{fig.Beamsize} shows the size of the 30\,m beam on top of an interferometric map of SiO emission \citep{Burillo00}. 
A $K$-band image of NGC\,253 \citep{Engel98} is displayed in grey-scale to illustrate the size of the observed region 
relative to the galaxy.
\clearpage
\clearpage

The observations were pointed at the position  $\rm \alpha_{J2000}=00^h47^m33\fs 3,\,\delta_{J2000}=-25^\circ17'23''$.
This position is $\sim 6''$ south-east of the dynamical center of NGC\,253.
A pointing accuracy of $\sim 3''$ was achieved by measuring cross-scans on nearby continuum sources every $\sim 2$ hours. 
The observations were carried out in a wobbler-switching mode with a symmetrical beam throw of
$4'$ in azimuth and a switching frequency of 0.5\,Hz.

Each of the two available SIS receivers working at 2\,mm with orthogonal polarizations provides a 1\,GHz bandwidth. 
The receivers were tuned to adjacent frequencies, with an overlap of 100\,MHz, to cover
an instantaneous bandwidth of 1.9\,GHz for each frequency setup.
As spectrometers we used two $256\times4$\,MHz filterbanks, providing a velocity resolution between
7 and 9\,${\rm km\,s}^{-1}$ at the observed frequencies. 

\begin{figure}[!th]
\begin{center}
	\includegraphics[width=0.5\linewidth, angle=0]{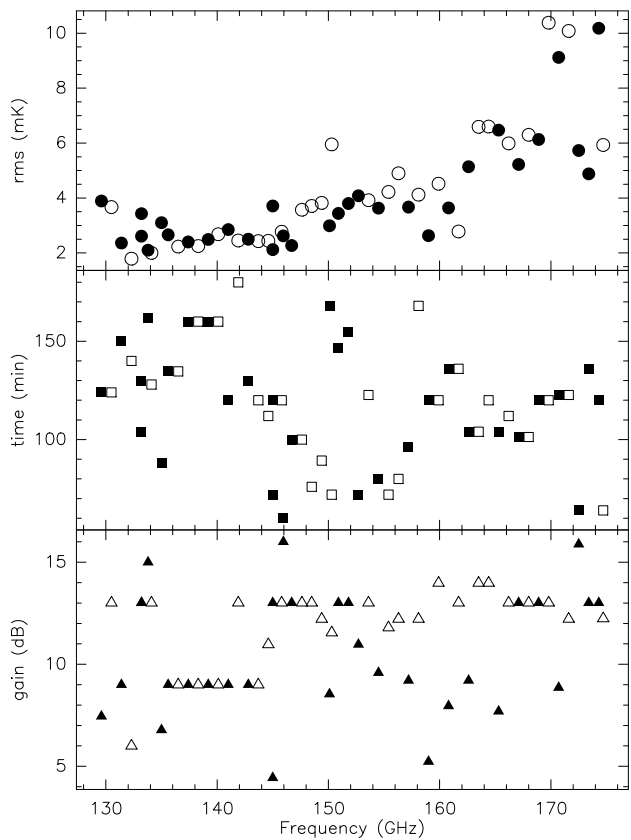}
\caption {Summary of the observational parameters of individual frequency setups. 
Open and filled symbols refer to each of the 2\,mm SIS receivers.
{\it Upper panel:} rms of the spectrum in 7--9\,km\,s$^{-1}$ wide channels in $T_{\rm mb}$ after subtracting a linear baseline; 
{\it Central panel:} Total integration time which includes the time spent on the reference position; 
{\it Lower panel:} Measured rejection of the image (upper) sideband.}
\label{fig:datastatistics}
\end{center}
\end{figure}

The rejection of the image band (the upper side band) was typically $\sim$10\,dB so that only the very few strongest lines were also
detected from the image band. 
The image sideband rejection was calculated for each frequency setup (lower panel in Fig.~\ref{fig:datastatistics}) by measuring the difference 
in power between a hot and a cold load, first with full rejection of the upper side band and then with full rejection of the lower side 
band by using a Martin-Puplett interferometer. 
The image gain is then computed as the ratio between these two measurements.
We assume this gain ratio to be constant throughout the whole 1\,GHz frequency band covered by each backend.

The spectra were calibrated with a standard dual load system. The temperature scale of the spectra is in $T_{\rm MB}$
obtained as
\begin{equation}
T_{\rm MB}=(F_{\rm eff}/B_{\rm eff}) T_{\rm A}^*,
\label{eq:ta2tmb}
\end{equation}
where the forward efficiency ($F_{\rm eff}$) is 0.93 at 2\,mm. The beam efficiency was calculated 
for each frequency with the Ruze function,
\begin{equation}
B_{\rm eff} = 1.2 \epsilon \exp{[-(4\pi R \sigma/ \lambda)^2]},
\end{equation}
using $\epsilon=0.69$, $R \sigma=0.07$ and $\lambda$ in mm.

At the latitude of Pico Veleta Observatory, NGC\,253 is at elevations $>20^\circ$ only for about four hours.
A complete observing session with about 120 minutes effective observing time was spent on each frequency setup, 
out of which half the time was spent on source.
The integration time for each frequency setup is shown in the central panel of Fig.~\ref{fig:datastatistics}.

The observing mode combined with the stability of the system provided high quality baselines. Only
linear baselines were subtracted from the spectra. 
The rms of the residuals after subtracting the fitted line profiles on each frequency setup is shown in the upper panel of
Fig.~\ref{fig:datastatistics}. The values are mostly of order of 2--4\,mK in $\sim 8\, \rm km \,s^{-1}$ wide channels, 
increasing 
up to $\sim$6--10\,mK at the highest measured frequencies.

In the scanned 46\,GHz wide band we have detected a total of 111 lines, the weakest of which have an
intensity of $\sim 3$\,mK. 
The spectral density of detected lines is 2.4 features/GHz.
Since the Doppler linewidth in NGC\,253 is about 200\,km\,s$^{-1}$, or 100\,MHz, 
we are not limited by line blending, but by a lack of sensitivity.
For targeted searches with longer integration time it may still be possible to detect and identify lines at levels of $\leq 1$\,mK.
 
\begin{figure}[!p]
\centering
	\includegraphics[angle=0,width=0.4\linewidth]{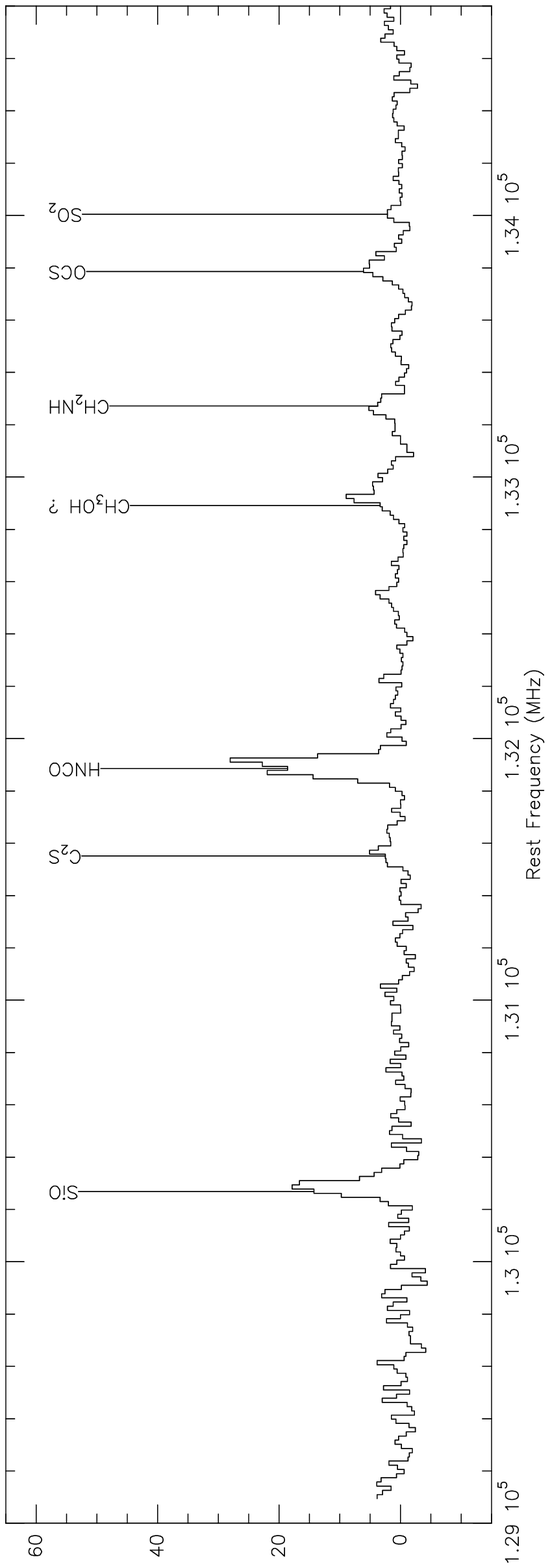}
	\includegraphics[angle=0,width=0.4\linewidth]{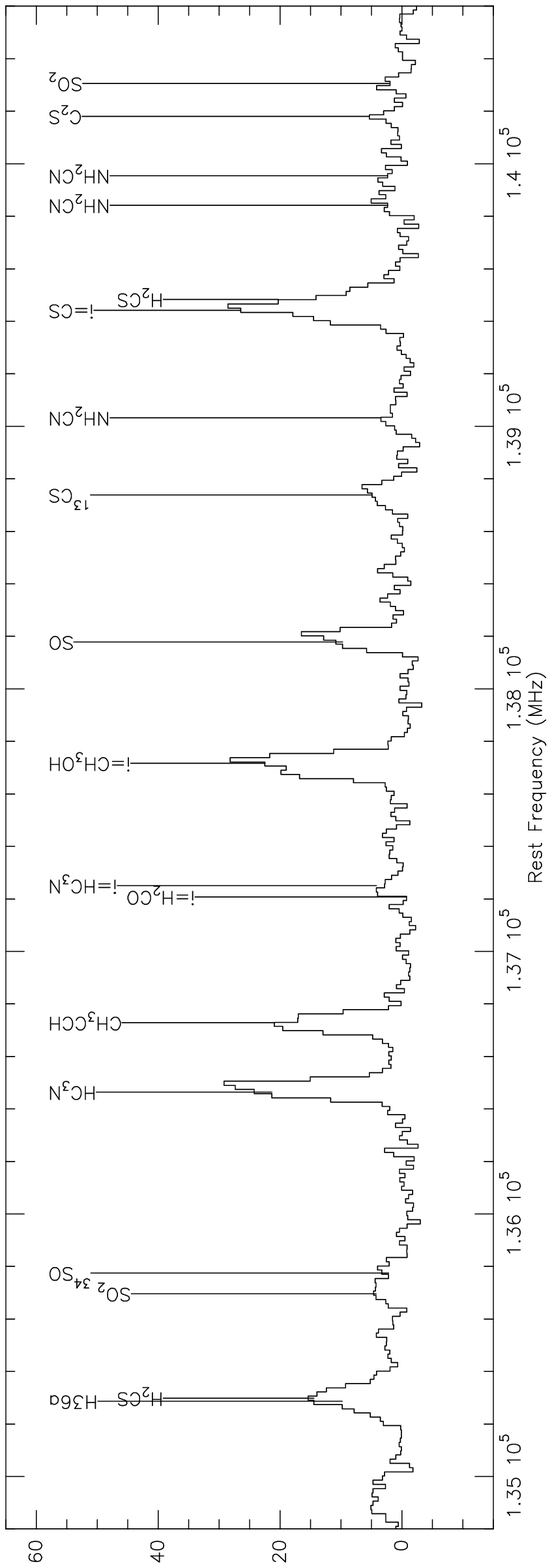}
	\caption{
        2\,mm spectral line survey composite towards the 
	nuclear region of NGC\,253. The resolution has been smoothed to four channels which is equivalent to velocity
	resolutions of 35 to 27\,km\,s$^{-1}$.
	Line identifications are plotted
	on top of the spectra. Species labeled with $\it i=molecule$ correspond
	to lines observed from the image band.}
	\label{fig.Survey}
\end{figure}

Fig.~\ref{fig.Survey} shows the full 2\,mm spectral line scan of NGC\,253 with the line identifications superposed on the spectra.
In Table~\ref{tab.fits} we are listing the parameters derived from Gaussian fits to the detected lines.
We identify 25 molecular species and 3 hydrogen recombination lines.
\clearpage

\clearpage
\begin{center}
	\includegraphics[angle=0,width=0.4\linewidth]{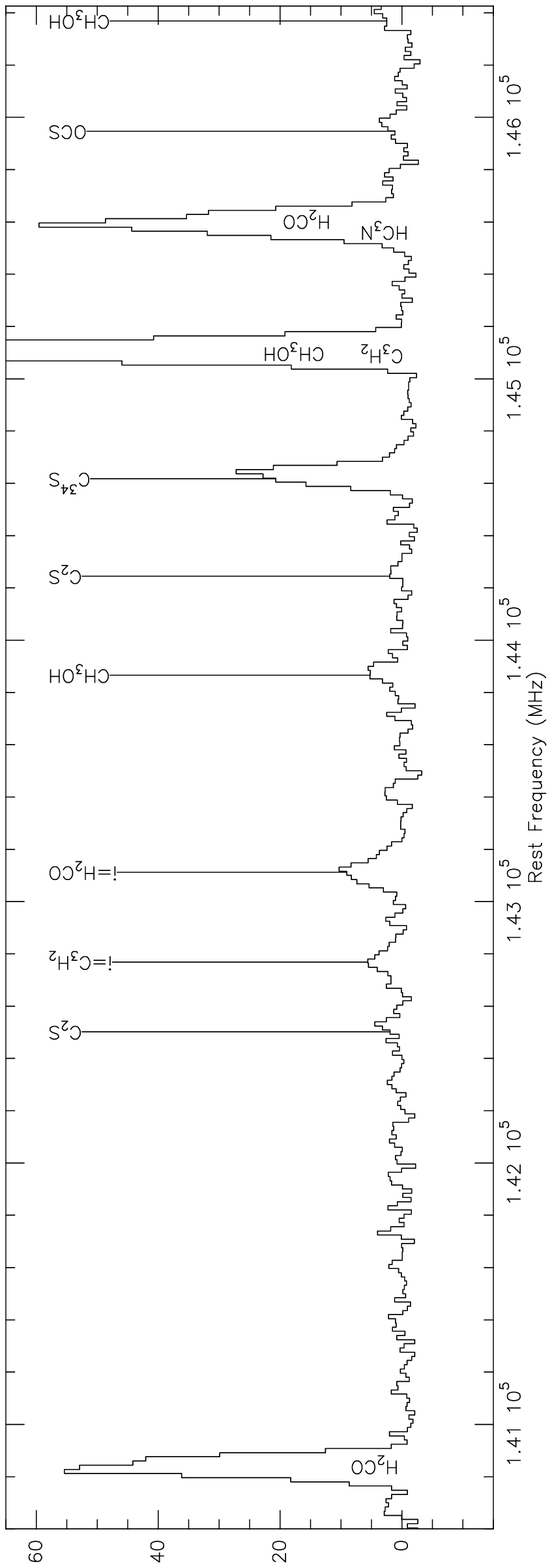}
	\includegraphics[angle=0,width=0.4\linewidth]{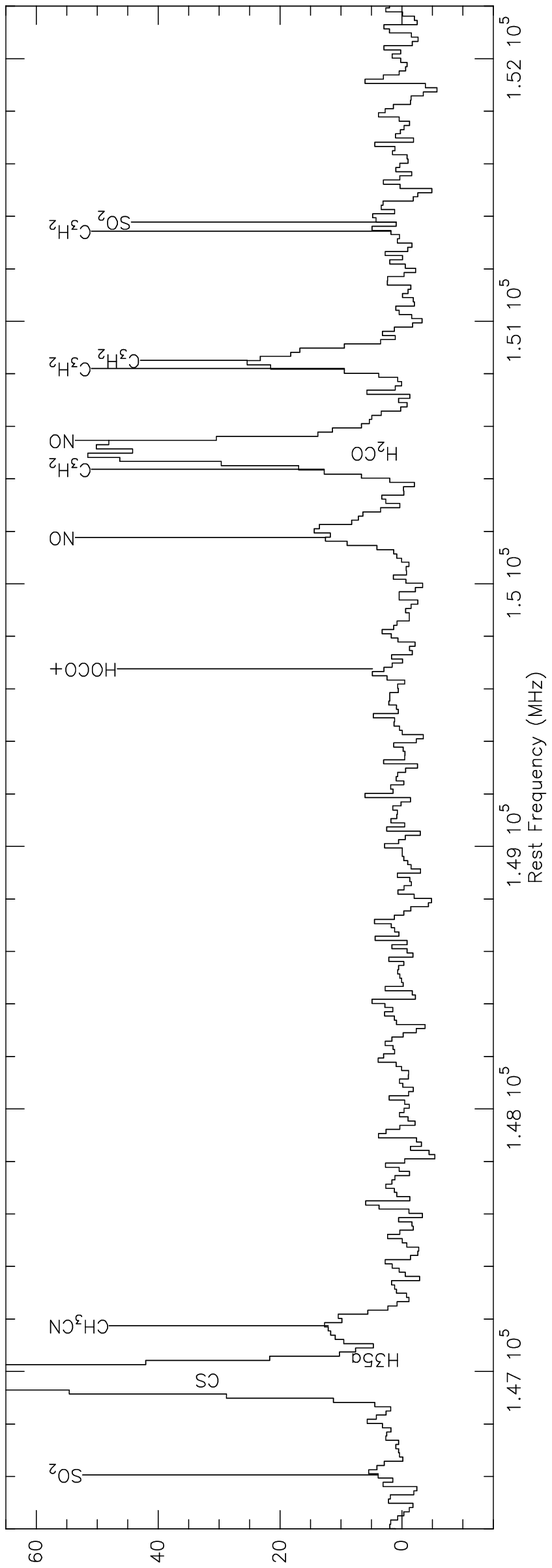}
\\Fig. 4. --- (Cont.)\clearpage
	\includegraphics[angle=0,width=0.4\linewidth]{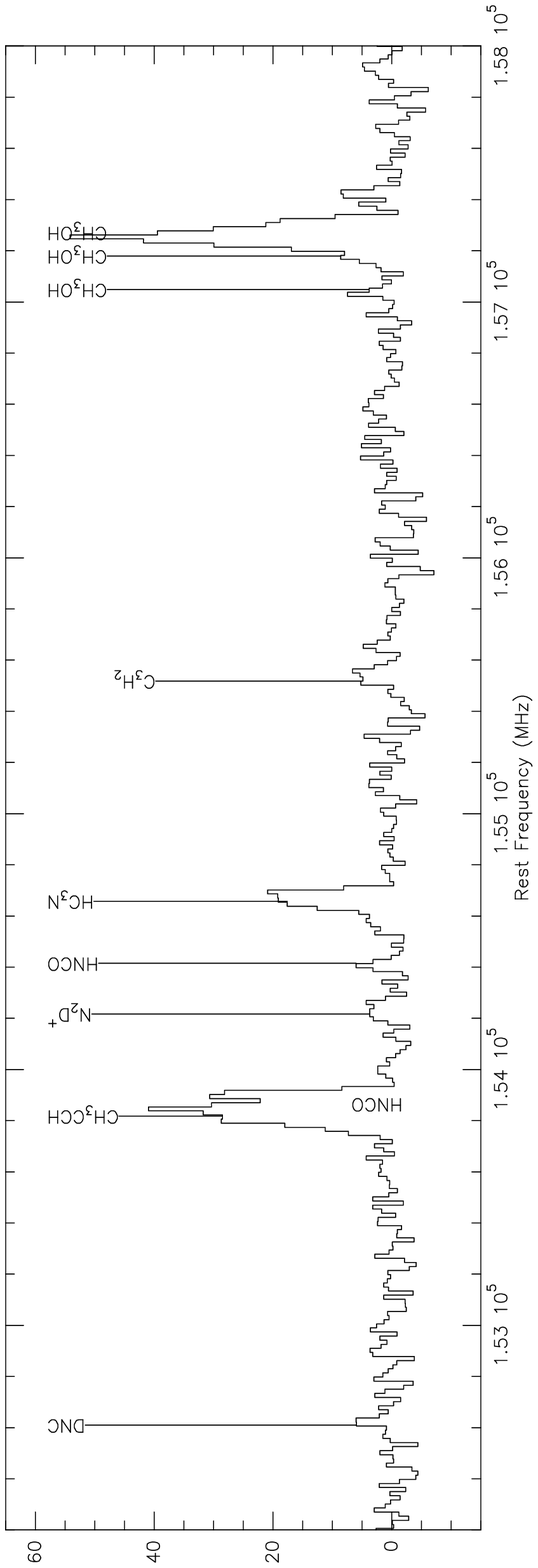}
	\includegraphics[angle=0,width=0.4\linewidth]{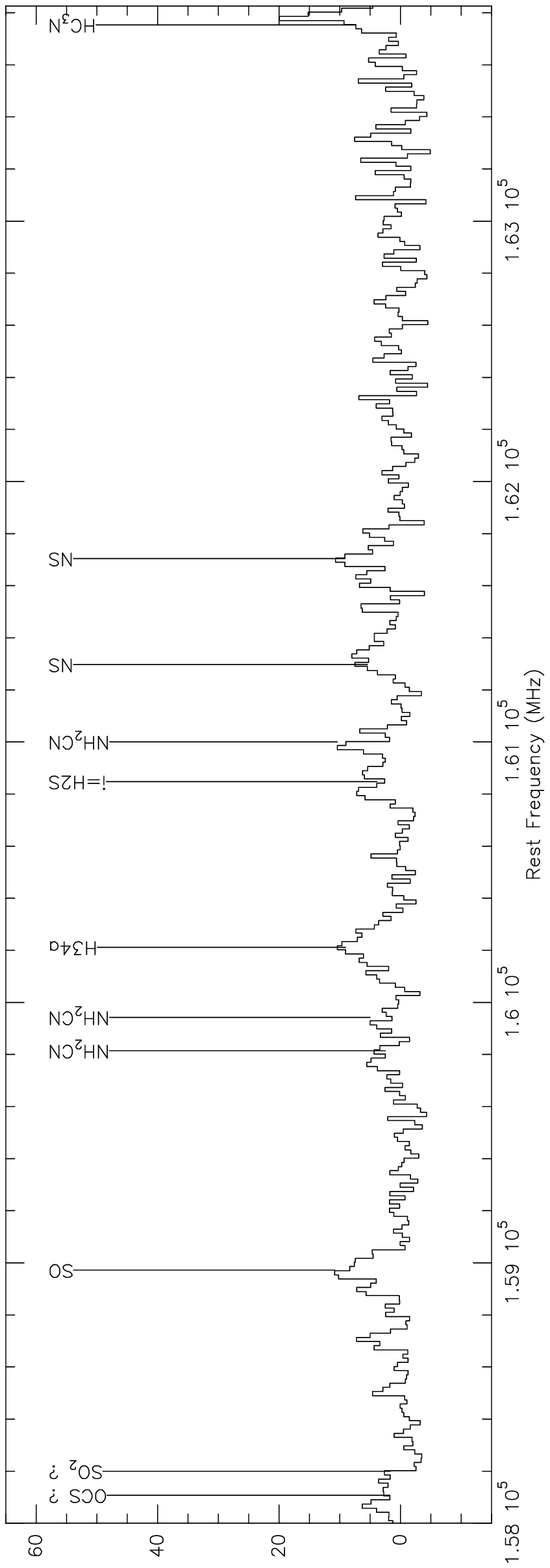}
\\Fig. 4. --- (Cont.)\clearpage
	\includegraphics[angle=0,width=0.4\linewidth]{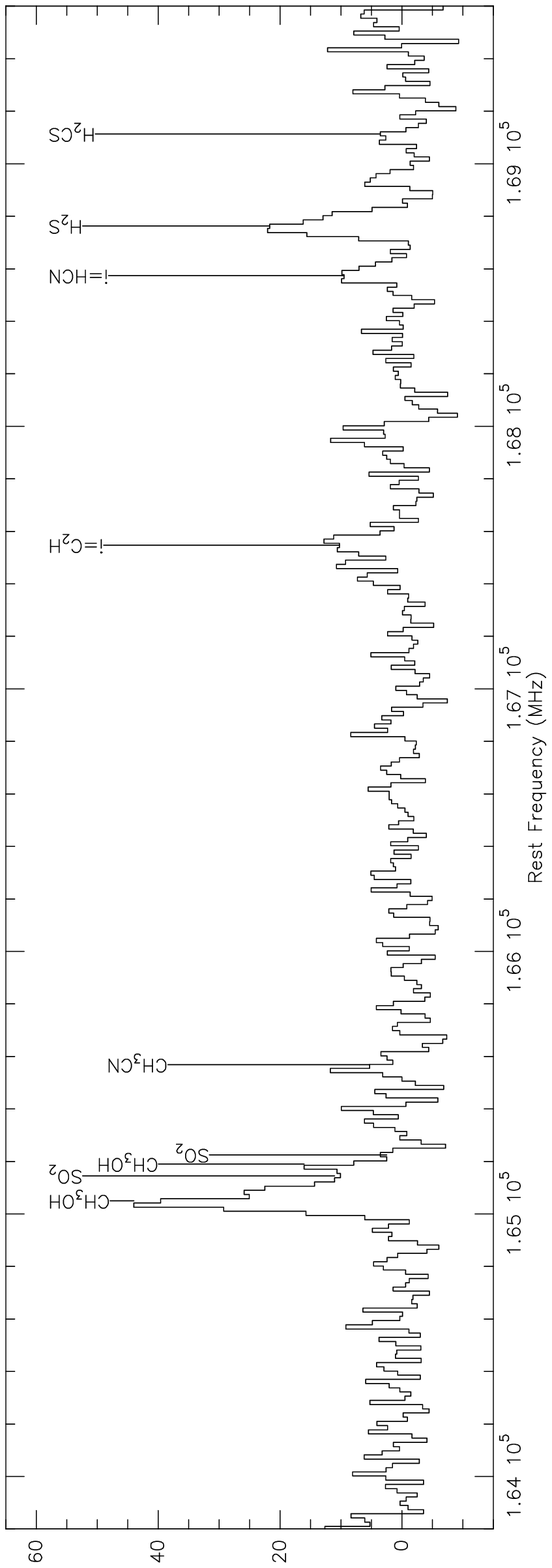}
	\includegraphics[angle=0,width=0.4\linewidth]{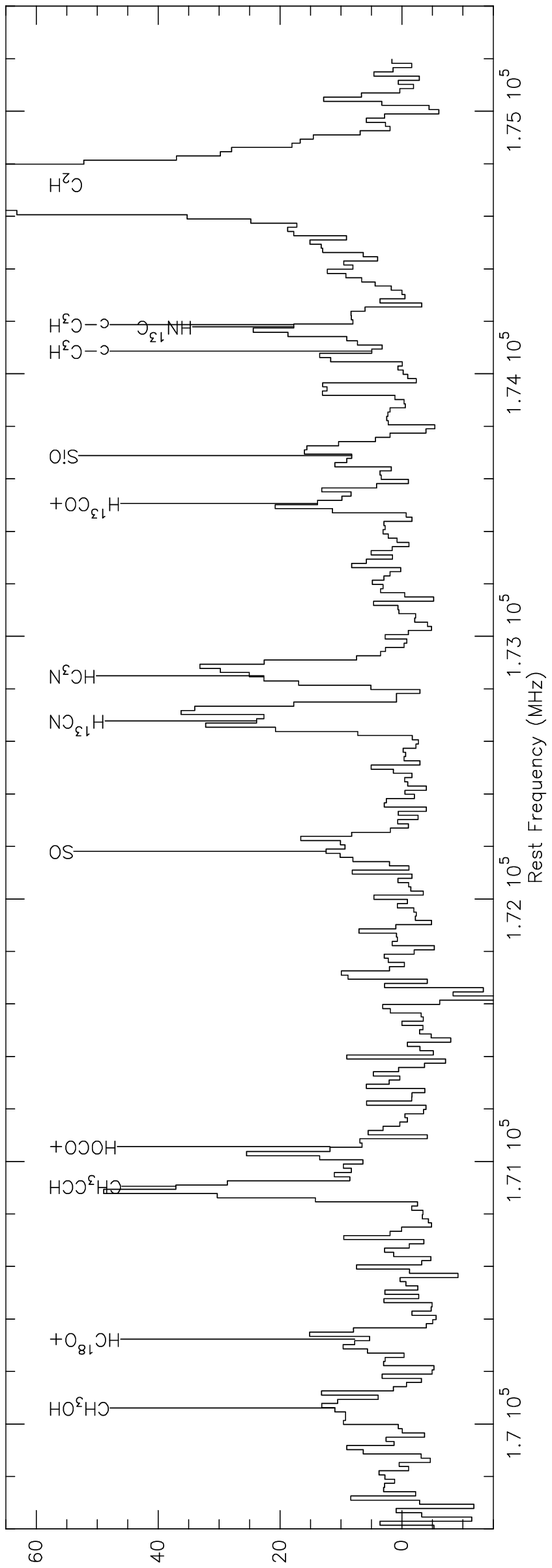}
\\Fig. 4. --- (Cont.)
\end{center}
\clearpage

\begin{table}[!p]
\caption{Results from Gaussian fits to the observed lines}
\begin{center}
\scriptsize
\begin{tabular}{l l c r r r r}
\tableline
\tableline
Molecule                                   &$\nu$        &   note      & $\int{T_{\rm MB}{\rm d}v}$ & $V_{\rm LSR}$             &  $\Delta v_{1/2}$               & $T_{\rm MB}$ \\
Transition                                 &(MHz)        &             &  mK\,km\,s$^{-1}$          & km\,s$^{-1}$              &  km\,s$^{-1}$                   & mK         \\
\tableline
{\bf SiO} $3-2$\,v=0                       & 130268.6    &             & 1580 ( 50)                 & 183                       &   76                            &  18 \\
                                           &             &             & 1170 ( 50)                 & 260                       &   91                            &  12 \\
{\bf C$_2$S} $11_{10}-10_9$                & 131551.9    &             &  640 ( 90)                 & 230                       &  151                            &   4.0  \\
{\bf HNCO} $6_{0,6}-5_{0,5}$               & 131885.7    &             & 2400 (150)                 & 182                       &  72                             &  32  \\
                                           &             &             & 2000 (180)                 & 284                       &  80                             &  24  \\
{\bf CH$_2$NH}  $2_{1,1}-1_{1,0}$          & 133272.1    & hf          &  420 ( 40)                 & 187                       &  83\tablenotemark{b}            &   4.8  \\
                                           &             &             &  450 ( 40)                 & 283                       &  83\tablenotemark{b}            &   5.1  \\
{\bf OCS} $11-10$                          & 133785.9    &             & 1220 (130)                 & 212                       &  200                            &   5.8  \\
{\bf SO$_2$} $8_{2,6}-8_{1,7}$             & 134004.8    &             & 300  ( 70)                 & 242                       &   91                            &  3.1  \\
{\bf H36$\alpha$}                          & 135286.0    & b s         & 2800 (...)                 & 250\tablenotemark{a}      &  282\tablenotemark{a}           &  9.3  \\  
{\bf H$_2$CS} $4_{1,4}-3_{1,3}$            & 135297.8    & b           & 1040 ( 90)                 & 221                       &  158                            &   6.2  \\
{\bf SO$_2$} $5_{1,5}-4_{0,4}$             & 135696.0    &             & 740  (160)                 & 245\tablenotemark{a}      &  140\tablenotemark{a}           &  4.2  \\
{\bf $^{34}$SO} $4_3-3_2$                  & 135775.3    & b           &  300 ( 60)                 & 180\tablenotemark{a}      &   73\tablenotemark{a}           &   3.8  \\
                                           &             & b           &  200 ( 70)                 & 274\tablenotemark{a}      &   88\tablenotemark{a}           &   2.1  \\
{\bf HC$_3$N} 15-14                        & 136464.4    &             & 2410 (130)                 & 183                       &   77                            &  29 \\
                                           &             &             & 2000 (300)                 & 271                       &   85                            &  22 \\
{\bf CH$_3$CCH} $8_k-7_k$                  & 136728.0    &             & 3410 (110)                 & 255                       & 158                             &  20  \\
{\bf SO} $4_3-3_2$                         & 138178.5    &             & 1400 (200)                 & 180                       &   73                            &  18  \\
                                           &             &             & 1100 (200)                 & 274                       &   88                            &  12  \\
{\bf $^{13}$CS} $3-2$                      & 138739.3    &             &  550 ( 70)                 & 188                       &   70\tablenotemark{b}           &   7.4  \\
                                           &             &             &  390 ( 50)                 & 274                       &   70\tablenotemark{b}           &   5.2  \\
{\bf NH$_2$CN} $7_{1,7}-6_{1,6}$           & 139032.0    &             &  580 (110)                 & 280\tablenotemark{a}      & 155\tablenotemark{a}            &   3.5  \\
{\bf H$_2$CS} $4_{1,3}-3_{1,2}$            & 139483.4    & b           & 1340 (150)                 & 221                       &  176                            &   7.2  \\
{\bf NH$_2$CN} $7_{0,7}-6_{0,6}$           & 139842.1    &             &  490 (160)                 & 280\tablenotemark{a}      & 155\tablenotemark{a}            &   3.0  \\
{\bf NH$_2$CN} $7_{0,7}-6_{0,6}$           & 139954.4    & m           &  520 (160)                 & 279                       & 140                             &   3.5  \\
{\bf C$_2$S} $10_{11}-9_{10}$              & 140180.7    &             &  470 (130)                 & 250                       &   85                            &   5.2  \\
{\bf SO$_2$} $6_{2,4}-6_{1,5}$             & 140306.1    &             & 510  ( 60)                 & 248                       &  115                            &  4.1  \\
{\bf H$_2$CO} $2_{1,2}-1_{1,1}$            & 140839.5    &             & 4250 (140)                 & 186                       &  104\tablenotemark{b}           &  38 \\
                                           &             &             & 6050 (150)                 & 288                       &  104\tablenotemark{b}           &  55 \\
{\bf C$_2$S} $11_{11}-10_{10}$             & 142501.7    &             &  430 ( 70)                 & 192                       &   85\tablenotemark{a}           &   4.7  \\
{\bf CH$_3$OH}  $3_{1,3}-2_{1,2}$ A$+$     & 143865.7    &             & 1000 (130)                 & 225                       &  168                            &   5.6  \\
{\bf C$_2$S} $12_{11}-11_{10}$             & 144244.8    &             &  290 (120)                 & 210                       &  112                            &   2.4  \\
{\bf C$^{34}$S} $3-2$                      & 144617.1    &             & 3000 (300)                 & 188                       &  105                            &  27  \\
                                           &             &             & 1500 (300)                 & 284                       &   86                            &  16  \\
{\bf c-C$_3$H$_2$} $3_{1,2}-2_{2,1}$       & 145089.5    & b s         & 1750 (...)                 & 234\tablenotemark{a}      &  156\tablenotemark{a}           &  11 \\
{\bf CH$_3$OH}  3-2                        & 145103.2    & b m         & 9400 (200)                 & 185                       &  115                            &  77  \\
                                           &             &             & 5760 (190)                 & 294                       &   96                            &  57  \\
{\bf HC$_3$N} 16-15                        & 145560.9    & b s         & 2100 (...)                 & 183\tablenotemark{a}      &   77\tablenotemark{a}           &  26 \\
                                           &             & b s         & 1700 (...)                 & 271\tablenotemark{a}      &   85\tablenotemark{a}           &  19 \\
{\bf H$_2$CO} $2_{0,2}-1_{0,1}$            & 145602.9    & b           & 3400 (100)                 & 177                       &   90\tablenotemark{b}           &  35 \\
                                           &             & b           & 3300 (100)                 & 283                       &   90\tablenotemark{b}           &  34 \\
\tableline
\end{tabular}
\end{center}
\label{tab.fits}
\end{table}

\begin{table}[!p]
\addtocounter{table}{-1}
\caption{(Cont.)}
\begin{center}
\scriptsize
\begin{tabular}{l l c r r r r}
\tableline
\tableline
Molecule     		                   &$\nu$        &   note      & $\int{T_{\rm MB}{\rm d}v}$ & $V_{\rm LSR}$             &  $\Delta v_{1/2}$               & $T_{\rm MB}$ \\
Transition   		                   &(MHz)        &             &  mK\,km\,s$^{-1}$          & km\,s$^{-1}$              &  km\,s$^{-1}$                   & mK         \\
\tableline
{\bf OCS} $12-11$                          & 145946.8    &             &  580 (170)                 & 212\tablenotemark{a}      &  210\tablenotemark{a}           &   2.6  \\
{\bf CH$_3$OH}  $3_{1,2}-2_{1,2}$ A$- -$   & 146368.3    &             & 1020 (150)                 & 196                       &  139                            &   6.9  \\
{\bf SO$_2$} $4_{2,2}-4_{1,3}$             & 146605.5    &             & 800  (170)                 & 241                       &  148                            &  5.1  \\
{\bf CS} $3-2$                             & 146969.0    & b           &11900 (200)                 & 185                       &  100                            & 111  \\
                                           &             & b           &13700 (200)                 & 288                       &  117                            & 110  \\
{\bf H35$\alpha$}                          & 147046.8    & b s         & 2800 (...)                 & 250\tablenotemark{a}      &  282\tablenotemark{a}           &  9.3  \\
{\bf CH$_3$CN} $8_k-7_k$                   & 147174.5    & b m         & 2040 (180)                 & 258                       &  165                            &  12  \\
{\bf HOCO$^+$} $7_{0,7}-6_{0,6}$           & 149675.8    &             &  410 (110)                 & 281                       &  68                             &   5.6  \\
{\bf NO} $\frac{3}{2}-\frac{1}{2}\,\Pi^+$  & 150176.5    & hf          & 3400 (500)                 & 244                       &  158                            &  10 \\
{\bf c-C$_3$H$_2$} $2_{2,0}-1_{1,1}$       & 150436.5    & b s         &  600 (...)                 & 234\tablenotemark{a}      &  156\tablenotemark{a}           &   3.6 \\
{\bf H$_2$CO} $2_{1,1}-1_{1,0}$            & 150498.3    & b           & 5300 (100)                 & 180                       &   99\tablenotemark{b}           &  50 \\
                                           &             & b           & 5100 (100)                 & 286                       &   99\tablenotemark{b}           &  48 \\
{\bf NO} $\frac{3}{2}-\frac{1}{2}\,\Pi^-$  & 150546.5    & hf b s      & 3400 (...)                 & 244\tablenotemark{a}      &  158\tablenotemark{a}           &  10 \\
{\bf c-C$_3$H$_2$} $4_{0,4}-3_{1,3}$       & 150820.6    & b           & 1150 (...)                 & 234\tablenotemark{b}      &  156\tablenotemark{b}           &   7.0 \\
{\bf c-C$_3$H$_2$} $4_{1,4}-3_{0,3}$       & 150851.9    & b           & 3380 (110)                 & 234\tablenotemark{b}      &  156\tablenotemark{b}           &  20 \\
{\bf c-C$_3$H$_2$} $5_{1,4}-5_{0,5}$       & 151343.8    & b s         &  165 (...)                 & 234\tablenotemark{a}      &  156\tablenotemark{a}           &   1.0 \\
{\bf SO$_2$} $2_{2,0}-2_{1,1}$             & 151378.6    & b           & 590  (170)                 & 225                       &  158                            &  3.5  \\
{\bf DNC} $2-1$                            & 152609.7    &             &  490 (160)                 & 225                       &  61                             &  7.5  \\
{\bf CH$_3$CCH} $9_k-8_k$                  & 153817.2    & b           & 4300 (200)                 & 271                       & 157                             &  26  \\
{\bf HNCO} $7_{0,7}-6_{0,6}$               & 153865.0    &             & 2660 (150)                 & 183                       &  72                             &  34  \\
                                           &             & b s         & 2240 (...)                 & 284\tablenotemark{a}      &  80\tablenotemark{a}            &  26  \\
{\bf N$_2$D$^+$}  $2-1$                    & 154217.0    &             &  570 (170)                 & 219                       & 120                             &  4.5  \\
{\bf HNCO} $7_{1,6}-6_{1,5}$               & 154414.7    &             &  540 (110)                 & 283\tablenotemark{a}      &  80\tablenotemark{a}            &   6.4  \\
{\bf HC$_3$N} 17-16                        & 154657.3    &             & 1640 (120)                 & 191                       &   72\tablenotemark{b}           &  21 \\
                                           &             &             & 1400 (110)                 & 272                       &   72\tablenotemark{b}           &  18 \\
{\bf c-C$_3$H$_2$} $3_{2,2}-2_{1,1}$       & 155518.2    &             &  820 (180)                 & 205                       &  118                            &   6.6 \\
{\bf CH$_3$OH}  $6_{0,6}-6_{-1,6}$ E       & 157048.6    &             &  540 (120)                 & 275                       &  103\tablenotemark{b}           &   5.0  \\
{\bf CH$_3$OH}  $5_{0,5}-5_{-1,5}$ E       & 157178.9    & b           &  960 (150)                 & 242                       &  103\tablenotemark{b}           &   8.8  \\
{\bf CH$_3$OH}  $J_{0,k}-J_{-1,k}$ E       & 157270.7    & b m         & 2110 (160)                 & 183                       &  103\tablenotemark{b}           &  19  \\
              $J=1..4; k=J$                &             &             & 5700 (200)                 & 290                       &  103\tablenotemark{b}           &  52  \\
{\bf SO} $3_4-2_3$                         & 158971.7    &             &  700 (200)                 & 180\tablenotemark{a}      &   92                            &   7.2  \\
                                           &             & b           & 1000 (200)                 & 274\tablenotemark{a}      &   84                            &  11  \\
{\bf NH$_2$CN} $8_{0,8}-7_{0,7}$           & 159814.6    &             &  750 (180)                 & 283\tablenotemark{b}      & 155\tablenotemark{b}            &   4.5  \\
{\bf NH$_2$CN} $8_{0,8}-7_{0,7}$           & 159942.7    & m           &  700 (200)                 & 283\tablenotemark{b}      & 158\tablenotemark{b}            &   4.1  \\
{\bf H34$\alpha$}                          & 160211.5    &             & 2800 (200)                 & 250                       &  282                            &  9.3  \\
{\bf NH$_2$CN} $8_{1,7}-7_{1,6}$           & 161000.3    & b           & 1430 (170)                 & 280                       & 155\tablenotemark{a}            &   8.6  \\
{\bf NS} $\frac{7}{2}-\frac{5}{2}\,e$      & 161297.2    & hf          & 1800 (500)                 & 199                       &  224                            &  3.1 \\
{\bf NS} $\frac{7}{2}-\frac{5}{2}\,f$      & 161697.2    & hf          & 2400 (400)                 & 249                       &  278                            &  3.4 \\
{\bf HC$_3$N} 18-17                        & 163753.4    &             & 1500 (400)                 & 185\tablenotemark{a}      &   73                            &  20 \\
                                           &             &             &  700 (300)                 & 272\tablenotemark{a}      &   98                            &   6.6 \\
\tableline                                                                                                         
\end{tabular}                                                                                                                                  
\end{center}                                                                                                                                   
\end{table}

\begin{table}[!p]
\addtocounter{table}{-1}
\caption{(Cont.)}
\begin{center}
\scriptsize
\begin{tabular}{l l c r r r r}
\tableline
\tableline
Molecule                                   &$\nu$        &   note      & $\int{T_{\rm MB}{\rm d}v}$ & $V_{\rm LSR}$             &  $\Delta v_{1/2}$               & $T_{\rm MB}$ \\
Transition                                 &(MHz)        &             &  mK\,km\,s$^{-1}$          & km\,s$^{-1}$              &  km\,s$^{-1}$                   & mK         \\
\tableline
{\bf CH$_3$OH}  $J_{1,k-1}-J_{0,k}$ E      & 165050.1    & b m         & 1700 (400)                 & 169                       &   72                            &  22  \\
              $J=1..3; k=J$                &             &             & 4700 (400)                 & 273                       &  100                            &  45  \\  
{\bf SO$_2$} $5_{2,4}-5_{1,5}$             & 165144.6    & b s         & 640  (...)                 & 248\tablenotemark{a}      &  115\tablenotemark{a}           &  5.2  \\
{\bf CH$_3$OH}  $4_{1,3}-4_{0,4}$ E        & 165190.5    & b           &  900 (200)                 & 291                       &   90\tablenotemark{a}           &   9.1  \\
{\bf SO$_2$} $7_{1,7}-6_{0,6}$             & 165225.4    & b s         & 900  (...)                 & 248\tablenotemark{a}      &  115\tablenotemark{a}           &  7.4  \\
{\bf CH$_3$CN} $9_k-8_k$                   & 165568.9    & b m         & 1200 (300)                 & 258\tablenotemark{a}      &  166\tablenotemark{a}           &   7.0  \\
{\bf H$_2$S}  $1_{1,0}-1_{0,1}$            & 168762.7    &             & 1520 (180)                 & 180                       &  100                            &  14 \\
                                           &             &             & 2250 ( 80)                 & 275                       &   88                            &  24 \\
{\bf H$_2$CS} $5_{1,5}-4_{1,4}$            & 169113.5    &             &  970 (200)                 & 270                       &  165\tablenotemark{a}           &   5.5  \\
{\bf CH$_3$OH}  $3_{2,3}-2_{1,3}$          & 170060.6    &             & 2400 (400)                 & 248                       &  178                            &  12  \\
{\bf HC$^{18}$O$^+$} $2-1$                 & 170322.7    &             &  500 (300)                 &  206                      &  46                             &  11  \\
                                           &             &             &  500 (200)                 &  285                      &  49                             &  10  \\
{\bf CH$_3$CCH} $10_k-9_k$                 & 170905.6    &             & 5100 (500)                 & 275                       & 102                             &  47  \\
{\bf HOCO$^+$} $8_{0,8}-7_{0,7}$           & 171055.9    & b           & 1700 (300)                 & 295                       &  66                              &  25  \\
{\bf SO} $4_4-3_3$                         & 172181.4    &             & 1300 (300)                 & 180\tablenotemark{a}      &   87                            &  14  \\
                                           &             &             &  900 (300)                 & 274\tablenotemark{a}      &   74                            &  11  \\
{\bf H$^{13}$CN} $2_k-1_k$                 & 172677.9    & hf          & 3200 (200)                 & 181                       &  74                             &  40  \\
                                           &             &             & 2400 (200)                 & 279                       &  66                             &  34  \\
{\bf HC$_3$N} 19-18                        & 172849.3    &             & 2800 (400)                 & 184                       &   74                            &  35 \\ 
                                           &             &             & 1800 (400)                 & 269                       &   75                            &  22 \\ 
{\bf H$^{13}$CO$^+$} $2-1$                 & 173506.7    &             &  860 ( 90)                 & 170                       &  60\tablenotemark{b}            &  14  \\
                                           &             &             & 1450 (100)                 & 269                       &  60\tablenotemark{b}            &  23  \\
{\bf SiO} $4-3$\,v=0                       & 173688.3    &             & 1070 (110)                 & 184\tablenotemark{a}      &   76\tablenotemark{a}           &  13 \\
                                           &             &             & 1000 (120)                 & 270\tablenotemark{a}      &   91\tablenotemark{a}           &  10 \\
{\bf c-C$_3$H} $3_{1,2}-2_{1,1}$           & 174086.1    & hf          &  900 ( 200)                & 180\tablenotemark{a}      &  41\tablenotemark{b}            &   6.5  \\
                                           &             &             & 1600 ( 500)                & 280\tablenotemark{a}      &  40                             &  12  \\
{\bf HN$^{13}$C} $2-1$                     & 174179.4    &             & 1050 (170)                 & 271                       &  60                             &  16  \\
{\bf C$_2$H} $2-1$                         & 174663.2    & hf          &30000 (6000)                & 162                       & 108                             &  82  \\       
                                           &             &             &27000 (6000)                & 282                       & 104                             &  88  \\      
\tableline                                                                                                         
\end{tabular}                                                                                                                                  
\end{center}                                                                                                                                   
\scriptsize
\tablecomments{(see Appendix~\ref{apen.Blended} for details):\\
b - Blended line.\\
s - Synthetic Gaussian to isolate a given feature. See text for details on the parameters used for deriving these profiles for each species.\\
hf - Hyperfine structure. Frequency and intensity refers to the main component of the group.\\
m - Multi-transition line. Frequency refers to the main component of the group.\\ }
\tablenotetext{a}{Parameter fixed in the Gaussian fit}
\tablenotetext{b}{Parameters forced to have the same value in the Gaussian fit}
\end{table}

\begin{table}[!p]
\caption{Physical parameters derived for all detected species. Source averaged column densities, rotation temperature and abundances
relative to H$_2$.}
\begin{center}
\scriptsize
\begin{tabular}{l l l r r}
\tableline
\tableline
\multicolumn{2}{c}{Molecule}                   & $N$ \tablenotemark{a}   & $T_{\rm rot}$  & [X]/[H$_2$] \tablenotemark{b}              \\
         &                                     & (cm$^{-2}$)             & (K)            &   ($\times 10^{-9}$)             \\
\tableline
SiO                        &180\,km\,s$^{-1}$  & $5.0(1.0)\,10^{12}$     &  7.4(0.7)    &     0.07              \\
                           &270\,km\,s$^{-1}$  & $3.6(0.9)\,10^{12}$     &  8.7(1.2)    &     0.05              \\
C$_2$S                     &                   & $1.4(1.2)\,10^{13}$     &  24(9)       &     0.2              \\
NH$_2$CN                   &                   & $1.2(0.5)\,10^{13}$     &  67(13)      &     0.2            \\  
CH$_3$CN                   &                   & $2.0(0.6)\,10^{13}$     &  9.6(0.7)    &     0.3              \\
c-C$_3$H                   & 130\,km\,s$^{-1}$ & $8.1(1.6)\,10^{12}$     &  12          &     0.1           \\ 
                           & 280\,km\,s$^{-1}$ & $1.5(0.3)\,10^{13}$     &  12          &     0.2           \\
c-C$_3$H$_2$               &                   & $3.0(6.0)\,10^{13}$     &   9(8)       &     0.4              \\
NS $^{c}$                  &                   & $4.3(0.6)\,10^{13}$     &   7.2(1.0)   &     0.6              \\
HC$_3$N                    &180\,km\,s$^{-1}$  & $1.8(0.5)\,10^{13}$     &  33(4)       &     0.3              \\
                           &270\,km\,s$^{-1}$  & $1.9(0.6)\,10^{13}$     &  24(3)       &     0.3              \\
HOCO$^+$                   &                   & $2.6(1.8)\,10^{13}$     &  12          &     0.4            \\
H$_2$CS                    &                   & $4.4(4.5)\,10^{13}$     &  11(4)       &     0.6              \\
SO$_2$                     &                   & $5.0(2.0)\,10^{13}$     &   15(2)      &     0.7              \\
H$_2$S                     &180\,km\,s$^{-1}$  & $2.3(1.1)\,10^{13}$     &  12          &     0.3              \\
                           &275\,km\,s$^{-1}$  & $3.5(1.4)\,10^{13}$     &  12          &     0.5              \\
CH$_2$NH                   & 190\,km\,s$^{-1}$ & $2.5(1.1)\,10^{13}$     &  12          &     0.4            \\
                           & 300\,km\,s$^{-1}$ & $2.8(0.9)\,10^{13}$     &  12          &     0.4            \\
HNC \tablenotemark{c}      & 270\,km\,s$^{-1}$ & $7.2(1.2)\,10^{13}$     &  12          &     1             \\
SO                         &180\,km\,s$^{-1}$  & $4.5(3.3)\,10^{13}$     &  40(24)      &     0.7              \\
                           &270\,km\,s$^{-1}$  & $3.1(2.8)\,10^{13}$     &  34(24)      &     0.5              \\
HNCO                       & 180\,km\,s$^{-1}$ & $5.7(2.7)\,10^{13}$     &  23(6)       &     0.8            \\
                           & 280\,km\,s$^{-1}$ & $4.8(2.5)\,10^{13}$     &  23          &     0.7            \\
HCO$^+$ \tablenotemark{c}  &170\,km\,s$^{-1}$  & $3.9(0.8)\,10^{13}$     &  12          &     0.6              \\
                           &270\,km\,s$^{-1}$  & $6.4(1.2)\,10^{13}$     &  12          &     1                \\
H$_2$CO                    &180\,km\,s$^{-1}$  & $6.9(0.9)\,10^{13}$     &  27(2)       &     1               \\
                           &285\,km\,s$^{-1}$  & $9.0(1.7)\,10^{13}$     &  34(4)       &     1               \\
OCS \tablenotemark{d}      &                   & $2.5(0.3)\,10^{14}$     &  17(2)       &     4               \\
HCN \tablenotemark{c}      & 180\,km\,s$^{-1}$ & $1.8(0.3)\,10^{14}$     &  12          &     2              \\
                           & 280\,km\,s$^{-1}$ & $1.4(0.2)\,10^{14}$     &  12          &     2             \\
CS \tablenotemark{c,d}     &180\,km\,s$^{-1}$  & $2.0(0.2)\,10^{14}$     &  9.7(0.4)    &     3               \\
                           &280\,km\,s$^{-1}$  & $1.4(0.2)\,10^{14}$     & 10.0(0.2)    &     2              \\
CH$_3$CCH                  &                   & $4.3(0.2)\,10^{14}$     &  63(20)      &     6             \\
CH$_3$OH                   &                   & $8.3(0.3)\,10^{14}$     &  11.6(0.2)   &     12              \\
C$_2$H                     & 160\,km\,s$^{-1}$ & $6.8(1.1)\,10^{14}$     &  12          &     10            \\
                           & 280\,km\,s$^{-1}$ & $5.6(0.9)\,10^{14}$     &  12          &     8             \\
NO \tablenotemark{d}       &                   & $4.0(0.4)\,10^{15}$     &   6(2)       &     60              \\
\tableline
\end{tabular}
\label{tab.MolecDensity}
\end{center}
\scriptsize
\tablenotetext{a}{Errors derived from the rotation diagrams without beam filling factor considerations (see Sect.~\ref{sect.uncertainty}).}
\tablenotetext{b}{Assumed $N({\rm H}_2)=6.7\,10^{22}{\rm cm}^{-2}$. See Table~\ref{tab:galaxh2}.}
\tablenotetext{c}{Calculated from the observed $^{13}$C isotope transition with $^{12}{\rm C}/^{13}{\rm C}\sim40$ from \citet{Henkel93}.}
\tablenotetext{d}{Using additional transitions measured by \citet{Martin03,Martin05}.}
\end{table}
\clearpage
\section{ Data analysis}

\subsection{Line identification and fitting}

In order to identify and fit the observed features we have used the rest frequencies from the molecular line catalogs 
of \citet{Lovas92,Lovas04} and \citet{Pickett}.
The existing molecular line surveys towards the Galactic center sources Sgr\,B2(OH) \citep{Cummins,Turner89,Turner91} and Sgr\,B2(N, M and NW)
\citep{Nummelin98,Nummelin00} were also used as additional information to estimate the expected intensity of the fainter identified lines
not included in the \citet{Lovas04} catalog.

One of the main difficulties in identifying lines from single dish observations of extragalactic sources
stems from their large width,
typically $\geq100$\,km\,s$^{-1}$. 
In addition, the emission lines from the nuclear region of NGC\,253 show two velocity components, roughly at 
$180$ and $280\,\rm km\,s^{-1}$, with linewidths of $\rm \sim 100\,and\,110\, km\,s^{-1}$ respectively 
\citep[from CS,][]{Martin05}.
These components arise from the two main molecular lobes separated by $10''$ and located on both sides of the nucleus.
As a consequence, some of the observed transitions will appear partially or totally blended.
Table~\ref{tab.fits} shows the results of the Gaussian fits to all the molecular transitions identified 
towards the NGC\,253 nuclear region.
In the third column of Table~\ref{tab.fits} we have included a note to indicate whether the transition is affected by blending.
Appendix~\ref{apen.Blended} describes the fitting procedures used for the different types of blending found in our spectral line survey.

\subsection{Column density determination}
From the observed main beam brightness temperature of the measured molecular lines
one can estimate column densities and rotation temperatures ($T_{\rm rot}$) for each species
(see Appendix~\ref{apen.PopDiag} for a detailed discussion).
All the necessary spectroscopic information required to derive these parameters 
(i.e. $A_{ul}, \nu, g_u, E_u\,and\, Z$ in Eq.~\ref{eq.PopDiag}) were extracted or derived from the JPL catalog \citep{Pickett}.

In order to derive $T_{\rm rot}$ and to extrapolate the column densities in the observed states to a total column density for a 
given molecule, more than one transition has to be measured.
In the case that only one transition was detected in the 2\,mm scan, 
an educated guess of the rotation temperature was made (see Appendix~\ref{apen.notes}).

Table~\ref{tab.MolecDensity} shows the derived total column densities and rotation temperatures for the detected species in NGC\,253.
Source averaged abundances are also given, where a source averaged molecular hydrogen column density 
$N({\rm H_2})=6.7\,10^{22}\,{\rm cm}^{-2}$ has been assumed (see Sect.~\ref{GalaxComp}).
\clearpage
\begin{figure}[p]
\centering
	\includegraphics[angle=0,width=0.30\linewidth]{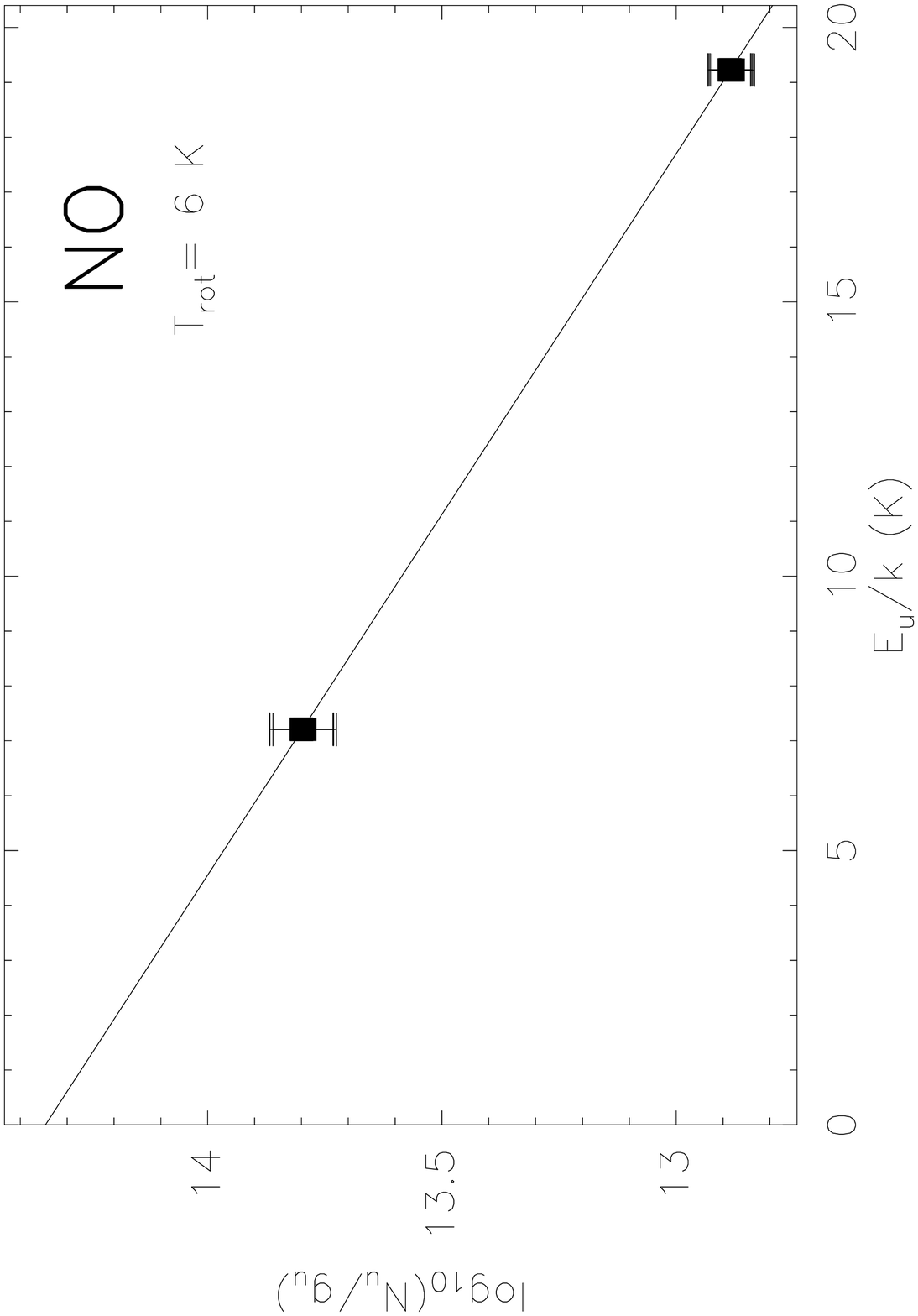}
	\includegraphics[angle=0,width=0.30\linewidth]{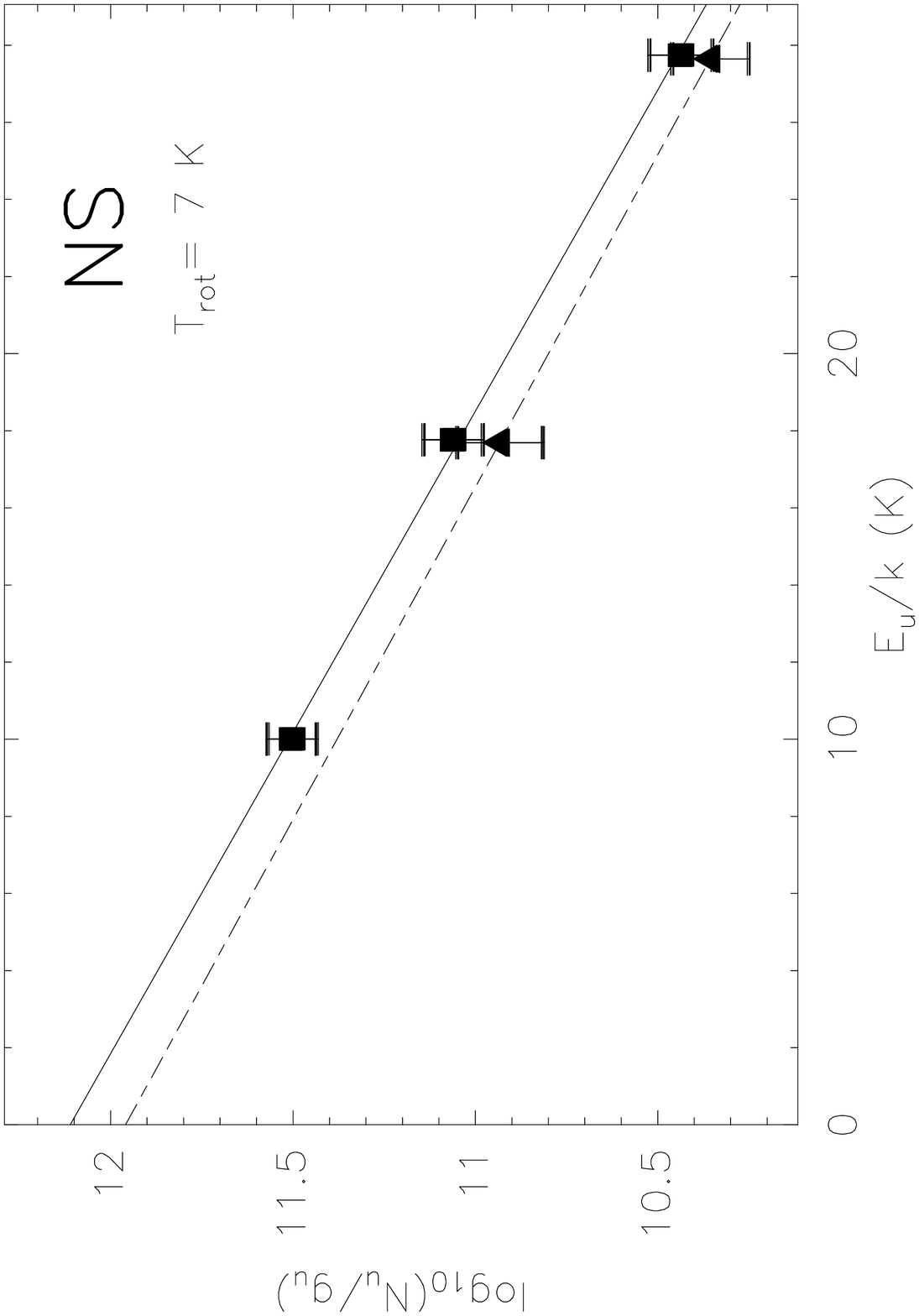}
	\includegraphics[angle=0,width=0.30\linewidth]{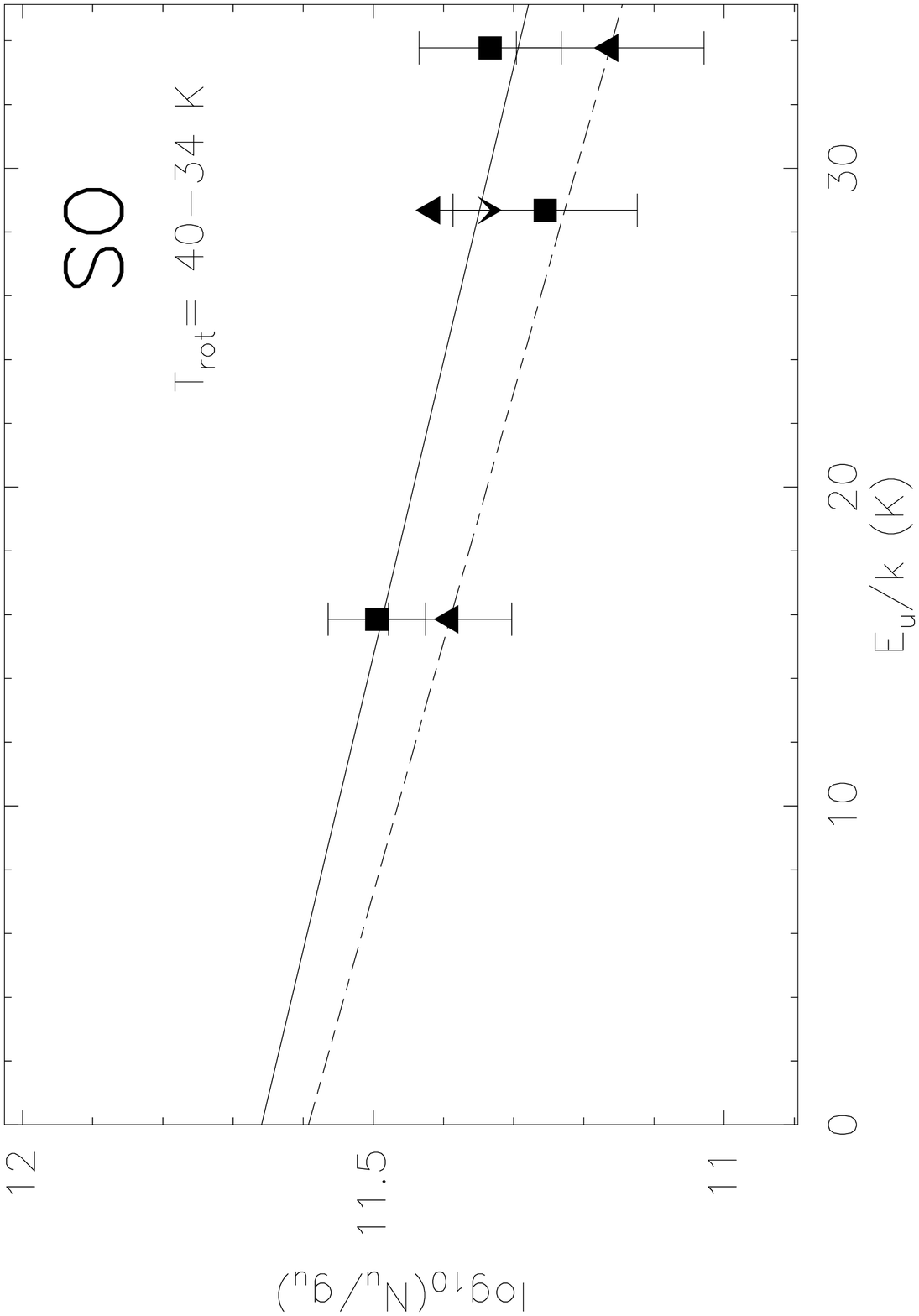}
	\includegraphics[angle=0,width=0.30\linewidth]{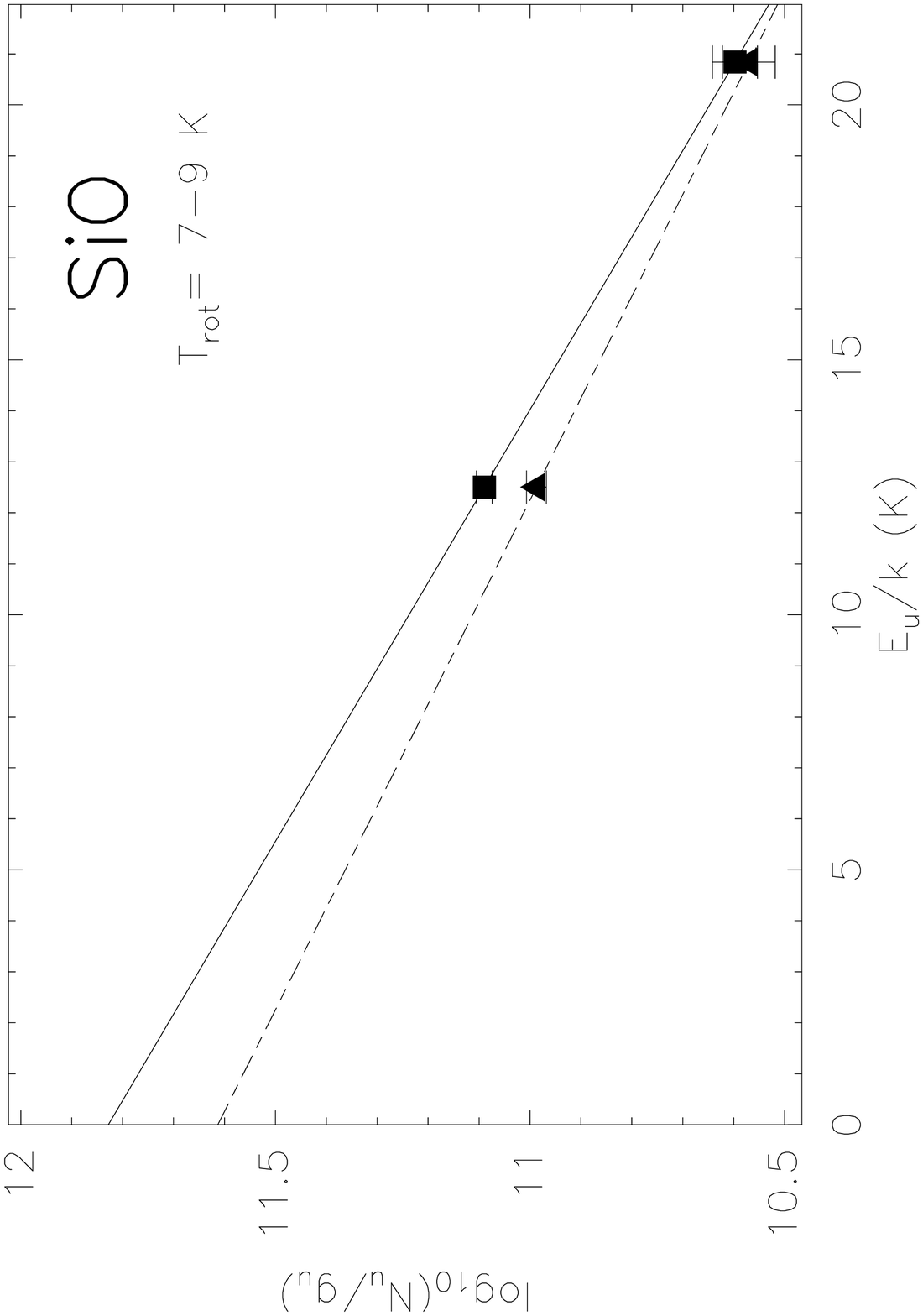}
	\includegraphics[angle=0,width=0.30\linewidth]{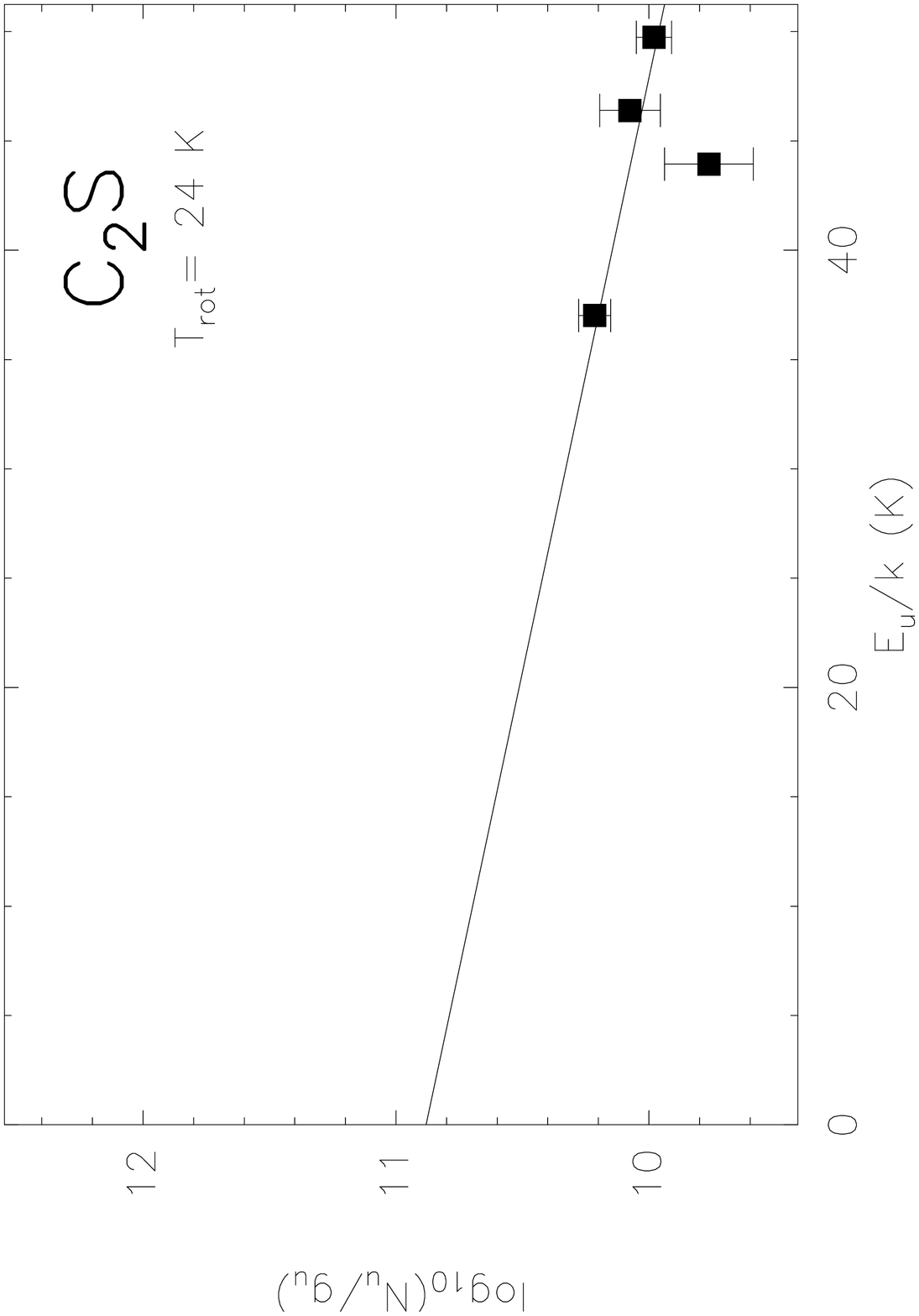}
	\includegraphics[angle=0,width=0.30\linewidth]{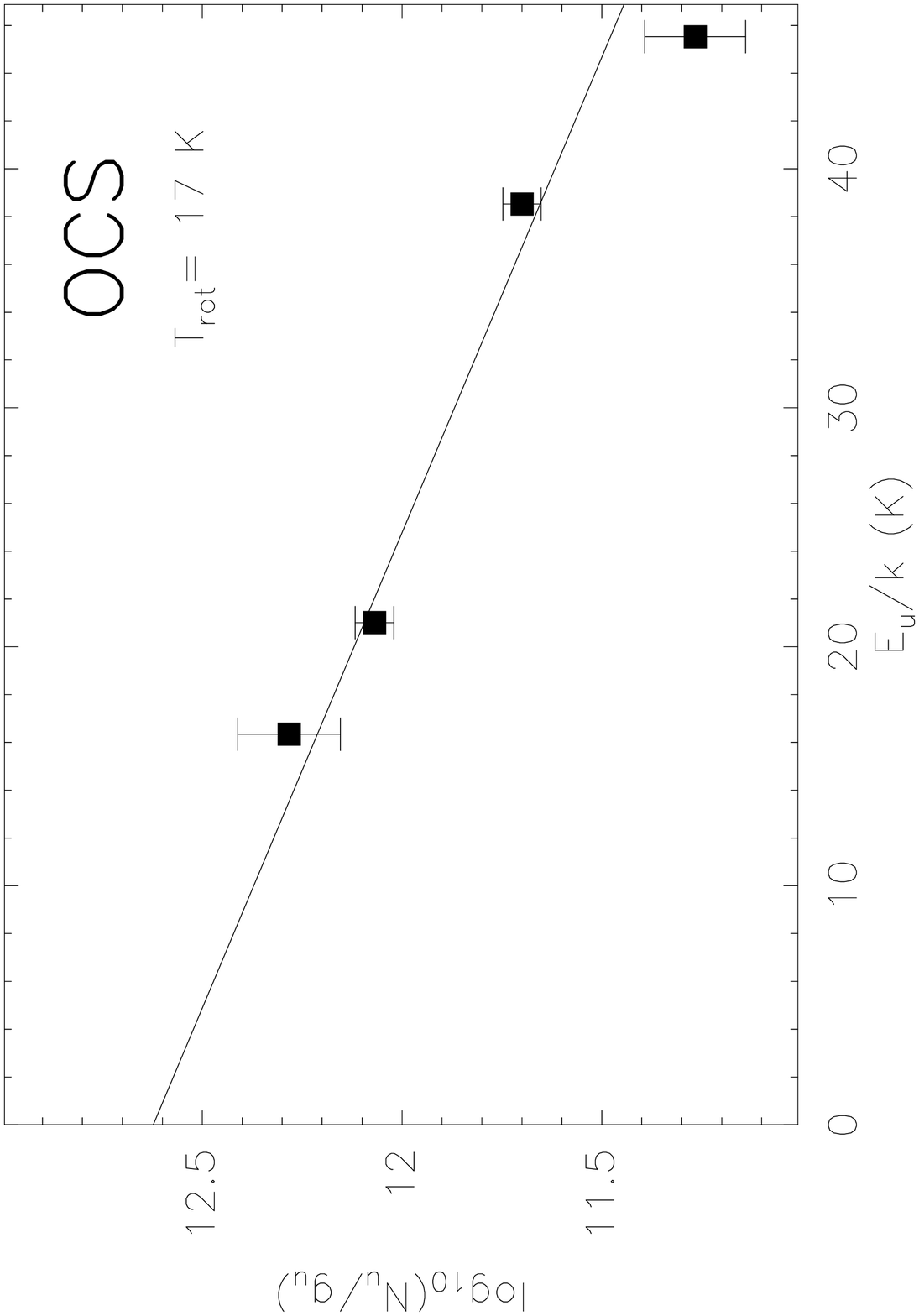}
	\caption{
        Rotation diagrams of detected species.
	When differentiated, velocity components are represented as filled squares and solid lines (180\,km\,s$^{-1}$) 
	and filled triangles and dashed lines (270\,km\,s$^{-1}$).
	Unfilled markers represent blended observed transitions whose intensities have been estimated by inter- or extrapolation 
	from the diagrams (see Appendix~\ref{apen.Blended} for details).
	The NS, NO and OCS diagrams include additional transitions observed by \citet{Martin03,Martin05}.
	}
	\label{fig.PopDiag}
\end{figure}
\clearpage
\begin{center}
	\includegraphics[angle=0,width=0.30\linewidth]{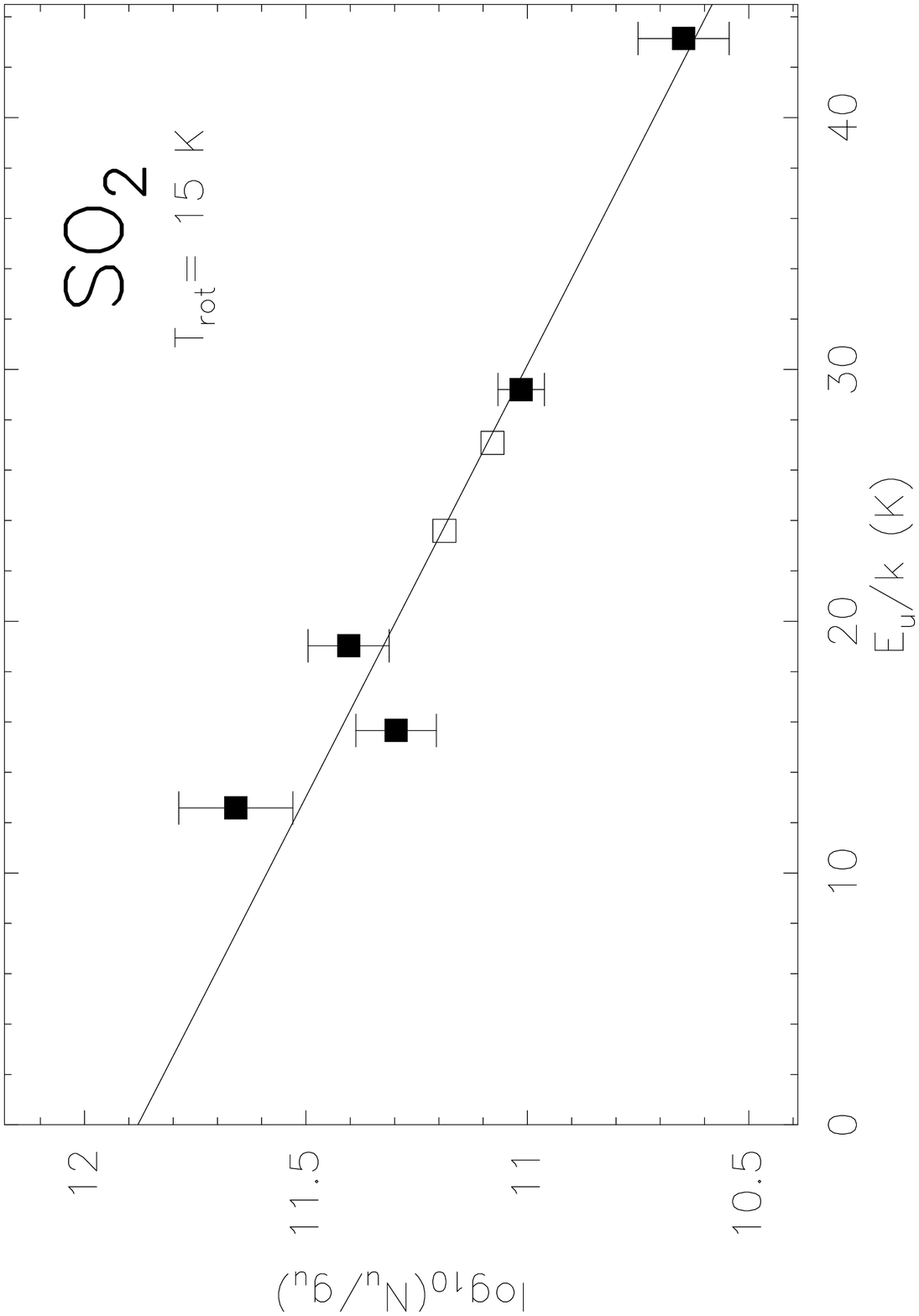}
	\includegraphics[angle=0,width=0.30\linewidth]{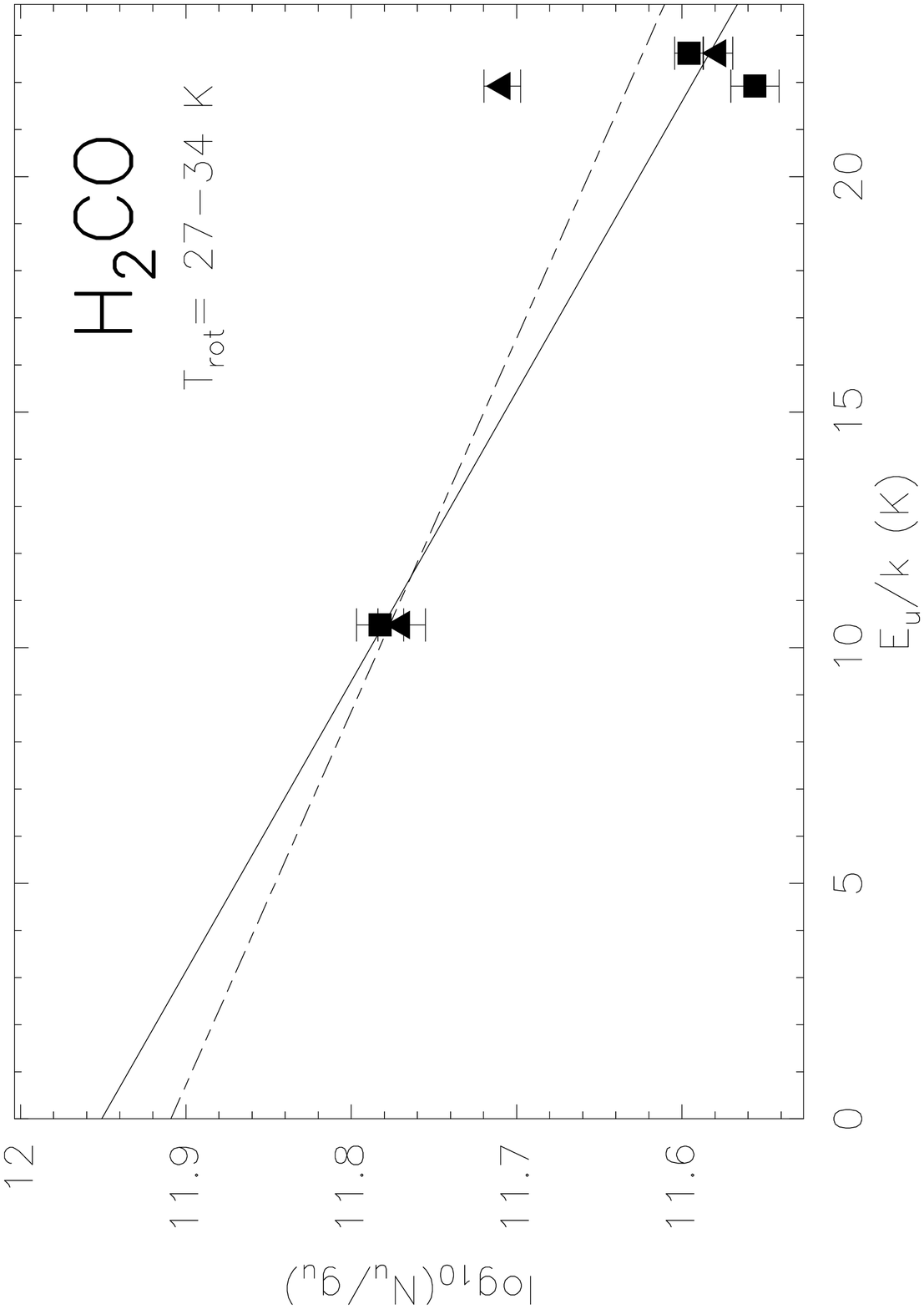}
	\includegraphics[angle=0,width=0.30\linewidth]{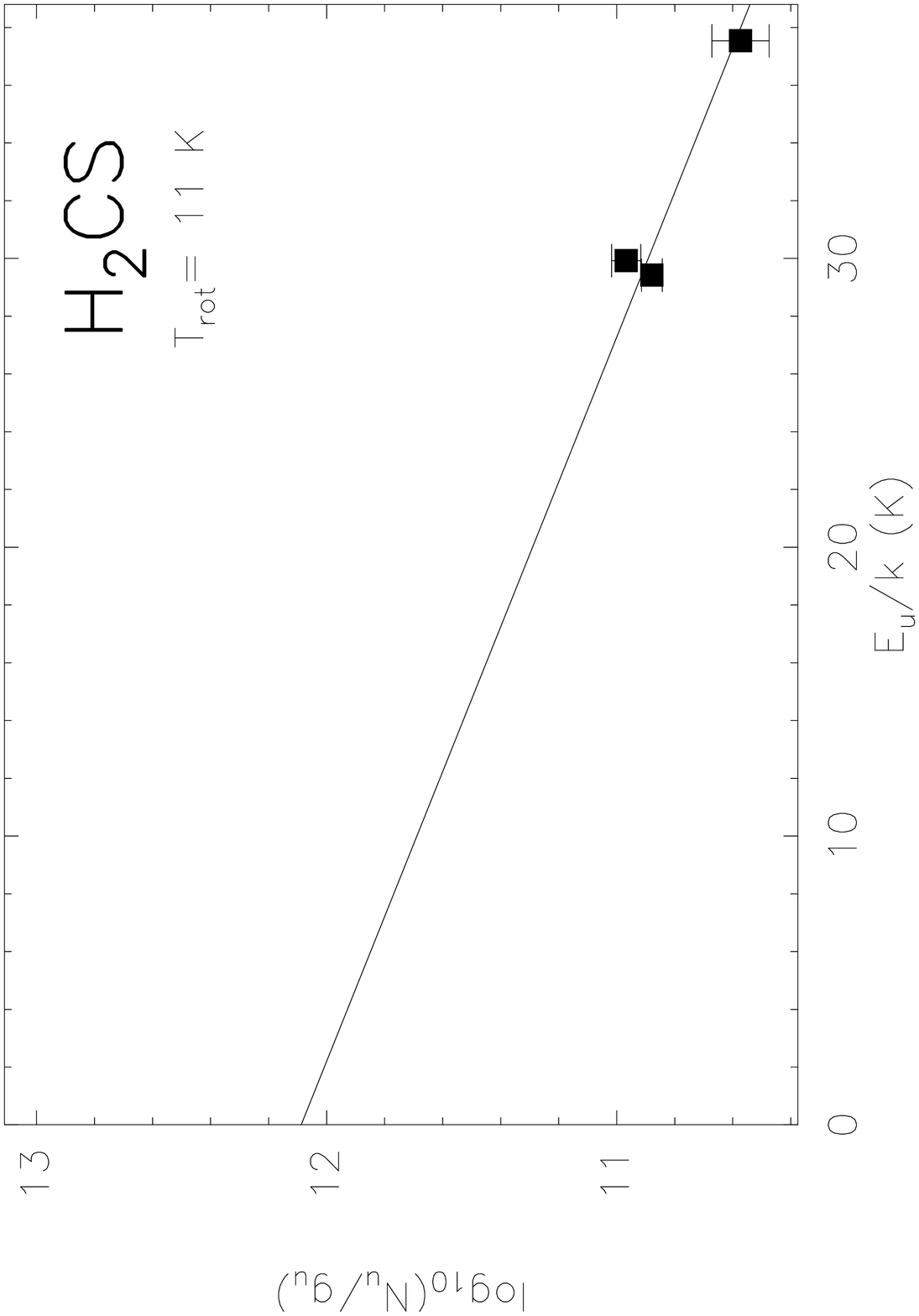}
	\includegraphics[angle=0,width=0.30\linewidth]{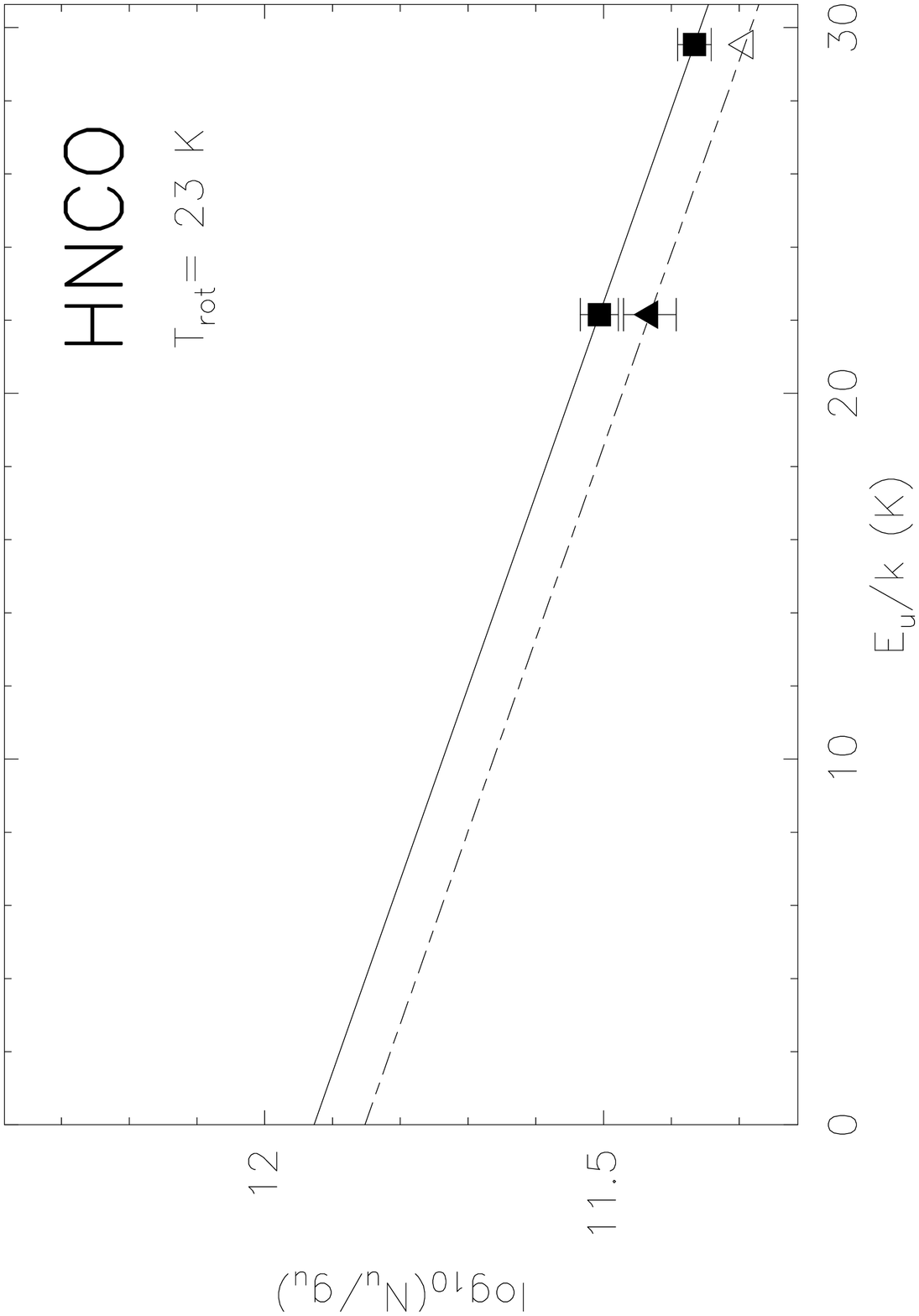}
	\includegraphics[angle=0,width=0.30\linewidth]{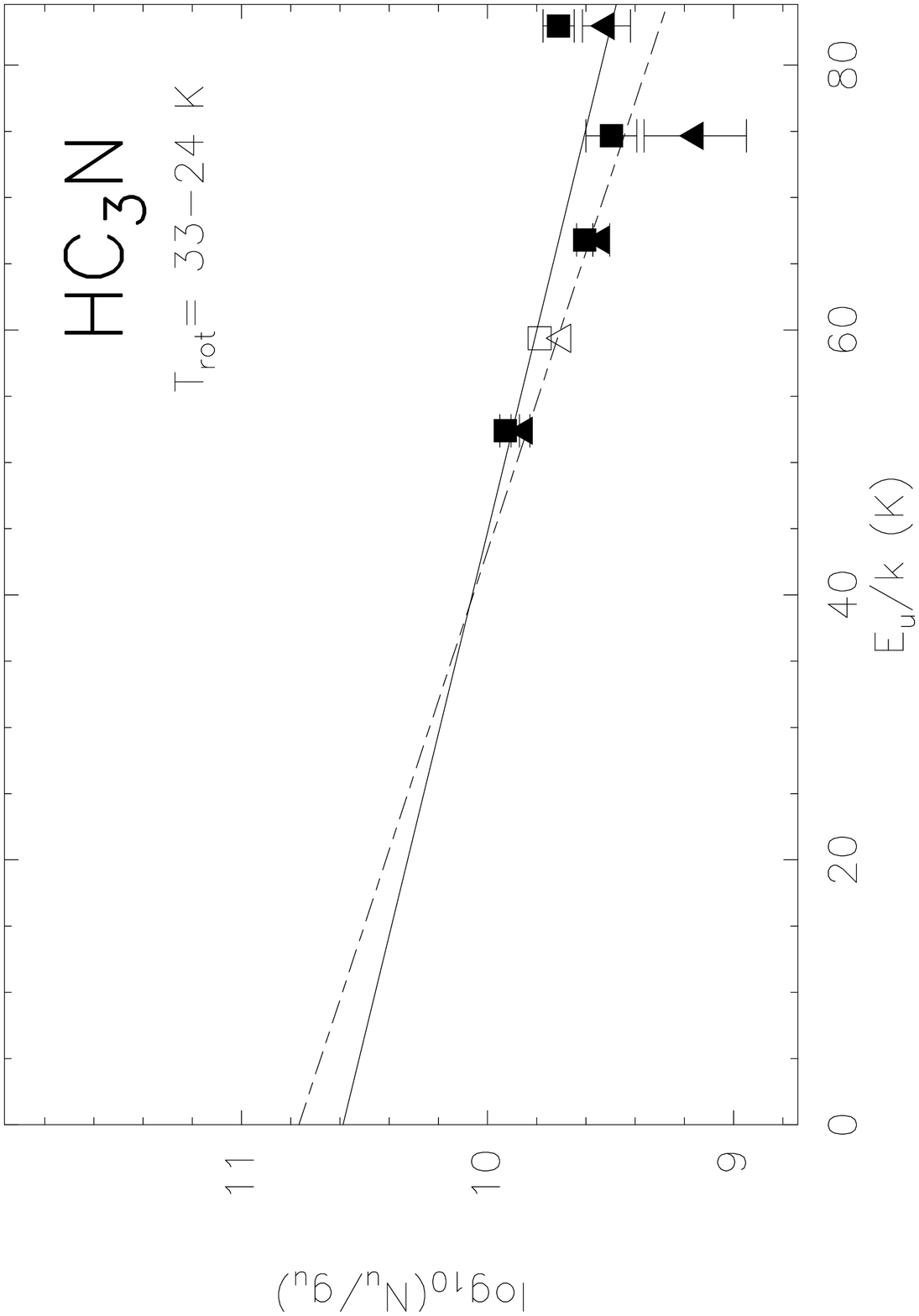}
	\includegraphics[angle=0,width=0.30\linewidth]{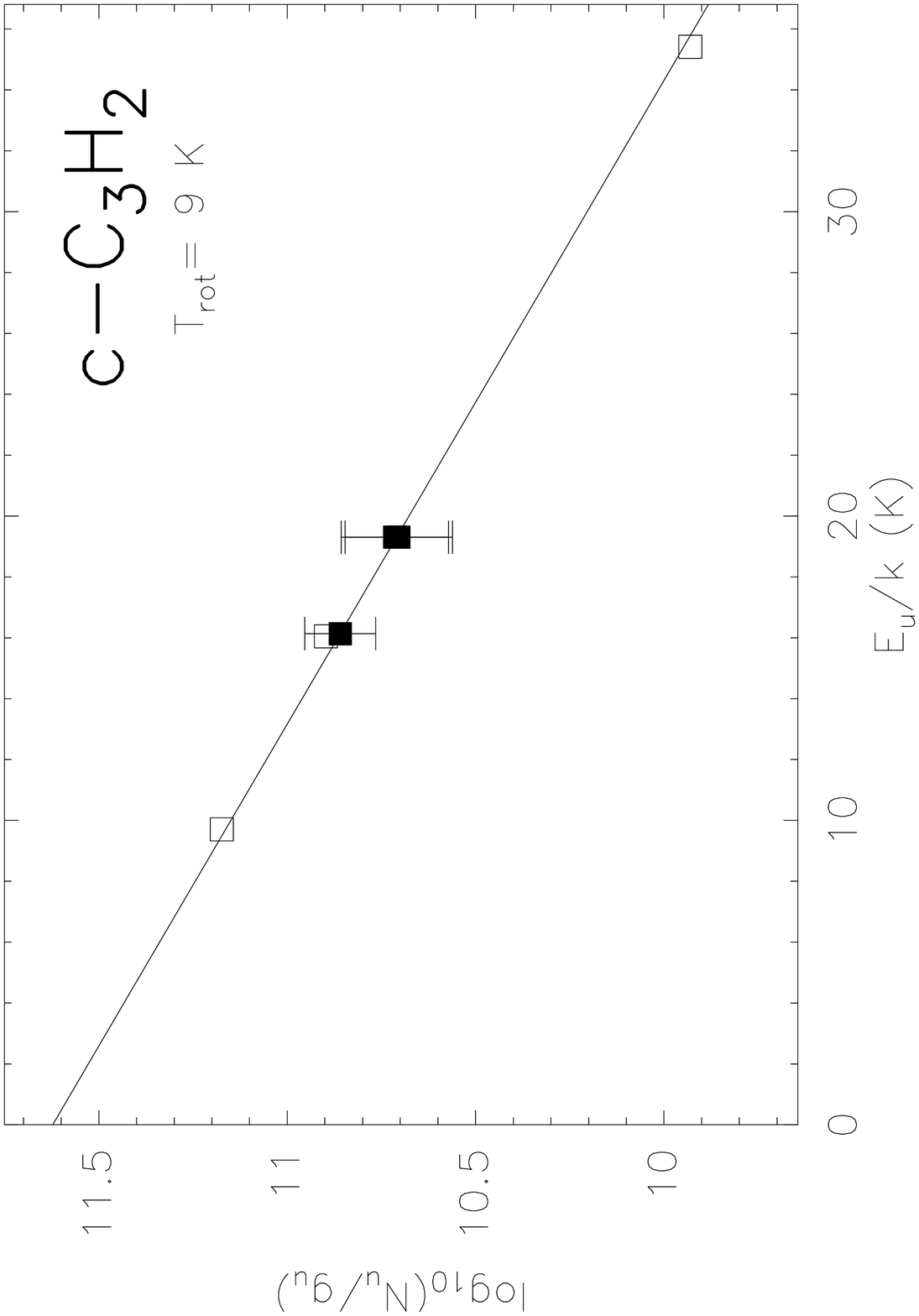}
\\Fig. 5. --- (Cont.)\clearpage
	\includegraphics[angle=0,width=0.30\linewidth]{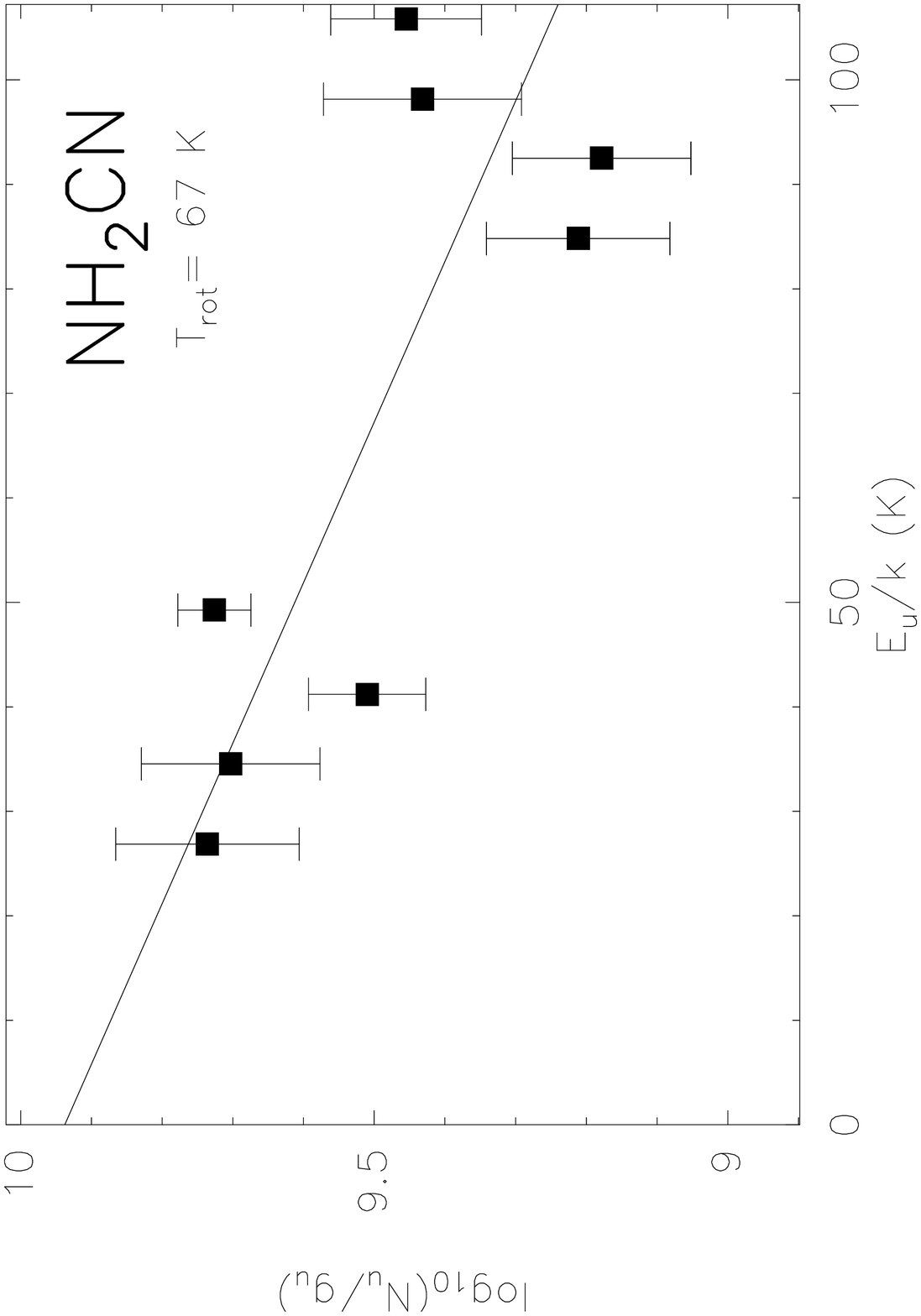}
	\includegraphics[angle=0,width=0.30\linewidth]{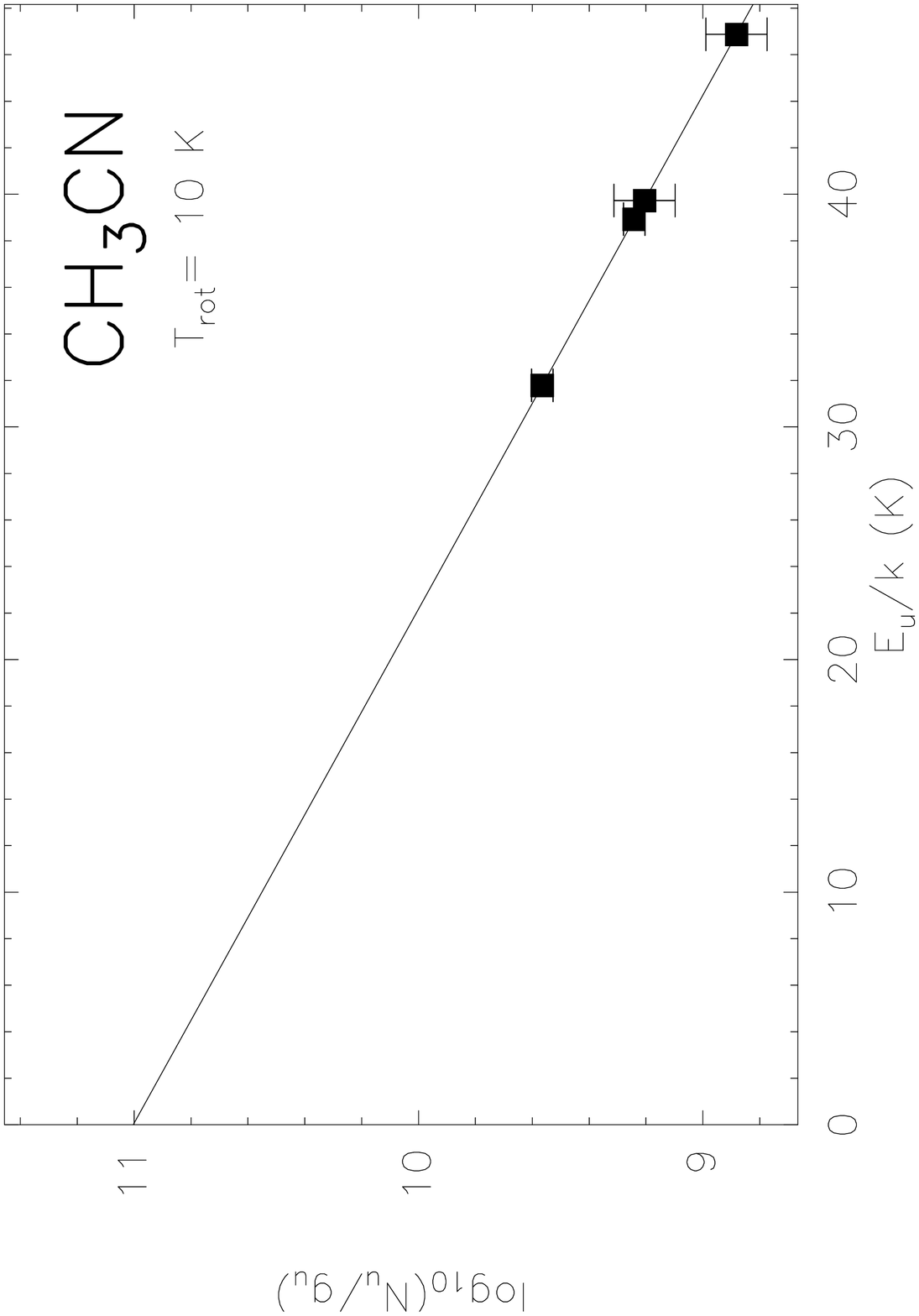}
	\includegraphics[angle=0,width=0.30\linewidth]{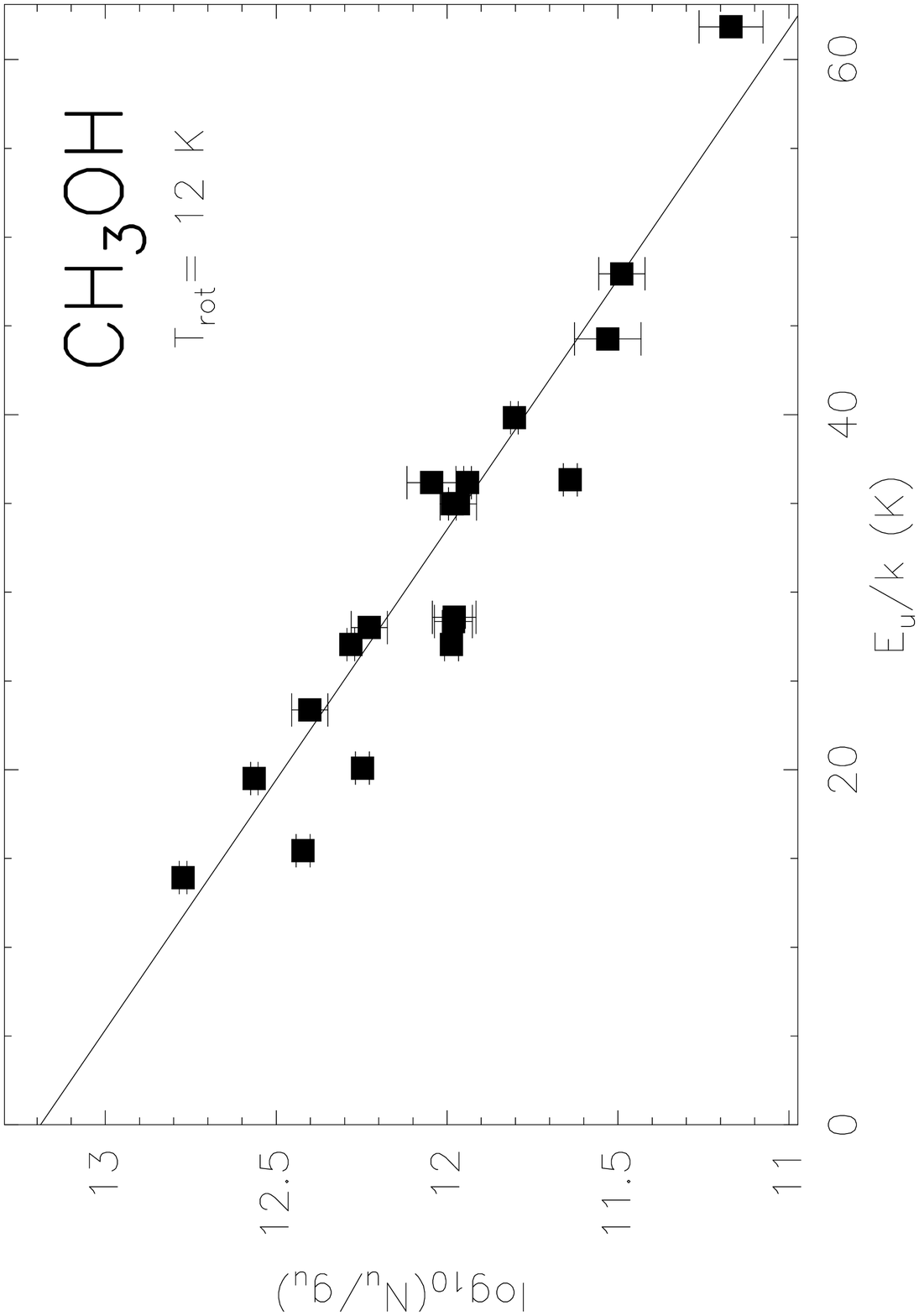}
	\includegraphics[angle=0,width=0.30\linewidth]{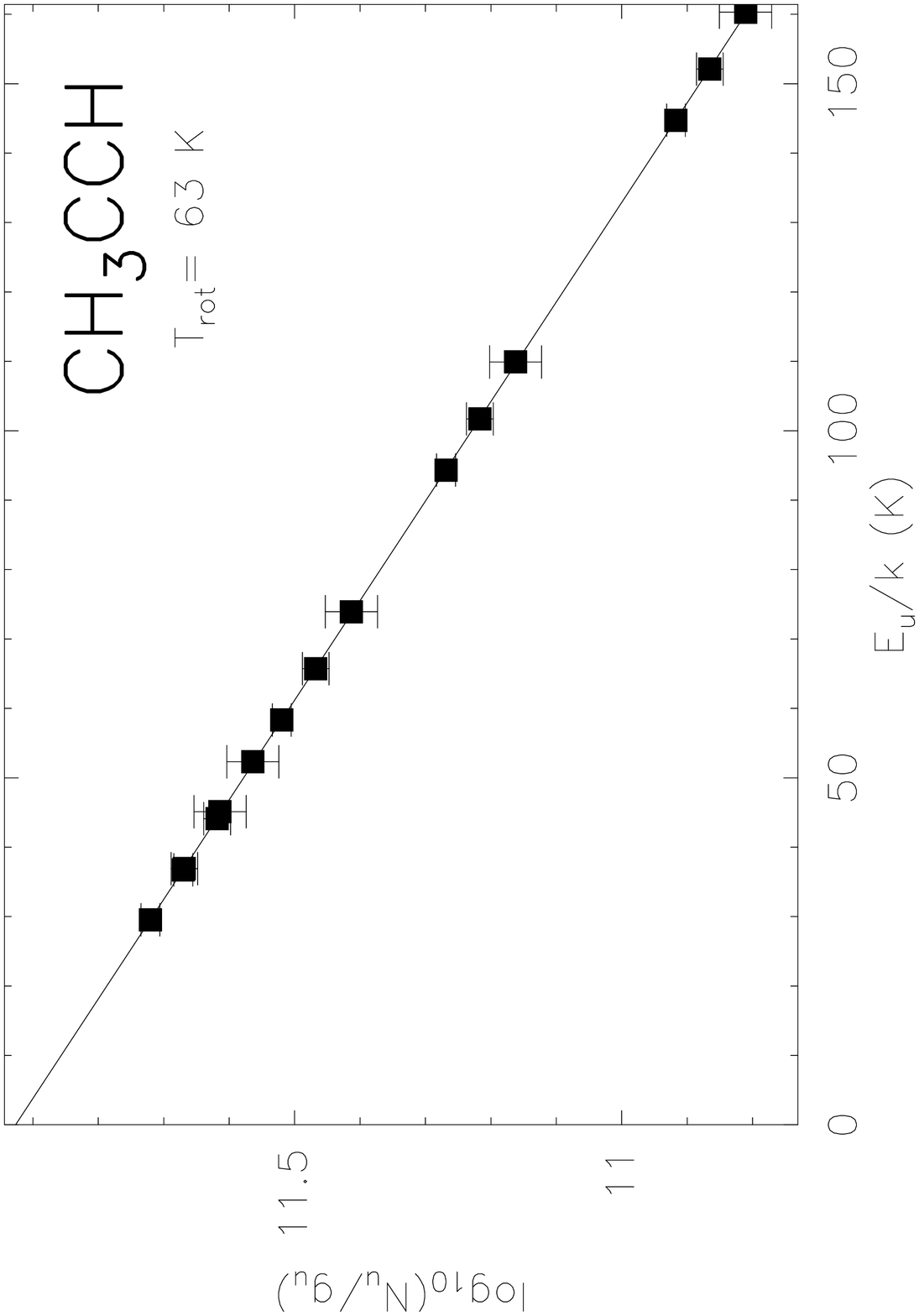}
\\Fig. 5. --- (Cont.)
\end{center}
\clearpage

\subsection{Uncertainty in column densities and rotation temperatures}
\label{sect.uncertainty}

The main source of uncertainty, when deriving the column densities and rotation temperatures,
arises from the uncertainty of the extent of the emitting regions. 

The source averaged brightness temperature ($T_{\rm B}$) can be estimated from the measured main beam brightness temperature as 
\begin{equation}
T_{\rm B}=\eta_{bf}^{-1} \, T_{\rm MB}
\label{eq.Tb}
\end{equation}
where, in the approximation of a Gaussian source distribution of size $\theta_s$ observed with a Gaussian beam of size $\theta_b$,
the beam filling factor $\eta_{bf}=\theta_s^2/(\theta_s^2 + \theta_b^2)$ accounts for the dilution effect due to the coupling 
between the source and the telescope beam.

In order to correct for beam dilution in the nuclear region of NGC\,253, \citet{Mauers03} smoothed the high resolution 
interferometric CS map from \citet{Peng96} to convert the three observed transitions of CS by \citet{Mauers89} to a common angular
resolution of $32''$. Since we know the intensities of the CS lines with the original resolution ($T_{\rm MB}$) 
and those corrected to a $32''$ beam ($T^{32''}_{\rm MB}$) we can derive the beam filling factor. 
From the expression $\eta^{32''}_{bf} T_{\rm MB} = \eta_{bf} T^{32''}_{\rm MB}$ we 
estimate the extent of the emitting region to be $\theta_s=23'',\, 21''\,{\rm and}\,19''$ for the CS $J=2-1,\,J=3-2$ and $J=5-4$
transitions respectively. 
In our analysis we will then consider an equivalent source size of $20''$ to convert our main beam brightness 
temperature to source averaged brightness temperature.

The other main source of uncertainty stems from the assumption that all the observed species arise from the same volume.
The only way to confirm this would be high-resolution observations of different transitions of several species. 
The interferometric map of the $J=2-1$ line of CS by
\citet{Peng96} shows that most of the emission of this molecule is concentrated within a radius of $\sim 20''$. 
The CS emission is seen to be clumped into four large main molecular cloud complexes symmetrically distributed with respect to the
dynamical center and roughly aligned with the molecular bar.
A recent high-resolution multiline study carried out towards IC\,342 \citep{Meier05} shows clear differences in the distribution
of several important molecular tracers. 
Towards NGC\,253, however, high-resolution maps of molecular transitions requiring different excitation conditions such
as HCN, CO, SiO, H$^{13}$CO$^{+}$ and NH$_3$ \citep{Paglione95,Paglione04,Burillo00,Ott05} 
show very similar distributions, in good agreement with that of CS.
Therefore it is plausible that for NGC\,253, as long as the bulk of the emission arises from the area confined by
the starburst, the 
extent of the emission of different molecules is similar.

\begin{figure}[!h]
\centering
	\includegraphics[angle=-90,width=0.5\linewidth]{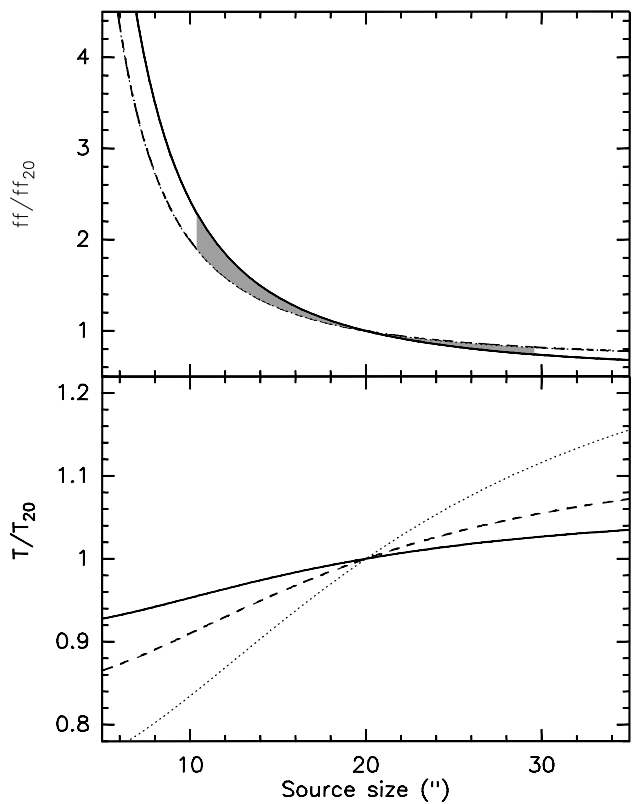}
	\caption{{\it Upper panel:} Variation of the filling factor normalized to that corresponding to a source extent of $20''$
	as a function of source size.
	The continuous line corresponds to the beam at 129\,GHz and the dashed line to the beam at 175\,GHz. 
	The shaded area highlights values with source sizes between 10 and 30$''$.
	{\it Lower panel:} Effect of the assumed source size on the rotation temperature determined from rotation diagrams 
	relative to the assumed $T_{\rm rot}=15$K for $\theta_s=20''$ with two transitions measured at 129 and 175\,GHz where the 
	involved levels are separated by 60 (continuous line), 30 (dashed line) and 15\,K (dotted line).}
	\label{fig.ffac}
\end{figure}

We now address the question of how an error in the assumed source size will affect our determination of physical parameters.
As derived from Eq.~\ref{eq.Nup} (see Appendix~\ref{apen.PopDiag}), the calculated total column density is affected 
by the filling factor through $T_{\rm B}$.
The upper panel in Fig~\ref{fig.ffac} shows the variation of the inverse of the filling factor (i.e. the conversion factor from $T_{\rm MB}$ to $T_{\rm B}$
from Eq.~\ref{eq.Tb}) as a function of source size, normalized to the value for a source of $20''$. 
This value represents, in a first approximation, the factor by which we should multiply our derived column densities 
if the source extent would be different from that assumed. 
We have considered the extreme cases of the beam sizes at 129 and 175\,GHz.
The grey shaded area corresponds to source sizes between $10''$ and $30''$ where we expect the source size to be confined. 
In case of a source size larger than $20''$ column densities would be overestimated by only $\sim 20\%$, while 
for sizes down to $10''$, the column density would be underestimated by less than a factor of 2.
Only in case of an extremely clumped and compact emission where source sizes were well below $10''$, column densities would be 
dramatically affected.
We have also considered the effect of source size on the derived rotation temperature from rotation diagrams. 
We will assume a case in which we derive $T_{\rm rot}=15$\,K (for a $20''$ source) by observing two transitions of a given species 
at 129 and 175\,GHz respectively.
In the lower panel in Fig.~\ref{fig.ffac} we have plotted the relative change in the derived $T_{\rm rot}$ for those cases where 
the transitions would have upper levels with an energy difference of 60, 30 and 15\,K.
It is obvious that the more affected $T_{\rm rot}$  is that derived from the data with the smaller dynamical range in energies.
From Fig.~\ref{fig.ffac} we see that the largest expected uncertainty in the determination of $T_{\rm rot}$ is smaller than $20\%$. 
\clearpage
\clearpage

\section{Discussion}

\subsection{Extragalactic molecular census}
\label{sect.census}

Table~\ref{tab.census} lists the detection of all the species known to date in the extragalactic ISM in chronological order.
During the last three decades, 29 molecules and 11 rare isotopic substitutions have been detected outside the Galaxy.
As a result of the frequency scan presented in this paper \citep[see also][]{Martin03,Martin05}, 8 new molecules and 2 isotope 
(shown in boldface in Table~\ref{tab.census}) have been added to the census of known extragalactic species, which represents a $\sim30\%$ 
increase in the total number of identified molecules.
While H$_2$S and H$_2$CS were detected towards the nearby Large Magellanic Cloud \citep{Heikkila99}, 
their identification in our survey \citep[see][]{Martin05} represents the first detections towards a starburst galaxy.
Note that $\rm CH_2NH$, $\rm HOCO^+$ and $\rm C_3H$ (Table~\ref{tab.census}) are tentatively
detected and need confirmation by observations of additional transitions.
In Table~\ref{tab.census} the known extragalactic species not detected in the 2\,mm band are shown in italics.
\clearpage
\begin{table}
\caption{Known extragalactic molecules (see Sect.~\ref{sect.census})}
\begin{center}
\scriptsize
\begin{tabular}{l c r r}
\tableline
\tableline
Molecule                &       Publication      &    Telescope             &       Ref.         \\
                        &        Year            &                          &                    \\
\tableline
$OH$                    &       1971             &     OVRO                 &        1           \\
$\rm H_2CO$             &       1974             &      64\,m               &        2           \\
$CO$                    &       1975             &      11\,m               &        3           \\
$^{13}CO$               &       1975             &      11\,m               &        4           \\
$H_2O$                  &       1977             &     100\,m               &        5           \\
$HCN$                   &       1977             &      11\,m               &        6           \\
$H_2$                   &       1978             &     2.3\,m               &        7           \\
$NH_3$                  &       1979             &     100\,m               &        8           \\
$HCO^+$                 &       1979             &       7\,m               &        9           \\ 
$CH$                    &       1980             &      64\,m               &       10           \\
$\rm CS$                &       1985             &       7\,m               &       11           \\
$\rm C_3H_2$            &       1986             &      43\,m               &       12           \\
$CH^+$                  &       1987             &     1.4\,m               &       13           \\
$\rm CH_3OH$            &       1987             &      30\,m               &       14           \\
$CN$                    &       1988             &      30\,m               &       15           \\
$\rm C_2H$              &       1988             &      30\,m               &       15           \\
$HNC$                   &       1988             &      30\,m               &       15           \\
$\rm HC_3N$             &       1988             &      30\,m               &       15,16        \\
$\rm HNCO$              &       1989             &      30\,m               &       17,18        \\
$\rm C^{34}S$           &       1989             &      30\,m               &       19           \\
$C^{18}O$               &       1991             &      11\,m/30\,m         &       20           \\
$C^{17}O$               &       1991             &      11\,m/30\,m         &       20           \\  
$\rm SO$                &       1991             &      15\,m               &       21           \\
$N_2H^+$                &       1991             &      30\,m               &       22           \\
$\rm SiO$               &       1991             &      30\,m               &       22           \\
$\rm H^{13}CO^+$        &       1991             &      30\,m               &       22           \\  
$\rm HN^{13}C$          &       1991             &      30\,m               &       22           \\
$H^{13}CN$              &       1991             &      30\,m               &       22           \\    
$\rm CH_3CCH$           &       1991             &      30\,m               &       23           \\
$\rm CH_3CN$            &       1991             &      30\,m               &       23           \\
$\rm ^{13}CS$           &       1993             &      30\,m               &       24           \\
$\rm OCS$               &       1995             &      30\,m               &       25           \\
$HCO$                   &       1995             &      12\,m               &       26           \\
$\rm DCO^{+}$           &       1996             &      15\,m               &       27           \\
$\rm DCN$               &       1996             &      15\,m               &       27           \\
$HC^{15}N$              &       1999             &      15\,m               &       28           \\
$\rm     H_2S$          &       1999             &      15\,m               &       29           \\
$\rm     H_2CS$         &       1999             &      15\,m/30\,m         &       29,30        \\
$CO^+$                  &       2000             &      30\,m               &       31           \\
\tableline
\end{tabular}
\end{center}
\end{table}

\begin{table}
\addtocounter{table}{-1}
\caption{(Cont.)}
\begin{center}
\scriptsize
\begin{tabular}{l c r r}
\tableline
\tableline
Molecule                &       Publication      &    Telescope             &       Ref.         \\
                        &        Year            &                          &                    \\
\tableline
$\rm \bf SO_2$          &       2003             &      30\,m               &       32           \\
$\rm \bf NO$            &       2003             &      30\,m               &       32           \\
$\rm \bf NS$            &       2003             &      30\,m               &       32           \\
$\rm \bf ^{34}SO$       &       2003             &      30\,m               &       32           \\
$HOC^+$                 &       2004             &      30\,m               &       33           \\
$\rm \bf C_2S$          &       2006             &      30\,m               &       34           \\
$\rm \bf CH_2NH$        &       2006             &      30\,m               &       34           \\
$\rm \bf NH_2CN$        &       2006             &      30\,m               &       34           \\
$\rm \bf HOCO^+$        &       2006             &      30\,m               &       34           \\
$\rm \bf C_3H$          &       2006             &      30\,m               &       34           \\
$\rm \bf HC^{18}O^+$    &       2006             &      30\,m               &       34           \\
\tableline                                                              
\end{tabular}                                                       
\label{tab.census}
\end{center}
\scriptsize
\tablecomments{
Table adapted and updated from \citet{Mauers93}\\
In roman: molecules observed in this survey; 
In {\bf boldface}: molecular species not previously identified outside the Milky Way; 
In {\it italics}: species not measured in this survey.\\
{\it Telescopes.-}
(OVRO) Owens Valley interferometer;
(64\,m) Parkes 64\,m;
(11\,m) NRAO Kitt Peak 11\,m;
(100\,m) Effelsberg 100\,m;
(2.3\,m) Steward observatory 2.3\,m;
(7\,m) AT\&T Bell Labs 7\,m;
(43\,m) NRAO Green Bank 43\,m;
(1.4\,m) ESO 1.4\,m Coud\'e;
(30\,m) IRAM Pico Veleta 30\,m;
(15\,m) ESO SEST 15\,m;}
\tablerefs{
(1) \citet{Weliachew71};
(2) \citet{Gardner74};
(3) \citet{Rickard75};
(4) \citet{Solomon75};
(5) \citet{Churchwell77};
(6) \citet{Rickard77};   
(7) \citet{Thompson78};  
(8) \citet{Martin79};    
(9) \citet{Stark79};
(10) \citet{Whiteoak80};
(11) \citet{Henkel85};  
(12) \citet{Seaquist86};
(13) \citet{Magain87};  
(14) \citet{Henkel87};  
(15) \citet{Henkel88};  
(16) \citet{Mauers90};
(17) \citet{Nguyen89};
(18) \citet{Nguyen91}; 
(19) \citet{Mauers89}; 
(20) \citet{Sage91};
(21) \citet{Johansson91};
(22) \citet{Mauers91a};
(23) \citet{Mauers91b};
(24) \citet{Henkel93};
(25) \citet{Mauers95};
(26) \citet{Sage95};
(27) \citet{Chin96};
(28) \citet{Chin99};
(29) \citet{Heikkila99};
(30) \citet{Martin05};
(31) \citet{Fuente00};
(32) \citet{Martin03};
(33) \citet{Usero04};
(34) This work.}
\end{table}
\clearpage
\subsection{Deuterated molecules in NGC\,253}
\label{sect.deut}
The observation of deuterated molecules in external galaxies is interesting for several reasons. 
The interstellar D/H ratio depends on the degree of processing of the gas, 
since D was mainly produced soon after the big bang while nuclear processes during stellar evolution rather destroy
than form deuterium. A low D/H ratio may therefore indicate that the observed gas component has been recycled 
in stars, and a higher ratio could be a hint for recent infall of poorly processed material.
It has been observed that D is strongly enriched (by a factor of $>10^4$ compared to the cosmic D/H ratio) in several molecules 
observable at cm- and mm-wavelengths. From model calculations, such a high degree of fractionation is expected if the molecules are 
being formed in a relatively cool ($T_{\rm kin}< 30 \,\rm K$) environment \citep[e.g.][and references therein]{Millar02}. 
The fact that a high enrichment of deuterated molecules is also observed in hot molecular cores can be explained by evaporation 
of molecules from grain mantles \citep[e.g.][]{Pineau89}. 
While a value of $3(\pm 0.4)\,10^{-5}$ has been estimated for the 
primordial D/H ratio \citep[][and references therein]{Burles02}, a lower ratio of $1.4(\pm 0.24)\,10^{-5}$ is found in the Galactic 
neighborhood, and variations are shown on larger scales \citep{Vidal02}. 
In molecular clouds close to the center
of the Milky Way, \citet{Jacq99} find DCN and DCO$^+$ enhancements which are compatible with a D/H an order of magnitude 
smaller than toward the local interstellar medium. This indicates that the gas in the Galactic center region is highly processed. 
The enhancement of deuterated molecules in the Large Magellanic Cloud, on the other hand, is compatible with the local value 
of the D/H ratio \citep{Chin96}. No deuterated molecules have been detected so far toward the central regions of other galaxies. 
\citet{Mauers95} give an upper limit for the DCN/HCN abundance ratio of 4\,10$^{-3}$ toward NGC\,253.

Our survey includes the $J=2-1$ lines of DCO$^+$, DCN, DNC and N$_2$D$^+$. 
All other 2\,mm lines of deuterated molecules are expected to be much weaker. 
Table~\ref{tab.deuterated} shows the frequencies of these lines, the rms noise and limits or values of the integrated intensities.
DCO$^+$ and DCN are not detected.
There is a tentative feature at the frequency of DNC and a more significant feature at the line frequency of N$_2$D$^+$. 
At the frequencies of DNC $J=2-1$ no line from another molecule is expected to be strong enough to explain the observed emission. 
Close to the frequencies of N$_2$D$^+$ $J=2-1$ is the transition frequency of OC$^{34}$S $J=13-12$ (at 154.242 GHz). 
However, from our detections of OCS and our estimates of the $^{32}$S/$^{34}$S ratio \citep{Martin05}, 
this emission is expected to be an order of magnitude weaker than what is observed. Another potentially blending line is the 
$7_{1,6}-6_{1,5}, v_6=1$ transition of vibrationally excited HNCO (at 154.228\,GHz). This line has been detected toward the Galactic 
hot core G10.47+0.03 with an intensity of about 1/5 that of the corresponding HNCO ground state transition, compatible with a 
kinetic temperature of about 380\,K \citep{Wyrowski99}. The feature we are observing here is about as strong as the 
HNCO $7_{1,6}-6_{1,5}$ ground vibrational state transition (Table~\ref{tab.fits}). 
If the feature at 154.228 GHz comes from vibrationally 
excited HNCO this would either indicate a higher kinetic temperature or maser emission. Both would be remarkable in view of 
the large volume observed in our beam toward NGC\,253. 
While all this suggests that we have detected, for the first time, one or two deuterated molecules beyond the Magellanic Clouds, we 
will nevertheless regard the measured features as tentative.
\clearpage
\begin{table}[!t]
\begin{center}
\caption{Parameters of selected deuterated species}
\label{tab.deuterated}
\begin{tabular}{l c l c c c}
\tableline
\tableline
Molecule          &   Transition    &   $\nu$      & rms \tablenotemark{a}  & $\int{T_{\rm MB}{\rm d}v}$ \tablenotemark{b}& [DX]/[HX]  \\   
                  &      $J-J'$     &   (MHz)      & (mK)                   &     (K\,km\,s$^{-1}$)                       &   \\
\tableline                                                                                               
DCO$^+$           &       $2-1$       &  144077.3    &  1.9                 &  $<$0.30                                    &  $<4\,10^{-3}$  \tablenotemark{c}         \\
DCN               &       $2-1$       &  144827.9    &  1.1                 &  $<$0.18                                    &  $<1\,10^{-3}$  \tablenotemark{c}         \\
DNC               &       $2-1$       &  152609.7    &  3.1                 &     0.49                                    &  $\leq 1\,10^{-2}$ \tablenotemark{c}      \\
$\rm N_2D^+$      &       $2-1$       &  154217.0    &  2.8                 &     0.57                                    &  $\leq 1.4\,10^{-3}$ \tablenotemark{d}    \\
\tableline
\end{tabular} 
\end{center}
\tablecomments{See Sect.~\ref{sect.deut} for details.}
\tablenotetext{a}{rms computed for 20\,km\,s$^{-1}$ wide channels.}
\tablenotetext{b}{$3\sigma$ limit to the integrated intensity assuming a linewidth of $150\,\rm km\,s^{-1}$.
DNC and N$_2$D$^+$ detections are tentative.}
\tablenotetext{c}{Using the $^{13}$C bearing species measured in this survey and assuming $^{12}{\rm C}/^{13}{\rm C}\sim40$ \citep{Henkel93}.}
\tablenotetext{d}{Using the $J=1-0$ N$_2$H$^+$ transition from \citet{Mauers91a}.}
\end{table}
\clearpage
Table~\ref{tab.deuterated} gives limits to the [DX]/[HX] column density ratios for HCO$^+$, HCN and HNC based on our limits for the $J=2-1$ 
transitions of the deuterated isotopomers, the integrated intensities of the $^{13}$C bearing isotopomers measured in this survey and 
the $^{12}$C/$^{13}$C ratio of 40 obtained by \citet{Henkel93}. 
We have assumed that the excitation conditions of the deuterated and main species are similar.
Intensities were corrected to account for beam dilution assuming a source size of $20''$ (see also Sect.~\ref{sect.uncertainty}).
For N$_2$D$^+$, we have only the N$_2$H$^+$ $J=1-0$ line for a comparison \citep{Mauers91a}. 
The value for the N$_2$D$^+$/N$_2$H$^+$ column density ratio given in Table~\ref{tab.deuterated} 
assumes
optically thin emission, making the same assumptions on the excitation as they have 
been made for DCN in the appendix of \citet{Mauers95}.

We obtain limits to the [DX]/[HX] values of 4\,10$^{-3}$ for HCO$^+$, of 1\,10$^{-3}$ for HCN and 11\,10$^{-3}$ for HNC. 
These limits are two orders of magnitude above the local interstellar
D/H ratio and an order of magnitude larger than the [DCN]/[HCN] 
ratios estimated for Sgr\,B2 \citep{Jacq99}, 
but an order of magnitude lower than the values determined by \citet{Chin96} in the Large Magellanic Cloud. 
The limits are consistent with the high kinetic temperatures determined toward the molecular interstellar medium of NGC\,253. 
We can exclude the  scenario found in many Galactic hot cores, where recent grain mantle evaporation leads to
high D/H ratios that were preserved from a time when the temperature of the gas was much lower than now. 

\subsection{Alcohols in NGC\,253}

In the nuclear region of the Milky Way widespread large abundances of the methanol (CH$_3$OH) and ethanol (C$_2$H$_5$OH) alcohols have been reported
\citep{Requena05,Pintado01}.
This fact has motivated 
a careful search for the presence of $\rm C_2H_5OH$ in the 2\,mm scan. 
Table~\ref{tab.complexmol} shows the frequencies and energies of its 14 brightest 2\,mm transitions assuming
$T_{\rm rot}=12$\,K, as that derived for methanol (see Table~\ref{tab.MolecDensity}).
We have computed the rms of the noise for 20\,km\,s$^{-1}$ velocity resolution whenever the transition appeared not to be contaminated
by other spectral features. 
From this rms, we estimate a $3\sigma$ detection limit to the integrated intensity of the transitions.
With the limit to the detection of the 152\,GHz transition we derive an upper limit to the ethanol column density of 
$N(\rm C_2H_5OH)<1.8\,10^{14}\,cm^{-2}$.
This 
limit to the column density 
can be improved if we add up all the transitions which appear not to be blended.
We estimate a total column density limit of $N(\rm C_2H_5OH)<9.6\,10^{13}\,cm^{-2}$.

The methanol to ethanol abundance ratio derived for NGC\,253 is $\geq$10. 
This lower limit is still a factor of 2 below the ratios derived for the GC clouds by \citet{Requena05}. 
Thus, further high sensitivity observations are required to establish if the methanol to ethanol abundance ratio is similar to 
that found in the GC clouds, as expected if alcohols are also ejected from icy grain mantles by shocks in this galaxy\citep{Charnley95}.
\clearpage
\begin{table}[!tb]
\begin{center}
\caption{Limits to the detection of $\rm C_2H_5OH$}
\label{tab.complexmol}
\begin{tabular}{l l c c}
\tableline
\tableline
$\nu$             &   $\rm E_u$   &  rms \tablenotemark{a}       & $\int{T_{\rm MB}{\rm d}v}$ \tablenotemark{b} \\   
(MHz)             &($\rm cm^{-1}$)&  (mK)                        &     (K\,km\,s$^{-1}$)             \\
\tableline
129665.7          &  16.6         &     3.0                      &     $<1.440$                      \\
131103.0          &  16.6         &     1.8                      &     $<0.864$                      \\
131502.7          &  20.1         & \multicolumn{2}{l}{C$_2$S}                            \\
132935.1          &   6.9         & \multicolumn{2}{l}{baseline}                          \\
133323.4          &  16.5         & \multicolumn{2}{l}{CH$_2$NH}                          \\
142285.0          &  25.8         &     1.8                      &     $<0.864$                      \\
144734.0          &   9.3         & \multicolumn{2}{l}{C$^{34}$S}                         \\
147427.4          &  21.0         &     2.9                      &     $<1.392$                      \\
152656.8          &   9.3         &     3.4                      &     $<1.632$                      \\
159781.8          &  12.2         & \multicolumn{2}{l}{NH$_2$CN}                          \\
160699.0          &  31.4         & \multicolumn{2}{l}{i=H$_2$S}                          \\
161609.0          &  26.1         & \multicolumn{2}{l}{NS}                                \\
173391.3          &  12.2         &     3.6                      &     $<1.728$                      \\
174232.9          &  15.7         & \multicolumn{2}{l}{c-C$_3$H}                          \\
\tableline
\end{tabular}
\end{center}
\tablenotetext{a}{rms computed for 20\,km\,s$^{-1}$ wide channels. If the transition is blended, the species is indicated.}
\tablenotetext{b}{$3\sigma$ limit to the integrated intensity assuming an emission width of $150\,\rm km\,s^{-1}$.}
\end{table}
\clearpage

\subsection{The nuclear environment in NGC\,253}

The detection of new molecules in the extragalactic ISM, far from being a mere census of known species, provides a
complete description of the chemical complexity within the nuclear environment of galaxies and a tool to establish the origin of 
this rich chemistry and the heating of the molecular gas.

In the following subsections we will compare the observed abundances in NGC\,253 with those available for five external galaxies 
(Sect.~\ref{GalaxComp}) and five prototypical sources within the Milky Way (Sect.~\ref{GalacComp}).
The comparison focuses mainly on the abundance ratios between the different molecular species compared to those observed
toward NGC\,253. 
This will minimize the systematic error in the absolute fractional abundances due to the H$_2$ column density determinations.
The relative abundances between molecules will provide us with a reliable chemical differentiation giving us
insights into the dominant chemical mechanism acting in each source.

\subsubsection{Comparison with other galaxies}
\label{GalaxComp}

We have selected the nearby galaxies NGC\,4945, M\,82 and IC\,342 where most of the known molecular species have also been observed,
and Maffei2 and NGC\,6946 for which fewer species have been detected.

Table~\ref{tab:galaxh2} presents the integrated intensities of the $^{13}$CO $J=2-1$ transition that has been used to estimate the
H$_2$ source averaged column density. We used the conversion $N({\rm H_2})= 3.3\,10^{20}{\rm cm^{-2}}\,I({\rm ^{13}CO}_{2-1})$
\citep{Mauers03}.
\clearpage
\begin{table}[!h]
\caption{Molecular hydrogen source averaged column density}
\begin{center}
\begin{tabular}{l c c c c}
\tableline
\tableline  
Source              &   Transition      &  $\int{T_{\rm MB}\,{\rm d}v}$             &    $\theta_{\rm beam}$   &   $N_{\rm H_2}$      \\
                    &   $J-J'$          &  K\,km\,s$^{-1}$                          &     ($''$)               &     cm$^{-2}$        \\ 
\tableline
NGC\,253            &  $^{13}$CO $2-1$  &  52.4\tablenotemark{a}                    &     34                   &   $6.7\,10^{22}$     \\
NGC\,4945           &  $^{13}$CO $2-1$  &  81.2\tablenotemark{b}                    &     23                   &   $6.4\,10^{22}$     \\
M\,82               &  $^{13}$CO $2-1$  &  29.4\tablenotemark{a}                    &     34                   &   $3.8\,10^{22}$     \\
IC\,342             &  $^{13}$CO $2-1$  &  17.2\tablenotemark{a}                    &     34                   &   $2.2\,10^{22}$     \\
Maffei\,2           &  $^{13}$CO $2-1$  &  13.9\tablenotemark{a}                    &     34                   &   $1.8\,10^{22}$     \\
NGC\,6946           &  $^{13}$CO $2-1$  &  22.2\tablenotemark{c}                    &     21                   &   $1.5\,10^{22}$     \\
\tableline  
\end{tabular}
\end{center}
\tablecomments{A 20$''$ source size was assumed for all targets.}
\tablenotetext{a}{10\,m HHT by \citet{Mauers03}.}
\tablenotetext{b}{15\,m SEST by \citet{Wang04}.}
\tablenotetext{c}{30\,m IRAM by \citet{Israel01}.}
\label{tab:galaxh2}
\end{table}
\clearpage
Table~\ref{tab:Galaxcompare} shows the derived fractional abundances relative to H$_2$ towards the selected galaxies.
Because of the frequency coverage of our survey, not all the species listed in Table~\ref{tab.census} could be measured.
Therefore, in Table~\ref{tab:Galaxcompare}, we have also included data of additional key molecular species from other studies, 
namely NH$_3$, N$_2$H$^+$ and CN.
H$_2$ and CO are not included as they are tracers of the total mass of the molecular material content.
Species such as OH and H$_2$O have also been detected towards NGC\,253 and many other galaxies.
However, H$_2$O exhibits maser emission that does not allow us to determine directly the column densities,
while OH shows extremely complex behavior involving a mixture of absorption, thermal and maser emission
\citep[e.g.][]{Turner85, Ho87, Frayer98,Bradford99, Henkel04, Goico05}.
Unfortunately, there is not enough data available on CH, CH$^+$, HCO, CO$^+$ and HOC$^+$  for this comparison.

In order to achieve the highest homogeneity in the derived abundances,
we have collected all the line profile parameters from the available observations in the literature. 
From these values we have made our own estimates of the column densities for each molecule and each source with the
assumptions used for NGC\,253.

We have assumed a source size $\theta_s=20''$ for {\it all} these sources (see e.g. Sect.~\ref{sect.uncertainty} for NGC\,253 
and \citealt{Wang04} for NGC\,4945). 
This is a fairly simple assumption for some of the targets
and may lead to considerable errors in the column density determination. As this factor is also included in the source
averaged column density determination of 
the molecular hydrogen, the error in the resulting fractional abundance will, however, cancel out to first order.

\begin{figure}[p]
\centering
	\includegraphics[angle=-90,width=0.8\linewidth]{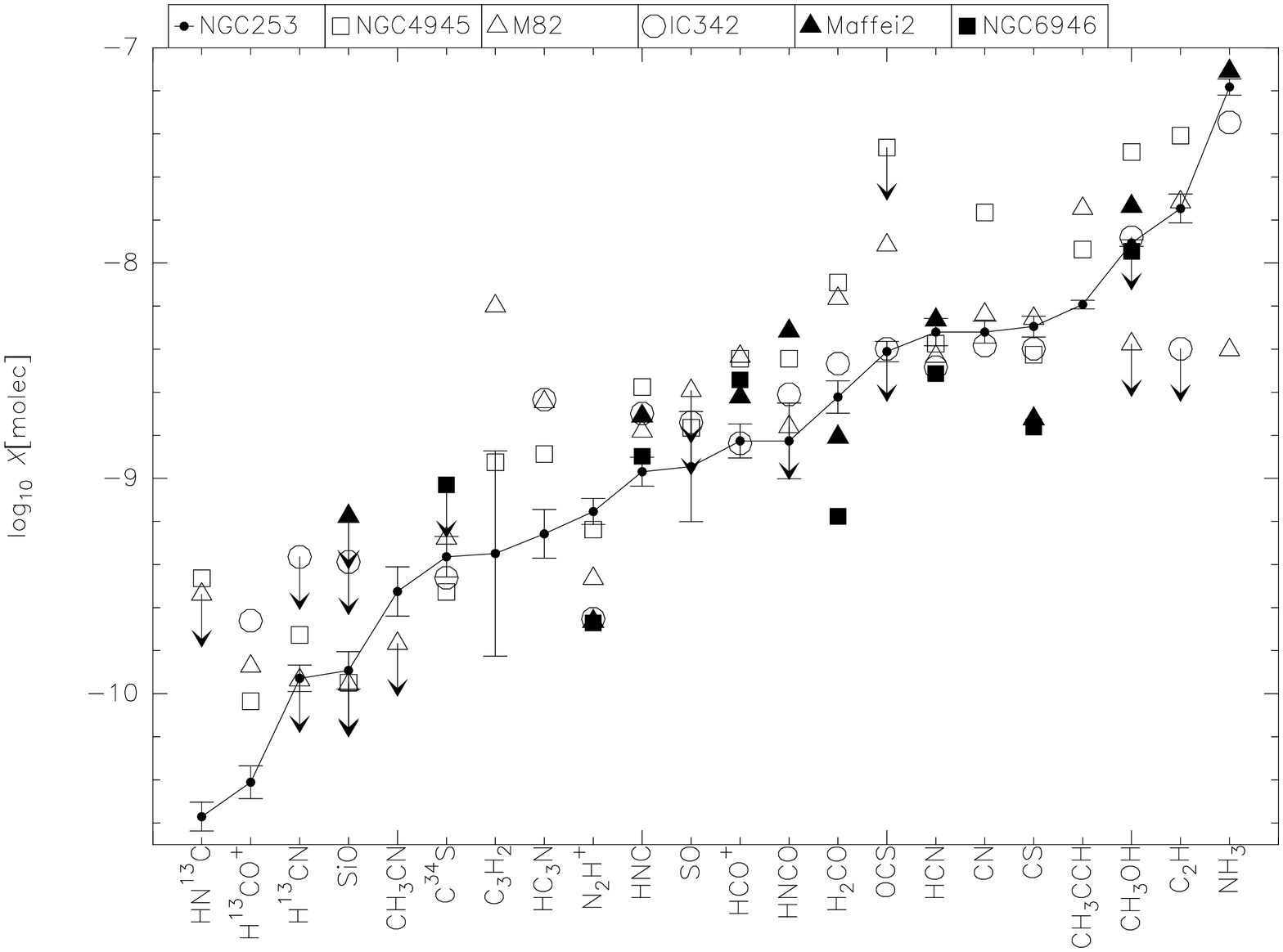}
	\includegraphics[angle=-90,width=0.8\linewidth]{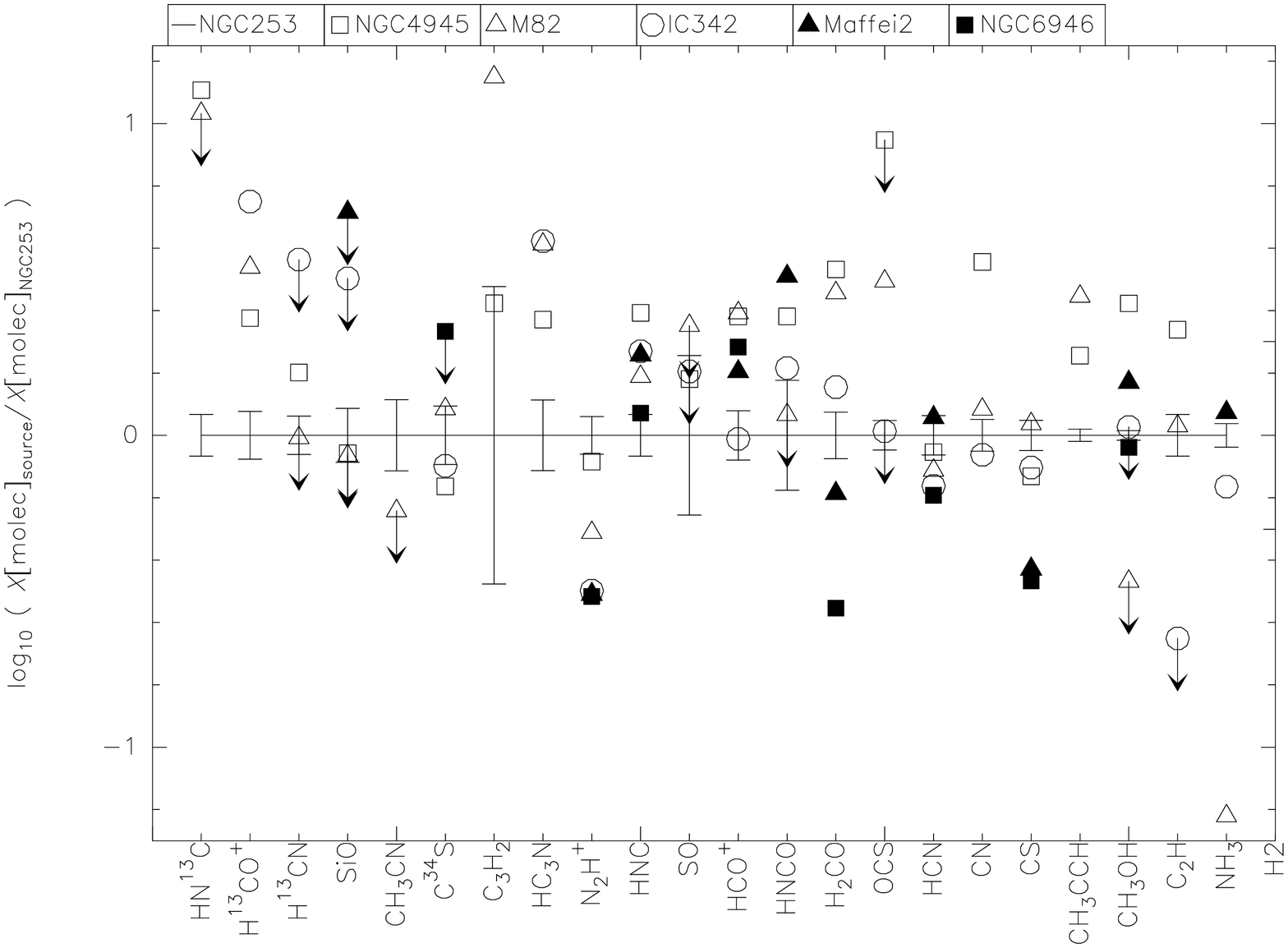}
	\caption{
        Molecular abundances ($X$) of selected extragalactic sources compared to those of NGC\,253. 
	Arrows represent upper limits. 
	A continuous line connects the abundances in NGC\,253.
	{\it Top panel:} Logarithmic fractional abundances relative to H$_2$ as presented in Table~\ref{tab:Galaxcompare}.
	{\it Bottom panel:} Logarithmic abundances for each species normalized to that measured in NGC\,253.
	}
	\label{fig.Galaxcompare}
\end{figure}

The top panel in Fig.~\ref{fig.Galaxcompare} shows a graphical representation of the data in Table~\ref{tab:Galaxcompare}. 
Molecular species are ordered based on increasing abundances in NGC\,253.
We have also considered the logarithmic difference of abundances in each selected galaxy with respect to the abundances derived in this work for NGC\,253 
as $\Delta X =  {\rm log}_{10}\,(X/X_{\rm NGC\,253})$.
This variable, $\Delta X$, is plotted in the bottom panel of Fig.~\ref{fig.Galaxcompare}.
The difference in the $Y$-axis between two species provides a measure of their abundance ratio
relative to that in NGC\,253.

{\it M\,82} shows relative abundances quite similar to those of NGC\,253 in many but not all of the observed species. 
Evident chemical differences between both nuclear starbursts
were already analyzed 
\citep{Mauers93,Takano95,Takano02,Wang04}, and have been interpreted as a difference in the evolutionary stage of both starbursts.
Compared to that in NGC\,253, the nuclear starburst in M\,82 represents an evolved state where PDRs dominate the heating of the 
molecular material \citep{Burillo02}.
From our comparison, the high relative abundance of C$_3$H$_2$, a common PDR tracer, is particularly outstanding.
On the other hand we note remarkably low abundances of CH$_3$OH and SiO, which are thought to be ejected from grain mantles
into the gas-phase \citep{Charnley95,Pintado92,Requena05}. 
This suggests that shocks do not play the dominant role within the nuclear region of M\,82. 
Molecules such as HNCO, apparently associated with shock chemistry \citep{Zinchen00}, also show low relative abundances compared to those in NGC\,253.
Furthermore, the nuclear environment of M\,82 is characterized by systematically lower relative abundances of CH$_3$CN, NH$_3$ and N$_2$H$^+$.
It is also interesting to note how other species used as PDR tracers, namely C$^{34}$S and C$_2$H \citep{Meier05}, do not show high relative
abundances in M\,82 compared to the other observed species. 
Moreover, they are a factor of $\sim2-3$ less abundant than most other the species.
If we compare the M\,82 relative abundances with those of the Orion Bar in Sect.~\ref{GalacComp}, we find that all the species with
low abundances in M\,82
show the same behavior in the Orion Bar. 
This comparison strongly supports the idea of M\,82 as an evolved nuclear starburst largely dominated by PDRs.

In {\it NGC\,4945}, relative abundances of all the detected molecular species resemble within a factor of 3 those of NGC\,253. 
The non-detection of SiO indicates a clear underabundance, similar to what is seen towards M\,82. On the other hand, unlike in M\,82, 
CH$_3$OH shows an abundance similar to that in NGC\,253. This resemblance between the abundances in NGC\,4945 and NGC\,253, altogether
with excitation considerations and a detailed study of species such as CN, HCN, and HNC, has been interpreted as an intermediate 
evolutionary state, between NGC\,253 and M\,82 \citep{Wang04}. 
From our comparison, the nuclear starburst of NGC\,4945 shows much closer resemblance to NGC\,253 than to M\,82.

{\it IC\,342} shows high relative abundances of HC$_3$N and H$^{13}$CO$^+$, while molecules such as 
C$_2$H and N$_2$H$^+$ have lower abundances as compared with NGC\,253.
Many of the available observations for IC\,342 (marked by an asterisk in Table~\ref{fig.Galaxcompare}) were obtained near the D position as labeled by
 \citet{Downes92}, $\sim 15''$ north of the main nuclear star formation region.
The abundance of C$_2$H given in Table~\ref{fig.Galaxcompare} has been taken from \citet{Meier05} towards the D position.
Its relative abundance is low compared to that in NGC\,253.
Even taking the C$_2$H abundance at the position where this molecule peaks \citep[i.e. a value an order of magnitude higher towards
the central position,][]{Meier05},
the relative abundances of this species with respect to the rest of molecules would be similar to that in NGC\,253.
High resolution maps from \citet{Meier05} clearly differentiate a region dominated by UV radiation from the central nuclear cluster 
and the region where the chemistry is dominated by shocks produced by collisions of clouds in a barred potential.
Given the position towards which many of the spectra were taken, away from the dynamical center, the molecular emission of IC\,342 should resemble 
more that of NGC\,253 than that of M\,82.
In fact IC\,342 does not show the relatively low HNCO abundance seen in M\,82
but does show a similar low relative abundance of C$_2$H.
Still, there is a large number of molecules not yet observed which are critical to fully evaluate the chemistry of IC\,342, 
such as SiO, SO, OCS and CH$_3$OH.

As far as {\it Maffei\,2 \rm and \it NGC\,6946} are concerned, the available molecular information is not sufficient to 
establish a clear chemical differentiation from the other sources. 
What can be noted is a remarkably high relative abundance of HNCO in 
Maffei\,2 compared to the other galaxies and a low relative abundance of H$_2$CO in NGC\,6946.
Both galaxies also show low relative abundances of N$_2$H$^+$.
\clearpage
\begin{table}[p]
\caption{Molecular fractional abundances toward NGC\,253 and other galaxies}
\label{tab:Galaxcompare}
\begin{center}
\scriptsize
\begin{tabular}{l c c c c c c}
\tableline
\tableline
Molecule         &   NGC      &  NGC         &  M\,82       &  IC                           &  Maffei2     & NGC          \\
                 &      253   &       4945   &              &      342                      &              &      6946    \\
\tableline
HN$^{13}$C       &  $-10.6$   &  $   -9.5 $  &   $   -9.5$  &   $    ...$                   &  $    ...$   &  $    ...$   \\
H$^{13}$CO$^+$   &  $-10.4$   &  $   -10.0$  &   $   -9.9$  &   $   -9.5$\tablenotemark{*}  &  $    ...$   &  $    ...$   \\
H$^{13}$CN       &  $-9.9 $   &  $   -9.7 $  &   $ < -9.9$  &   $ < -9.4$\tablenotemark{*}  &  $    ...$   &  $    ...$   \\
SiO              &  $-9.9 $   &  $ < -9.9 $  &   $ < -9.9$  &   $ < -9.3$\tablenotemark{*}  &  $ < -9.1$   &  $    ...$   \\
CH$_3$CN         &  $-9.5 $   &  $    ... $  &   $ < -9.7$  &   $    ...$                   &  $    ...$   &  $    ...$   \\
C$^{34}$S        &  $-9.4 $   &  $   -9.4 $  &   $   -9.3$  &   $   -9.5$\tablenotemark{*}  &  $    ...$   &  $ < -9.0$   \\
c-C$_3$H$_2$     &  $-9.3 $   &  $   -8.9 $  &   $   -8.1$  &   $    ...$                   &  $    ...$   &  $    ...$   \\
HC$_3$N          &  $-9.2 $   &  $   -8.8 $  &   $   -8.7$  &   $   -8.6$                   &  $    ...$   &  $    ...$   \\
N$_2$H$^+$       &  $-9.2 $   &  $   -9.2 $  &   $   -9.4$  &   $   -9.6$\tablenotemark{*}  &  $   -9.7$   &  $   -9.6$   \\
HNC              &  $-9.0 $   &  $   -8.6 $  &   $   -8.8$  &   $   -8.7$                   &  $   -8.7$   &  $   -8.8$   \\
SO               &  $-8.9 $   &  $   -8.7 $  &   $ < -8.5$  &   $ < -8.7$                   &  $    ...$   &  $    ...$   \\
HCO$^+$          &  $-8.8 $   &  $   -8.4 $  &   $   -8.4$  &   $   -8.8$                   &  $   -8.6$   &  $   -8.5$   \\
HNCO             &  $-8.8 $   &  $   -8.4 $  &   $ < -8.8$  &   $   -8.6$                   &  $   -8.3$   &  $    ...$   \\
H$_2$CO          &  $-8.6 $   &  $   -8.1 $  &   $   -8.2$  &   $   -9.4$                   &  $   -8.8$   &  $   -9.1$   \\
OCS              &  $-8.4 $   &  $ < -7.5 $  &   $   -7.9$  &   $ < -8.6$                   &  $    ...$   &  $    ...$   \\
CN               &  $-8.3 $   &  $   -7.7 $  &   $   -8.2$  &   $   -8.3$\tablenotemark{*}  &  $    ...$   &  $    ...$   \\
HCN              &  $-8.3 $   &  $   -8.3 $  &   $   -8.4$  &   $   -8.5$\tablenotemark{*}  &  $   -8.3$   &  $   -8.5$   \\
CS               &  $-8.2 $   &  $   -8.4 $  &   $   -8.2$  &   $   -8.4$\tablenotemark{*}  &  $   -8.7$   &  $   -8.7$   \\
CH$_3$CCH        &  $-8.2 $   &  $   -8.0 $  &   $   -7.7$  &   $    ...$                   &  $    ...$   &  $    ...$   \\
CH$_3$OH         &  $-7.9 $   &  $   -7.4 $  &   $ < -8.3$  &   $ < -7.8$                   &  $   -7.7$   &  $   -7.9$   \\
C$_2$H           &  $-7.7 $   &  $   -7.3 $  &   $   -7.6$  &   $ < -8.4$                   &  $    ...$   &  $    ...$   \\
NH$_3$           &  $-7.2 $   &  $    ... $  &   $   -8.4$  &   $   -7.3$                   &  $   -7.1$   &  $    ...$   \\
\tableline                                                                                                          
\end{tabular}
\end{center}
\tablenotetext{*}{
\scriptsize 
Observations taken $\sim 15''$ away from the central position (see Sect.~\ref{GalaxComp} for details).}
\tablerefs{
\scriptsize
NGC\,253, 
this paper, \citet{Martin03}[NS,NO], \citet{Martin05}[OCS], \citet{Mauers03}[NH$_3$], \citet{Henkel88}[CN], \citet{Mauers91a}[N$_2$H$^+$];\\
NGC\,4945, 
\citet{Wang04}, \citet{Henkel94}[SiO];\\
M\,82, 
\citet{Mauers89}[CS,C$^{34}$S], \citet{Mauers91a}[N$_2$H$^+$,SiO,H$^{13}$CO$^+$], \citet{Mauers91b}[C$_3$H$_2$,CH$_3$CN,CH$_3$CCH],
\citet{Mauers95}[OCS], \citet{Hutte97}[H$_2$CO,CH$_3$OH], \citet{Henkel88}[CN,HC$_3$N,C$_2$H], \citet{Henkel98}[HCO$^+$,HCN,H$^13$CN],
\citet{Petu92}[SO], \citet{Nguyen91}[HNCO], \citet{Nguyen92}[HCN,HCO$^+$,H$^{13}$CO$^+$], \citet{Wild90}[HCN,HCO$^+$],
\citet{Weiss01}[NH$_3$],\citet{Hutte95}[HNC];\\
IC\,342, 
\citet{Mauers89}[CS], \citet{Mauers91a}[N$_2$H$^+$,SiO,H$^{13}$CO$^+$], \citet{Mauers95}[C$^{34}$S,OCS,HC$_3$N,H$_2$CO],
\citet{Henkel88}[CN,HNC,HC$_3$N], \citet{Hutte95}[HNC], \citet{Hutte97}[H$_2$CO,CH$_3$OH], \citet{Petu92}[SO],
\citet{Wild90}[HCN,HCO$^+$], \citet{Nguyen91}[HNCO], \citet{Nguyen92}[HCN,HCO$^+$,H$^{13}$CO$^+$],
\citet{Henkel98}[HCN, H$^{13}$CN], \citet{Meier05}[C$_2$H], \citet{Mauers03}[NH$_3$];\\
Maffei2, \citet{Mauers89}[CS], \citet{Mauers91a}[SiO,N$_2$H$^+$], \citet{Hutte97}[H$_2$CO, CH$_3$OH], \citet{Nguyen91}[HNCO],
\citet{Nguyen92}[HCO$^+$,HCN], \citet{Henkel00,Mauers03}[NH$_3$];\\
NGC\,6946, \citet{Hutte97}[H$_2$CO, CH$_3$OH], \citet{Mauers89}[CS, C$^{34}$S], \citet{Nguyen92}[HCO$^+$, HCN], \citet{Mauers91a}[N$_2$H$^+$].
}
\end{table}
\clearpage
\clearpage
In order to quantify this comparison, Table~\ref{tab.GalaxcompareRMS} shows the mean and rms of $\Delta X$ for each source for two cases.
The first only uses the species detected in both sources. The number of molecules included in the calculations is also indicated.
The second also uses upper limits to the abundances derived for the comparison sources. To take these limits into 
account, we followed two criteria: upper limits higher than the abundances measured towards NGC\,253 were rejected as they do not
provide useful information for the comparison; for the calculations, the other observed limits, lower than the abundances in
NGC\,253, where divided by two as an estimated true value.
The second assumption may overestimate some abundances, but any other constraint might result in biased 
guesses.
While the mean value, $\overline {\Delta X}$, only gives an the overall absolute difference in the abundances of each source as
compared with NGC\,253, the rms of $\Delta X$ provides us with a fairly reliable measurement of differences in the abundance
pattern that defines the chemistry of each object.
M\,82, as pointed out before, clearly differs chemically from NGC\,253, showing the highest rms values.
NGC\,4945 and IC\,342 are much closer to NGC\,253.
\clearpage
\begin{table}[!t]
\begin{center}                     
\caption{Statistical comparison of abundances of selected galaxies with those in NGC\,253.}
\label{tab.GalaxcompareRMS}
\begin{tabular}{l c c c c c c}
\tableline
\tableline
            & \multicolumn{3}{c}{Without limits}            & \multicolumn{3}{c}{With limits}                \\
Source      & $\overline {\Delta X}$   & rms$(\Delta X)$ &   num. & $\overline {\Delta X}$   & rms$(\Delta X)$  &   num. \\
\tableline
NGC\,4945   &  0.30                    &  0.29           &   18   &  0.27                    &  0.32            &  19    \\
M\,82       &  0.19                    &  0.51           &   15   &  0.05                    &  0.54            &  19    \\
IC\,342     &  0.08                    &  0.34           &   12   &  0.00                    &  0.42            &  13    \\
Maffei\,2   &  0.02                    &  0.31           &   9    &  0.02                    &  0.31            &   9    \\
NGC\,6946   & -0.20                    &  0.30           &   7    & -0.20                    &  0.30            &   7    \\
\tableline
\end{tabular}
\tablecomments{$\Delta X =  {\rm log}_{10}\,(X/X_{\rm NGC\,253})$. See last paragraph in Sect.~\ref{GalaxComp} for details.}
\end{center}
\end{table}
\clearpage
\subsubsection{Comparison with prototype Galactic sources}
\label{GalacComp}

We have selected five Galactic sources, namely Sgr\,B2(N), Sgr\,B2(OH), TMC-1, L134N, and the Orion Bar, which are 
considered to be prototypes of their respective chemistry within the Galaxy. 
All these sources have been the target of multi-transition studies and some of them of frequency surveys.
Table~\ref{tab:Galaccompare} shows the abundances of all the molecules observed in the NGC\,253 survey compared to
those of these Galactic sources. 
In order to achieve consistency among ratios,
we have tried, when available, to use values from line surveys or multi-line studies for each source.

\begin{figure}[p]
\centering
	\includegraphics[angle=-90,width=0.8\linewidth]{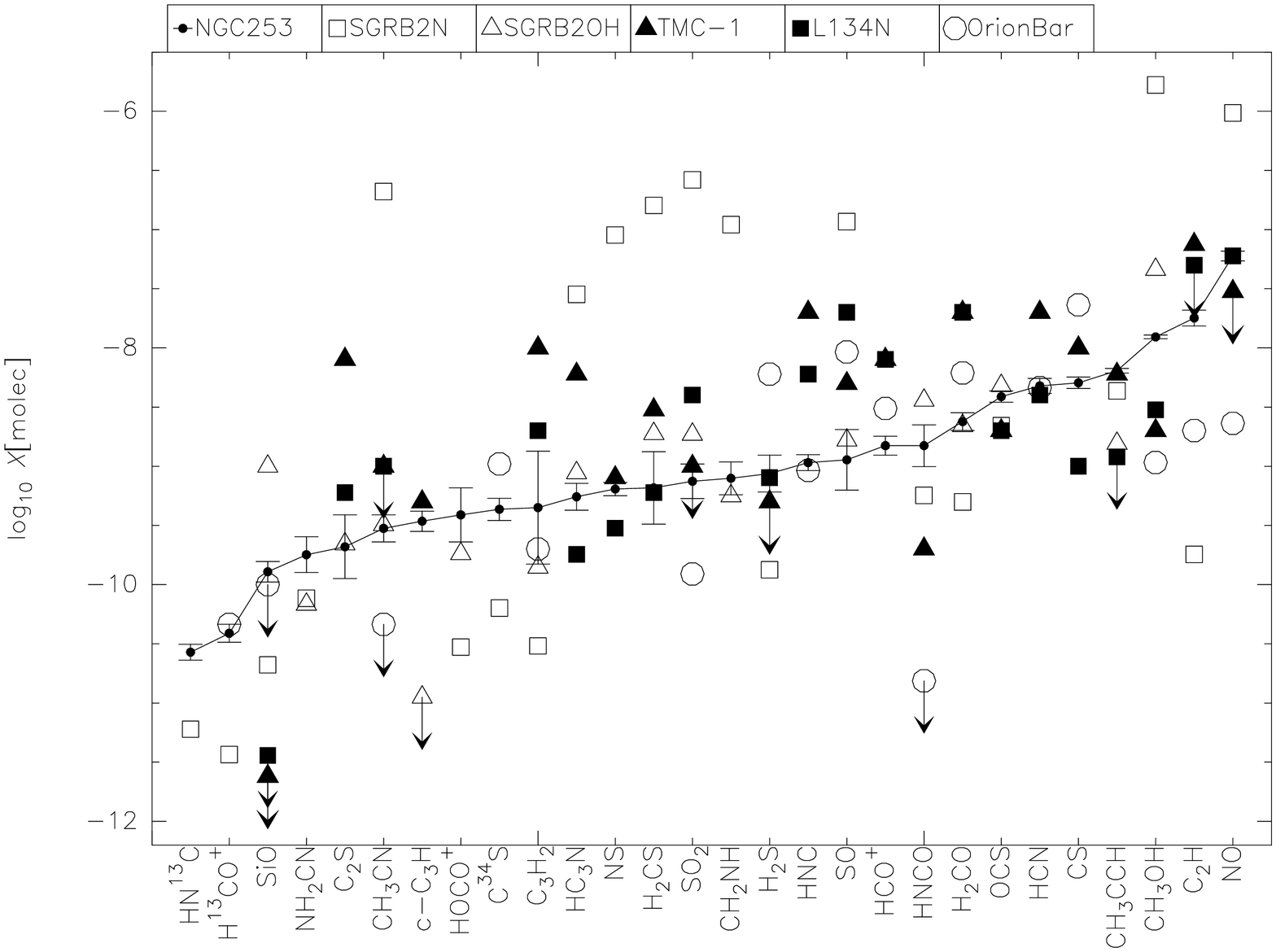}
	\includegraphics[angle=-90,width=0.8\linewidth]{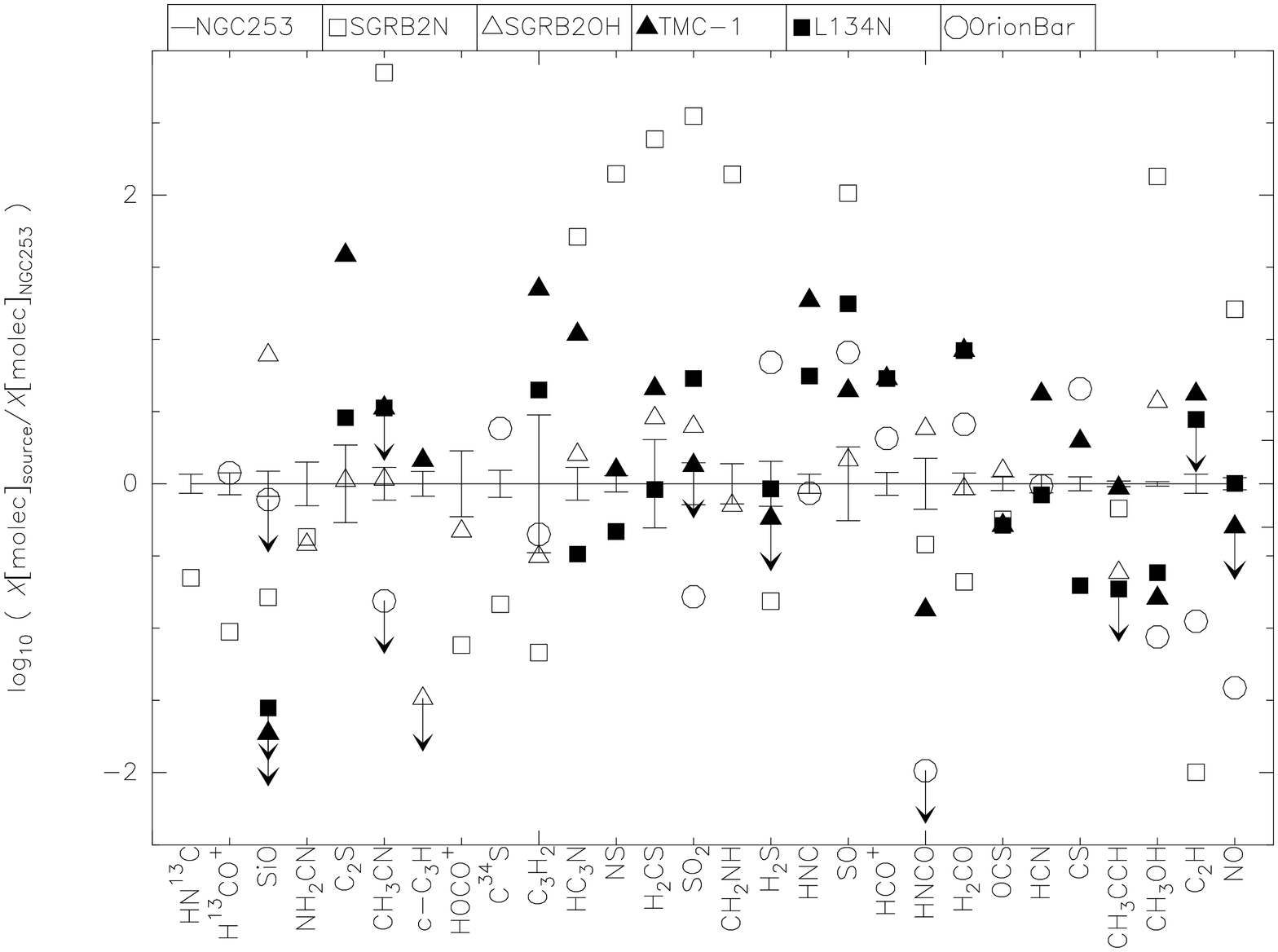}
	\caption{
        Molecular abundances ($X$) of selected Galactic sources compared to those of NGC\,253. 
	Arrows represent upper limits. A continuous line connects the abundances in NGC\,253.
	{\it Top panel:} Logarithmic fractional abundances relative to H$_2$ as shown in Table~\ref{tab:Galaccompare}.
	{\it Bottom panel:} Logarithmic abundances for each species normalized to that measured in NGC\,253.
	}
	\label{fig.Galaccompare}
\end{figure}

In the same way as in the extragalactic comparison, the top panel in Fig.~\ref{fig.Galaccompare} shows a graphical representation of 
the data in Table~\ref{tab:Galaccompare}, where molecular species are ordered based on increasing NGC\,253 abundances.
The bottom panel in Fig.~\ref{fig.Galaccompare} shows the logarithmic abundance of each molecule normalized to the abundance measured
in NGC\,253.

{\it Sgr\,B2(N)} is believed to represent the prototype of a {\it hot molecular core chemistry} associated with massive star formation
near the nucleus of the Milky Way.
A large number of molecular species have been identified towards this source which is the brightest molecular line emitter
within the Galaxy \citep[and references therein]{Friedel04}.
There are clear differences between the abundances found towards Sgr\,B2(N) and those in NGC\,253.
A set of molecules, namely CH$_3$CN, HC$_3$N, NS, H$_2$CS, SO$_2$, CH$_2$NH, SO, CH$_3$OH and NO, 
show abundances relative to the other species $\sim 2-3$ orders of magnitude larger than what we observe towards NGC\,253. 
Almost all the other species observed towards Sgr\,B2(N) have smaller abundances than in NGC\,253, but their relative abundances 
are, within an order of magnitude, similar to those in NGC\,253.
Exceptions are HOCO$^{+}$ and C$_2$H that show lower relative abundances by factors $\sim$30 and 100, respectively, in the Sgr\,B2(N) hot core.
Moreover, the rotation temperatures derived from the observed molecules in NGC\,253 are low compared to the typical temperatures of 
$>70$\,K observed towards hot cores.
Even if hot cores associated with massive protostars are present in the nuclear environment of NGC\,253, they clearly 
do not dominate the molecular emission.
While the angular size of the Sgr\,B2 molecular cloud complex is $\sim 18'$ \citep{Scoville75}, 
the emission of some large complex molecules towards Sgr\,B2(N) appears concentrated within a region of $\sim 5''$, 
corresponding to 0.2\,pc at a distance of 8.5\, kpc \citep[][]{Kerr86}. 
At the distance of NGC\,253, such a source would have a diameter of $<0\farcs 02$, and because of beam dilution, its detection 
is certainly out of reach for the 30\,m telescope.
A large number of such sources would be needed in order to be able to observe traces of their complex chemistry.

{\it The Orion Bar} has been selected as the prototype of {\it photodissociation regions} (PDRs).
A large fraction of the molecular gas in the Milky Way
and in external galaxies is expected to be affected by PDRs \citep{Hollenbach97}. 
In PDRs, molecular emission originates from the surface layers of interstellar 
molecular clouds exposed to intense far-ultraviolet photons from nearby young OB stars.
The non-detection of molecules such as CH$_3$CN and HNCO towards the Orion PDR clearly indicates a low relative abundance of these 
species with respect to other molecules, which is in contrast with the observed NGC\,253 abundances. 
Similarly, the relative abundances of SiO, SO$_2$, CH$_3$OH, C$_2$H and NO appear to be $\sim 1-2$ orders of magnitude lower in the
Orion bar than in NGC\,253.
High abundances of hydrocarbon radicals, such as C$_2$H and C$_3$H$_2$, are observed towards PDRs \citep{Fosse00,Fuente93,Fuente03,Pety05}.
In this context, it is surprising to see how C$_2$H and C$_3$H$_2$, commonly used as PDR tracers, show 
relatively low abundances in the Orion PDR when compared not only with NGC\,253 but also with the prototypical Galactic sources
except the Sgr\,B2(N) hot core.
The observations of the Orion Bar have been taken from the position of maximum H$_2$ column density \citep{Jansen95,Fuente96},
which does not match the position of highest C$_3$H$_2$ abundance \citep{Fuente03}.
However, the contribution of the position of the C$_3$H$_2$ emission peak to the total column density is smaller than that of the 
H$_2$ column density peak. 
Therefore, it is expected that the molecular abundance of the H$_2$ peak dominates the emission of PDRs, also in external galaxies.
Anyhow, even if we consider the position of the larger C$_3$H$_2$ abundance 
\citep[up to one order of magnitude,][]{Fuente03}, the relative abundance would still be close to that found in NGC\,253.
From this comparison as well as that to M\,82 (see Sect.~\ref{GalaxComp}) it is clear that the chemistry of the nuclear environment of NGC\,253 
is not dominated by photodissociation.

{\it Quiescent cold dark clouds} are represented by {\it TMC-1 and L134N}. The molecular composition at sites of 
low mass star formation is dominated by gas-phase ion-molecule chemistry due to a lack of embedded luminous sources.
In our comparison, both dark cloud complexes have in common low relative abundances of SiO and, to a lesser degree, of CH$_3$OH. 
The remaining species behave quite differently in each source.
SO$_2$, H$_2$S, HNCO and NO, show low abundances in TMC-1 while C$_2$S and C$_3$H$_2$ appear to have abundances larger than in NGC\,253 by up to
one order of magnitude.
On the other hand, L134N shows a high SO abundance and relatively low abundances of HC$_3$N and CH$_3$CCH.
The abundance pattern defined by these two dark clouds clearly does not follow the relative abundances found towards NGC\,253.
In addition, even the rotation temperatures derived for NGC\,253, which in general are smaller than $T_{\rm kin}$, are higher 
than the typical kinetic temperature ($T_{\rm kin}\sim 10$\,K) measured towards dark clouds. 

{\it Sgr\,B2(OH)}, located at the southern end of the Sgr\,B2 molecular envelope, is free from the emission steming from the hot cores in this cloud.
This position in the Sgr\,B giant molecular cloud is taken as the prototype of {\it Galactic center (GC) molecular cloud complexes} \citep{Pintado97}.
The H$_2$ column density of this source has been estimated to be roughly $N(\rm H_2)=5\,10^{23}$ 
\citep{Nummelin00}. 
The relative abundances of all the species measured towards this Galactic center molecular cloud complex appear to closely follow 
the abundance pattern in NGC\,253 within a factor $\sim5$.
The only species clearly underabundant in Sgr\,B2(OH) when compared to NGC\,253 is c-C$_3$H but this molecule is only tentatively
detected towards NGC\,253.
Thus, the ratios between any of the observed species are similar to those in NGC\,253 within one order of magnitude.
\clearpage
\begin{table}[p]                                      
\caption{Comparison of NGC\,253 fractional molecular abundances with those in selected Galactic sources}
\begin{center}
\scriptsize
\begin{tabular}{l c c c c c c}
\tableline
\tableline
Molecule         &   NGC        &  Sgr         &  Sgr        &  TMC-1      &  L134N      &  Orion      \\
                 &        253   &       B2(N)  &       B2(OH)&             &             &        Bar    \\
\tableline                                                                                                      
HN$^{13}$C       &    $ -10.6$  &   $  -11.0 $ &  $     ... $&  $     ...$ &  $     ...$ &    $    ... $ \\ 
H$^{13}$CO$^+$   &    $ -10.4$  &   $  -11.4 $ &  $     ... $&  $     ...$ &  $     ...$ &    $   -10.3$ \\   
SiO              &    $  -9.9$  &   $  -10.7 $ &  $     ... $&  $ < -11.6$ &  $ < -11.4$ &    $   -10.3$ \\   
NH$_2$CN         &    $  -9.7$  &   $  -10.1 $ &  $   -10.0 $&  $     ...$ &  $     ...$ &    $    ... $ \\   
C$_2$S           &    $  -9.7$  &   $   ...  $ &  $    -9.6 $&  $    -8.1$ &  $    -9.2$ &    $    ... $ \\   
CH$_3$CN         &    $  -9.5$  &   $  -6.7  $ &  $    -9.4 $&  $    -9.0$ &  $  < -9.0$ &    $ < -10.3$ \\   
c-C$_3$H         &    $  -9.5$  &   $ -10.5  $ &  $ < -10.9 $&  $    -9.3$ &  $     ...$ &    $    ... $ \\   
HOCO$^+$         &    $  -9.4$  &   $  -10.5 $ &  $    -9.7 $&  $     ...$ &  $     ...$ &    $    ... $ \\   
C$^{34}$S        &    $  -9.4$  &   $  -10.2 $ &  $     ... $&  $     ...$ &  $     ...$ &    $   -9.0 $ \\   
c-C$_3$H$_2$     &    $  -9.3$  &   $  -10.5 $ &  $    -9.8 $&  $    -8.0$ &  $    -8.7$ &    $   -9.7 $ \\   
HC$_3$N          &    $  -9.2$  &   $  -7.5  $ &  $    -9.0 $&  $    -8.2$ &  $    -9.7$ &    $    ... $ \\   
NS               &    $  -9.2$  &   $  -7.0  $ &  $     ... $&  $    -9.1$ &  $    -9.5$ &    $    ... $ \\   
H$_2$CS          &    $  -9.2$  &   $  -6.8  $ &  $    -8.7 $&  $    -8.5$ &  $    -9.2$ &    $    ... $ \\   
SO$_2$           &    $  -9.1$  &   $  -6.6  $ &  $    -8.7 $&  $  < -9.0$ &  $    -8.4$ &    $   -9.9 $ \\   
CH$_2$NH         &    $  -9.1$  &   $  -7.0  $ &  $    -9.2 $&  $     ...$ &  $     ...$ &    $    ... $ \\   
H$_2$S           &    $  -9.1$  &   $  -9.9  $ &  $     ... $&  $  < -9.3$ &  $    -9.1$ &    $   -8.2 $ \\   
HNC              &    $  -9.0$  &   $   ...  $ &  $     ... $&  $    -7.7$ &  $    -8.2$ &    $   -9.0 $ \\   
SO               &    $  -8.9$  &   $  -6.9  $ &  $    -8.7 $&  $    -8.3$ &  $    -7.7$ &    $   -8.0 $ \\   
HCO$^+$          &    $  -8.8$  &   $   ...  $ &  $     ... $&  $    -8.1$ &  $    -8.1$ &    $   -8.5 $ \\   
HNCO             &    $  -8.8$  &   $  -9.2  $ &  $    -8.4 $&  $    -9.7$ &  $     ...$ &    $< -10.8 $ \\   
H$_2$CO          &    $  -8.6$  &   $  -9.3  $ &  $    -8.6 $&  $    -7.7$ &  $    -7.7$ &    $   -8.2 $ \\   
OCS              &    $  -8.4$  &   $  -8.6  $ &  $    -8.3 $&  $    -8.7$ &  $    -8.7$ &    $    ... $ \\   
HCN              &    $  -8.3$  &   $   ...  $ &  $     ... $&  $    -7.7$ &  $    -8.4$ &    $   -8.3 $ \\   
CS               &    $  -8.2$  &   $   ...  $ &  $     ... $&  $    -8.0$ &  $    -9.0$ &    $   -7.6 $ \\   
CH$_3$CCH        &    $  -8.2$  &   $  -8.4  $ &  $    -8.8 $&  $    -8.2$ &  $  < -8.9$ &    $    ... $ \\   
CH$_3$OH         &    $  -7.9$  &   $  -5.8  $ &  $    -7.3 $&  $    -8.7$ &  $    -8.5$ &    $   -9.0 $ \\   
C$_2$H           &    $  -7.7$  &   $  -9.7  $ &  $     ... $&  $    -7.1$ &  $  < -7.3$ &    $   -8.7 $ \\   
NO               &    $  -7.2$  &   $  -6.0  $ &  $     ... $&  $  < -7.5$ &  $    -7.2$ &    $   -8.6 $ \\   
\tableline                                                                                                      
\end{tabular}                
\end{center}                     
\tablerefs{
\scriptsize
{\it NGC\,253}: this paper and \citet{Martin03}[NS,NO],\citet{Martin05}[OCS];\\
{\it Sgr\,B2N}: \citet{Nummelin00} and \citet{Minh91}[H$_2$S]. $N(\rm H_2)=3.0\,10^{24}\,cm^{-2}$ from \citet{Nummelin00};\\
{\it Sgr\,B2(OH)}: \citet{Cummins} and \citet{Turner91}. $N(\rm H_2)=5\,10^{23}\,cm^{-2}$ (see text for details on this estimate); \\
{\it TMC-1 and L134N}: \citet{Ohishi92}, \citet{McGonagle94}[NS], \citet{Ziurys89}[SiO]. $N(\rm H_2)=1.0\,10^{22}\,cm^{-2}$ from \citet{Ohishi92};\\
{\it Orion Bar}: \citet{Jansen95}, \citet{Fuente03}[C$_3$H$_2$], \citet{Schilke01}[SiO]; $N(\rm H_2)=6.5\,10^{22}\,cm^{-2}$ from \citet{Jansen95}.
}
\label{tab:Galaccompare}
\end{table}
\clearpage
\clearpage

In analogy to our statistical comparison of molecular abundances in prototypical nearby galaxies shown in Table~\ref{tab.GalaxcompareRMS}, 
Table~\ref{tab.GalaccompareRMS} presents a comparison between NGC\,253 and the selected Galactic sources.
To calculate the parameters, the available abundances in Table~\ref{tab:Galaccompare} have been taken into account
following the criteria explained at the end of Sect.~\ref{GalaxComp}. 
Given the uncertainty of its detection, c-C$_3$H was not included.
The computed rms of the logarithmic differences shows the envelope of the Sgr\,B2 complex as the one most closely 
resembling 
NGC\,253.
\clearpage
\begin{table}[t]
\begin{center}                     
\caption{Statistical comparison of abundances of selected Galactic sources with those in NGC\,253.}
\label{tab.GalaccompareRMS}
\begin{tabular}{l c c c c c c}
\tableline
\tableline
            & \multicolumn{3}{c}{Without limits}            & \multicolumn{3}{c}{With limits}                \\
Source      & $\overline {\Delta X}$   & rms$(\Delta X)$ &   num. & $\overline {\Delta X}$   & rms$(\Delta X)$  &   num. \\
\tableline
Sgr\,B2(N)  &   0.40                   &  1.50           &   22   &  0.40                    &   1.50           &   22   \\
Sgr\,B2(OH) &   0.07                   &  0.40           &   16   & -0.04                    &   0.58           &   17   \\
TMC-1       &   0.47                   &  0.66           &   18   &  0.26                    &   0.85           &   21   \\
L134N       &   0.18                   &  0.58           &   16   & -0.00                    &   0.76           &   18   \\
OrionBar    &  -0.07                   &  0.71           &   14   & -0.28                    &   0.85           &   17   \\
\tableline
\end{tabular}
\tablecomments{See Table~\ref{tab.GalaxcompareRMS} and Sect.~\ref{GalacComp} for details.}
\end{center}
\end{table}
\clearpage

The agreement in relative abundances between NGC\,253 and a typical Galactic center cloud is a remarkable and clear indication
that the heating and chemistry 
of the bulk of the ISM is dominated by the same processes.

The molecular envelope around the star forming region Sgr\,B2 is formed by a relatively dense and warm material 
\citep[$n_{\rm H_2}=2\,10^{5}\rm cm^{-3}$ and $T_{\rm kin}=40-60\,$K,][]{deVicente97}.
The large abundances of NH$_3$, SiO and $\rm C_2H_5OH$ observed in the gas phase of the Galactic center molecular clouds are claimed 
to be the result of grain chemistry and subsequent ejection onto the gas phase due to the disruption of grains by 
low velocity shocks \citep[$\rm v<20\,km\,s^{-1}$,][]{Flower95, Pintado97, Pintado01}.
In the Galactic center, the possible origin for these shocks is so far unclear. 
It has been claimed that they could be produced by the large-scale shocks associated with cloud-cloud 
collisions due to the orbital motion of molecular clouds in a barred potential \citep{Hasegawa94, Sato00} and the interaction of 
expanding bubbles due to supernova events and/or strong stellar winds from Wolf-Rayet stars 
\citep{Sofue90,deVicente97,Pintado99} associated with a burst of star formation that occurred $10^7$ years ago 
\citep{RodriFer04,RodriFer05}.

\section{Conclusions}

\begin{enumerate}
\item{ We present the first unbiased molecular line survey of an extragalactic source.
The survey covers the 2\,mm atmospheric window from 129.1 to 175.2\,GHz toward the inner 200\,pc of NGC\,253.
A total of 111 features are identified as transitions of 25 different molecular species, eight out of which (three tentatively)
have been detected for the first time outside the Milky Way.
The rare isotopes, $^{34}$SO and $\rm HC^{18}O^+$, were also detected for the first time in an extragalactic source.
Additionally, three hydrogen recombination lines and, tentatively, two deuterated species are identified, N$_2$D$^+$ and DNC.
The origin of the observed features of both N$_2$D$^+$ and DNC is still unclear and deserves further investigation.
If real, these would represent the first deuterated molecules observed beyond the Magellanic Clouds.
Column densities and rotation temperatures have been determined for each species under Local Thermodynamic Equilibrium (LTE) conditions.
}

\item{ As a result of this survey and by adding existing data from three molecules, namely NH$_3$, N$_2$H$^+$ and CN, 
we obtain the most complete description to date of the chemical complexity within the nuclear few hundred parsecs of
a starburst galaxy.
We have performed a comparison between the chemistry of NGC\,253 and those of five other outstanding nearby galaxies.
This comparison clearly shows the strong chemical differentiation between nuclei of galaxies.
Most prominent differences are observed between NGC\,253 and M\,82.
This can be interpreted
in terms of a more evolved stage of the nuclear starburst in M\,82
The chemistry of NGC\,4945, though claimed to be at an intermediate stage of evolution between NGC\,253 and M\,82, clearly 
resembles more that of NGC\,253 except for the lack of SiO.
Similarly, the position within the nuclear region of IC\,342, where most of the molecular observations have been made, show 
relative abundances close to those in NGC\,253. This position is far from the main region of star formation where PDRs 
are claimed to dominate the chemistry.

As far as Maffei2 and NGC\,6946 are concerned, the available molecular observations 
do not yet allow us to carry out a similarly detailed comparison.
}

\item{ A comparison of the molecular abundances of NGC\,253 with selected Galactic sources shows a striking
similarity between NGC\,253 and  Sgr\,B2(OH).
The latter source, located in the envelope of Sgr\,B2, is commonly taken as the prototype of the molecular cloud complexes in the Galactic center region.
This indicates that the chemistry of the nuclear molecular environment in NGC\,253 and the
Galactic center molecular clouds are driven by similar chemical processes. If this is the case, the chemistry and 
the heating of the nuclear molecular material in NGC\,253 is dominated by large-scale low velocity shocks.
}

\item{The comparison with Galactic sources also shows a close similarity between the abundances within the nuclear starburst
of M\,82 and those of the Orion Bar, the prototype of a photon dominated region. 
This resemblance fully supports the idea of the chemistry in M\,82 being dominated by PDRs.
}
\end{enumerate}

\acknowledgements
SM and RM were supported by the {\it Programas de Acciones Integradas} between Spain and Germany.
JMP has been partially supported by the Spanish Ministerio de Educaci\'on y Ciencia under projects ESP~2004-00665, AYA 2002-10113-E and AYA~2003-02785-E.
SGB acknowledges financial support from AYA~2003-07584 and ESP~2003-04957.
We would like to thank the referee, whose comments helped to significantly improve the paper.

\begin{appendix}

\section{Gaussian fitting to blended lines}
\label{apen.Blended}
Table~\ref{tab.fits} includes a label in the third column indicating if and how the respective spectral feature is affected by blending.
In the following the encountered different cases are described:
\begin{itemize}
\item{{\it b $\rightarrow$ blended}: 
Some observed features are composed of several transitions from different species. 
If the line is only partially blended, it is still possible to separate the individual components by fitting different 
Gaussian profiles. For these fits, sometimes it was necessary 
to fix the velocity and/or linewidth.}
\item{{$s \rightarrow synthetic$}: 
The blending may prevent us from making a reliable multiple Gaussian fitting to an observed feature. 
In this case, when we have observed several transitions of one of the species contributing to the blended feature, 
it is possible to use these to estimate the species' contribution to the blended feature
by interpolating or extrapolating the linear fit to the 
rotation diagrams (see Appendix~\ref{apen.PopDiag}).
This is equivalent to using Eq.~\ref{eq.PopDiag} to estimate the expected integrated intensity of a line from the derived 
column density and rotation temperature. 
Using the estimated line intensity and the linewidth and radial velocity of the other observed transitions we subtract 
the expected synthetic Gaussian profile from the spectrum.
The estimated integrated line intensities using this method have been plotted in the corresponding rotation 
diagrams with open symbols (Fig.~\ref{fig.PopDiag}).
The transitions from other species contributing to the blended feature have been fitted to the residual spectrum after the 
subtraction of the synthetic profile.}
\item{{\it hf $\rightarrow$ hyperfine}: 
Some species show transitions which consist of a group of unresolved hyperfine components. 
The relative intensities of these lines are fixed by their spectroscopic parameters, assuming LTE and
optically thin line emission.
Thus, the fitting has been carried out with a comb of Gaussian profiles at the rest frequencies of the hyperfine 
components with the same widths and with fixed line intensities relative to the main component which is taken as a free parameter. 
The derived parameters shown in Table~\ref{tab.fits} correspond to the 
integrated intensity of all the hyperfine components but the frequency, line intensity and velocity refers to the main 
hyperfine component.}
\item{{\it m $\rightarrow$ multi-transition}: 
Some features of species such as CH$_3$OH and CH$_3$CN consist of groups of overlapping transitions, 
involving different energy levels. 
In these cases, a single Gaussian profile has been fitted to the observed spectral feature.
The total integrated line intensity of all the blended transitions has been used to derive rotation diagrams 
as described in Appendix~\ref{Sect.BlendedPopDiag}.}
\end{itemize}

\section{Rotation Diagrams}
\subsection{Basics}
\label{apen.PopDiag}
Assuming optically thin emission for the observed molecular transitions, the column density in the upper level can be derived as
\begin{equation}
N_u=\frac{8 \pi k \nu^2}{h c^3 A_{ul}} \left(1-\frac{J_{\nu}(T_{\rm BG})}{J_{\nu}(T_{\rm ex})}\right)^{-1} \int{T_{\rm B}\, dv},
\label{eq.Nup}
\end{equation}

where $A_{ul}$, the Einstein coefficient for a transition connecting the levels $u$ and $l$, can be written as 
$A_{ul}=(64 \pi^4 \nu^3/3 h c^3)S\mu^2_{ul}/g_u$, and
$J_\nu(T)=(h \nu/k)[exp(h \nu/ k T)-1]^{-1}$, with $T_{\rm BG}$ denoting the temperature of the background,
$T_{\rm ex}$ the excitation temperature of the transition and $g_u$ the degeneracy of the upper level.

In the Rayleigh-Jeans approximation ($h \nu \ll k T$), and assuming $T_{\rm ex}\gg T_{\rm BG}$, eq.~\ref{eq.Nup} can be reduced to
\begin{equation}
N_u=\frac{8 \pi k \nu^2}{h c^3 A_{ul}} W,
\label{eq.Nup_red}
\end{equation}

where the integrated brightness temperature $\int{T_{\rm B}\, dv}$ is denoted as $W$.
If LTE is assumed, the population distribution of all the levels can be represented by
a single temperature, the rotation temperature ($T_{\rm rot}$), given by
\begin{equation}
\frac{N_u}{g_u}= \frac{N}{Z} e^{-E_u/k T_{\rm rot}},
\label{eq.LTE}
\end{equation}

where $N$ is the total column density, $E_u$ is the energy of the upper level, and $Z$ the partition function calculated as
\begin{equation}
Z=\sum_{\forall i}^{} g_i\,e^{-E_i/k T_{\rm rot}}.
\label{eq.Z}
\end{equation}

Using Eq.~\ref{eq.Nup_red} in Eq.~\ref{eq.LTE} we obtain
\begin{equation}
N=\frac{8 \pi k \nu^2 Z}{h c^3 A_{ul} g_u} W e^{E_u/k \, T_{\rm rot}}
\label{eq.Ntot}
\end{equation}

or the equivalent logarithmic expression
\begin{equation}
{\rm log} \frac{8 \pi k \nu^2}{h c^3 A_{ul} g_u} W  = {\rm log} \frac{N}{Z} - \frac{log\,e \, E_u}{k \, T_{\rm rot}},
\label{eq.PopDiag}
\end{equation}

which constitutes the basic equation for the rotation diagrams 
\citep[see][for a complete description of this method]{Blake87,Turner91,Goldsmith99}.
Plotting $N_u/g_u$ for each transition against the energy of its upper level in a logarithmic scale,
the population distribution can be fitted by a straight line that provides the
total column density divided by the partition function as well as $T_{\rm rot}$.

\subsection{Blended lines}
\label{Sect.BlendedPopDiag}
Whenever a molecular feature is composed of a number of unresolved blended transitions of a given molecular species, rotation diagrams cannot be directly 
used to determine the physical properties of the molecular emission. 
The spectroscopic parameters of the blended individual transitions within an observed feature allow us to estimate the contribution 
of each transition to the total integrated intensity, assuming LTE and optically thin emission.
Thus, using eq.~\ref{eq.Ntot} we can express the relative integrated intensity between two transition with Einstein coefficients
$A_{ul}'\,{\rm and}\,A_{ul}''$, energies of the upper level $E_u'\,{\rm and}\,E_u''$, and degeneracies $g_u'\,{\rm and}\,g_u''$ as
\begin{equation}
\frac{W'}{W''} = \frac{\nu''^2 A'_{ul}g'_u }{\nu'^2 A''_{ul} g''_u} e^{(E''_u-E'_u)/k \, T^0_{\rm rot}}.
\label{eq.relative}
\end{equation}
This allows us to compute the relative intensity of transition for a given $T^0_{\rm rot}$. We will differentiate between 
the temperature, $T^0_{\rm rot}$, we assume for estimating the contribution of each transition to the blended line, and the 
rotational temperature, $T_{\rm rot}$, derived from the rotation diagram.

If we consider the case of a molecule with hyperfine structure where all the transitions within a feature have the same upper level energy,
then we can make the assumptions of $E'_u \approx E''_u$ and $\nu' \approx \nu''$. Therefore, the contribution of each transition will
not depend on the assumed $T_{\rm rot}$ and will only depend on the spectroscopic parameters of each transition in the form
\begin{equation}
\left( \frac{W'}{W''} \right)_{\rm HF} = \frac{A'_{ul}g'_u }{A''_{ul} g''_u}.
\label{eq.relativeHF}
\end{equation}

When the contribution of each hyperfine transition to the total line is computed and plotted on the rotation diagrams, 
they will all lie in the same point in the rotation diagram (see NO and NS diagrams).

In the general case where $E'_u \ne E''_u$, the calculated contribution of each line for a given $T^0_{\rm rot}$, will lie on a straight 
line when included into the rotation diagram. The line traced by these points will have the slope ${\rm log}_{10}e/k\,T^0_{\rm rot}$.

The LTE approximation assumes that all the transitions will lie in a straight line in the rotation diagram, thus we can modify 
the assumed $T^0_{\rm rot}$ used to calculate the contributions within the blended lines to obtain the best fit to a straight 
line to all data in the rotation diagram. 
At this point we will obtain the situation in which $T^0_{\rm rot} \approx T_{\rm rot}$.

In this paper we have several examples where this method has been applied, i.e. the CH$_3$CN rotation diagram, where only two 
multiple-transition lines are plotted, the CH$_3$CCH diagram, where three blended lines are perfectly aligned to a straight line and 
the CH$_3$OH diagram, where a mixture of multi-transition lines and single lines are plotted altogether (Fig.~\ref{fig.PopDiag}).

We have to take special care when dealing with the symmetric top rotors such as CH$_3$CN and CH$_3$CCH.
In this case, the rotation temperature is not sufficient to describe the relative population of the levels given that 
both $T_{\rm kin}$ and $T_{\rm rot}$ are required
to evaluate the partition function as described by \citet{Turner91}. Thus $T_{\rm rot}$ would characterize the population of $J$ levels
within a $K$ ladder, while $T_{\rm kin}$ would describe the population between $K$-ladders.
Our observations do not resolve the transitions of different $K$-ladders. We assume both temperatures to be the same 
($T_{\rm kin}=T_{\rm rot}$) so that the population of all levels is described by a single rotation temperature.
Given that $T_{\rm rot}$ is a lower limit to $T_{\rm kin}$, with typically $T_{\rm rot} \ll T_{\rm kin}$, this assumption may result in an overestimation of the column 
density derived from rotation diagrams as the contribution 
to the total integrated intensity of the $K\geq 1$ levels may have been underestimated. 
On the other hand, the evaluation of the partition function will also be underestimated.        
Thus, the overall effect in the total column density determination is not expected to be more than a factor of $2-3$
while the derived rotation temperature will only be marginally affected.

\section{Notes on individual molecules}
\label{apen.notes}

In the following, we will discuss the fitting procedure of the identified molecular species in detail.

\subsection{Carbon monosulfide: CS}
\label{sect.CS}
Only the $J=3-2$ transition of carbon monosulfide at 147\,GHz is observed in the survey.
It appears to be partially contaminated by the H35$\alpha$ recombination line. 
The estimated Gaussian profile derived from the observed H34$\alpha$ (Sect.~\ref{sect.hrec}) has been subtracted to account for 
its contribution.

Two rare isotopes, $^{13}$CS and C$^{34}$S at 144.6 and 138.8\,GHz respectively, 
are also observed in the spectral scan.
\citet{Martin05} used these observations of CS and its isotopes, as well as additional CS lines,
 to derive isotope ratios of $^{32}$S/$^{34}$S$\sim 8\pm 2$ and $^{34}$S/$^{33}$S$>9$.
From the observation of additional CS transitions at 3, 1.5 and 1.2\,mm, \citet{Martin05} derive a $T_{\rm rot}\sim 10$\,K.

\subsection{Nitric oxide: NO}
\label{sect.NO}
One rotational transition of nitric oxide (150.2 and 150.5\,GHZ) has been identified in our survey. 
\cite{Martin03} confirmed this identification
by observing additional transitions at 250\,GHz. These transitions have also been included in the rotation diagram in
Fig.~\ref{fig.PopDiag}.
The topology of the NO energy levels is similar to that of NS. 
Transitions connecting upper and lower levels are labeled $\Pi^{+}$ and $\Pi^{-}$. 

It was only possible to fit the $\Pi^{+}$ hyperfine group of the transition at 150.2\,GHz, given that the $\Pi^{-}$ group
is blended with the more intense H$_2$CO line (Sect.~\ref{sect.H2CO}). 
As shown in Table~\ref{tab.fits}, the $\Pi^{-}$ group of transitions has been estimated by fixing width and radial velocity of each
component to the values derived from the fitting to the $\Pi^{+}$ transition and the line intensities determined by their relative
line strengths.

\subsection{Nitric sulfide: NS}
\label{sect.NS}
One rotational transition of nitric sulfide (161.2 and 161.7\,GHz) has been detected in our survey.
\cite{Martin03} confirmed the identification by detecting other transitions at 115 and 207\,GHz.
The rotation diagram in Fig.~\ref{fig.PopDiag} includes these additional transitions.

This molecule presents $\Lambda$-doubling and hyperfine splitting. Because of the
$\Lambda$-doubling, each rotational level is divided into two levels with opposite parity.
Transitions connecting the upper levels are denoted as $e$ and those connecting the lower levels as {\it f} (see Table~\ref{tab.fits}). 
We are able to resolve the split rotational levels due to $\Lambda$-doubling but the intrinsic linewidths prevent us from 
resolving the hyperfine structure.

\subsection{Sulfur monoxide: SO}
\label{sect.SO}
We detect three transitions of sulfur monoxide (138.2, 159 and 172.2\,GHz). 
The velocity components at 180\,km\,s$^{-1}$ and 275\,km\,s$^{-1}$ are clearly identified.
As seen in the rotation diagrams (Fig.~\ref{fig.PopDiag}), the 275\,km\,s$^{-1}$ component of the $3_4-2_3$ transition shows a much higher
intensity than expected if we assume both components to have the same rotation temperature. 
This is likely due to the contamination of this component by the $8_{1,8}-7_{1,7}$ transition of NH$_2$CN. The uncertainty in the 
contribution of this line due to the noise strongly affects the fitting of the high velocity component of 
this SO transition.

One of the isotopic substitutions of sulfur monoxide, $^{34}$SO (135.8\,GHz) is detected blended with a SO$_2$ transition.
This faint feature has been fitted with a double Gaussian, with the velocity and width of each component fixed to the values derived
from the same transition, $4_3-3_2$, of the main isotope.
We calculate a ratio $^{32}$SO/$^{34}$SO$=5.1\pm 1.2$, using both velocity components of the $4_3-3_2$ transition of both isotopes.
This ratio is consistent within the errors with $^{32}$S/$^{34}$S$=8\pm2$ as derived by \citet{Martin05} from CS data.

\subsection{Silicon monoxide: SiO}
\label{sect.SiO}
Two transitions of silicon monoxide in the ground vibrational state are observed at 130.3 and 173.7\,GHz ($J=3-2$ and $J=4-3$, respectively).
It is possible to identify two velocity components from the $J=3-2$ transition at 180 and 260 km\,s$^{-1}$.
Velocity and width 
of the fainter and more noisy $J=4-3$ transition were fixed to the values obtained from
the fit to the $J=3-2$ line.
The rotation diagram has been derived for each velocity component (solid and dashed lines in Fig.~\ref{fig.PopDiag}).

\subsection{Hydrogen sulfide: H$_2$S}
\label{sect.H2S}
Only one line of hydrogen sulfide at 168.8\,GHz, with velocity components at 180 and 275\,km\,s$^{-1}$, is detected in the 2\,mm survey
\citep[see also][]{Martin05}.
Physical parameters in Table~\ref{tab.MolecDensity} have been derived by assuming a rotation temperature of 12\,K
similar to that derived for many other species (see Table~\ref{tab.MolecDensity}).

\subsection{Hydrocyanic acid: HCN}
The $J=2-1$ transition of HCN at 177.2\,GHz lies outside the covered frequency range, but the $2-1$ line of 
its H$^{13}$CN isotope at 172.6\,GHz was detected. A fit to this transition allows us to estimate the abundance of HCN if we assume
a rotation temperature of 12\,K and the $^{12}$C/$^{13}$C ratio of 40 derived by \citet{Henkel93}.

\subsection{Oxomethylium: HCO$^+$}
\label{sect.HCO+}
The only 2\,mm transition of the main species of oxomethylium, with energy levels low enough to be observable in the ISM of NGC\,253, at
a frequency of 178.3\,GHz, lies a few GHz above the observed frequency range.
Nevertheless, one of its isotopes, H$^{13}$CO$^+$, 
has been detected in the $J=2-1$ transition at 173.5\,GHz, near the upper end of the spectral scan.
Two velocity components can be fitted to the observed profile.
The HCO$^{+}$ column density shown in Table~\ref{tab.MolecDensity} has been calculated from the integrated intensity of the measured 
H$^{13}$CO$^+$ line with an excitation temperature of 12\,K and an isotopic ratio of $^{12}{\rm C}/^{13}{\rm C}\sim40$ \citep{Henkel93}.

The $J=2-1$ transition of HC$^{18}$O$^+$ is also identified at 170.3\,GHz. 
The low velocity component shows a higher intensity than expected from the observed H$^{13}$CO$^+$ profile as well as a slightly higher
velocity.
If we compare both profiles we derive a H$^{13}$CO$^+$/HC$^{18}$O$^+$ ratio of $1.7\pm1.1$ and $2.9\pm1.2$ for the low and high velocity
components, respectively.
The value derived for the high velocity component agrees within the errors with the expected value of $\sim3.7$ if we assume the isotopic
ratios $\rm^{12}C/^{13}C=40$ \citep{Henkel93} and $\rm ^{16}O/^{18}O=150$ \citep{Harrison99}.

\subsection{Hydroisocyanic acid: HNC}
\label{sect.HNC}
Similar to HCN, the $J=2-1$ transition of HNC at 181.3\,GHz lies outside the frequency range of the survey. 
We identify a feature at 174.1\,GHz as the $J=2-1$ transition of HN$^{13}$C.
The detection is uncertain and thus tentative as the line is strongly blended with the also tentatively identified $3_{1,2}-2_{1,1}$ 
transition of c-C$_3$H (Sect.~\ref{sect.C3H}).

\subsection{Ethynyl: C$_2$H}
The $J=2-1$ group of hyperfine transitions of C$_2$H, observed at 174.7\,GHz near the upper frequency cutoff of the survey, is the second 
brightest feature observed in the 2\,mm window after CS.
Even though the hyperfine structure is unresolved, it is still possible to differentiate the two velocity components at 160 and
280\,km\,s$^{-1}$. 

\subsection{Thioxoethenylidene: C$_2$S}
\label{sect.C2S}
The survey shows four faint lines of thioxoethenylidene (131.6, 140.2, 142.5 and 144.2\,GHz). 
The low intensity of the observed transitions makes the fitting uncertain
(see Table~\ref{tab.fits}).
Only the $11_{10}-10_9$ and $10_{11}-9_{10}$ transitions at 131.6\,GHz and  140.2\,GHz are
reliably detected, while the transitions at 142.5\,GHz and 144.2\,GHz are tentative.

\subsection{Carbon oxide sulfide: OCS}
\label{sect.OCS}
Two transitions of carbon oxide sulfide (133.8 and 145.9\,GHz) are detected in the survey. 
In the rotation diagram in Fig.~\ref{fig.PopDiag}
these observations are complemented by the 3\,mm OCS transitions presented by \citet{Martin05}.

The $J=13-12$ transition of OCS at 158.1\,GHz might have been detected close to a SO$_2$ line, but the signal-to-noise
ratio is not high enough for a reliable Gaussian fit.

\subsection{Sulfur dioxide: SO$_2$}
\label{sect.SO2}
Five transitions of sulfur dioxide (134.0, 135.7, 140.3, 146.6 and 151.4 \,GHz) have been identified. The upper energy levels of the 
transitions range from 12 to 43\,K providing a good estimate of the rotation temperature.

The $5_{1,5}-4_{0,4}$ transition at 135.7\,GHz is blended with a $^{34}$SO line (Sect.~\ref{sect.SO}). Velocity
and width of the line were fixed to perform a two Gaussian fit for the SO$_2$ and $^{34}$SO transitions.

The $2_{2,0}-2_{1,1}$ line at 151.4\,GHz is contaminated by emission of a much fainter c-C$_3$H$_2$ line which, according to its estimated intensity 
(Sect.~\ref{sect.cC3H2}), would account for $\sim15\%$ of the observed feature. The contribution of the c-C$_3$H$_2$
strongly depends on the assumed $T_{\rm rot}$. If we consider the uncertainty in the $T_{\rm rot}$ derived from c-C$_3$H$_2$
and estimate the contribution of this transition to the observed SO$_2$ line, the effect on the derived SO$_2$ parameters
would be of $\pm 1$\,K in $T_{\rm rot}$ and $\pm 5\%$ in the column density.

The $5_{2,4}-5_{1,5}$ and $7_{1,7}-6_{0,6}$ transitions are blended with a CH$_3$OH group of lines at 165.1\,GHz. 
We have estimated their contribution to the observed feature by assuming a $T_{\rm rot}=15$\,K as derived from the rotation diagram.

For the derived excitation conditions of SO$_2$, its $3_{2,2}-3_{1,3}$ line at 158.1\,GHz is expected to have an 
intensity of $\sim 5$\,mK, which is below the noise level at this frequency.

\subsection{Formaldehyde: H$_2$CO}
\label{sect.H2CO}
Two ortho (140.8 and 150.5\,GHz) and one para (145.6\,GHz) $J=2-1$ transitions of formaldehyde are detected. 
In each of the profiles, the two velocity components at 180 and 285\,km\,s$^{-1}$ are clearly separated.

The $2_{0,2}-1_{0,1}$ transition at 145.6\,GHz and the $2_{1,1}-1_{1,0}$ at 150.4\,GHz are blended with an HC$_3$N and a c-C$_3$H$_2$ 
line, respectively. Before performing the Gaussian fits to the H$_2$CO features, the estimated contributions from HC$_3$N 
(Sect.~\ref{sect.HC3N}) and c-C$_3$H$_2$ (Sect.~\ref{sect.cC3H2}) were subtracted from the spectra.

Rotation diagrams are plotted for each velocity component (Fig.~\ref{fig.PopDiag}) where rotation temperatures of 27\,K and 34\,K 
are derived for the 180 and 285\,km\,s$^{-1}$ components, respectively.
The 285\,km\,s$^{-1}$ component of the $2_{1,2}-1_{1,1}$ transition appears to have a significantly higher intensity than expected from the rotation
diagram. 
The large intensity of the $2_{1,2}-1_{1,1}$ line of H$_2$CO could be due to the contamination by an unidentified line.
If this measurement is not taken into account, the resulting temperature derived for the 285\,km\,s$^{-1}$ component
would be $28\pm2$\,K while the estimated column density would be lower by $\sim30\%$, 
closer to the parameters derived from the 180\,km\,s$^{-1}$ component.

\subsection{Thioformaldehyde: H$_2$CS}
\label{sect.H2CS}
Three transitions, all belonging to ortho-thioformaldehyde (135.3, 139.5 and 169.1\,GHz), 
are tentatively detected.

The $4_{1,4}-3_{1,3}$ transition is blended with the H36$\alpha$ recombination line. 
A Gaussian profile similar to that of the observed H34$\alpha$ line at 160\,GHz (Sect.~\ref{sect.hrec}) has been subtracted. 
We assume that the residual emission is due to H$_2$CS.
We also find the $4_{1,3}-3_{1,2}$ line blended with the CS $J=3-2$ emission from the image band.
With the known image sideband rejection at this frequency (Fig.~\ref{fig:datastatistics}), we can estimate the contribution of the 
CS line measured at 146.9\,GHz (Sect.~\ref{sect.CS}).
The resulting feature after the subtraction of the CS line is fitted  by a Gaussian profile.
There is a part of this feature which is not properly fitted by emission from CS and H$_2$CS. 
This may be caused by
a slight difference in the observing position between this spectrum and that containing CS in the signal band,
which would cause an appreciable change in the shape of the subtracted line. Contamination by a third species is unlikely.
The $5_{1,5}-4_{1,4}$ line at 169.1\,GHz is not blended, but the relatively high noise level at this 
frequency and the presence of nearby lines makes the fit uncertain.

A rotation temperature of 11\,K is determined from the rotation diagram (Fig.~\ref{fig.PopDiag}) and is consistent with those derived from 
other molecules.
With this rotation temperature, the $4_{0,4}-3_{0,3}$ transition of para-H$_2$CS at 137.3\,GHz is expected 
to have an intensity of $\sim 6$\,mK.
The non-detection of this line could be explained by an excitation temperature of $>20$\,K.

\subsection{Isocyanic acid: HNCO}
\label{sect.HNCO}
The two brightest lines of isocyanic acid in the covered frequency band are clearly detected
(131.9 and 153.9\,GHz). 
The 280\,km\,s$^{-1}$ component of the $7_{0,7}-6_{0,6}$ transitions at 153.9\,GHz is blended
with a group of CH$_3$CCH lines.
Thus the contribution of this component has been calculated by assuming the 180 and 280\,km\,s$^{-1}$ velocity components 
have the same excitation temperature. This assumption is supported by the good fit of the CH$_3$CCH line to the residual feature
(Sect.~\ref{sect.CH3CCH}).
The line observed at 154.4\,GHz is tentatively identified as the 280\,km\,s$^{-1}$ velocity component of the HNCO $7_{1.6}-6_{1,5}$
transition. 
The $7_{1,7}-6_{1,6}$ transition at 153.3\,GHz, which should have a similar intensity, is not detected. 
Due to the uncertainty of this identification, this line has not been used in the analysis of the HNCO excitation conditions.
Including this transition in the rotation diagram (Fig.~\ref{fig.PopDiag}) would result in a derived $T_{\rm rot} \sim 29$\,K, 
slightly higher than that derived for the 180\,km\,s$^{-1}$ velocity component, and a column density $\sim 16$\,\% lower than that in
Table~\ref{tab.MolecDensity}.

\subsection{Protonated carbon dioxide: HOCO$^+$}
We tentatively identified two transition (149.6 and 171.0\,GHz) of protonated carbon dioxide, a quasilinear molecular ion and nearly symmetric top.
These are the brightest observable transitions in the 2\,mm range.
Only the high velocity component of both lines are detected.
The large intensity observed in the $8_{1,8}-7_{1,7}$ transitions, expected to be similar or lower than the $7_{0,7}-6_{0,6}$, is mainly due to the
uncertainties of the baseline substraction.
A rotational temperature of 12\,K have been assumed to estimate an average column density with both detected transitions.

Further observations of other HOCO$^+$ transitions are needed to confirm the detection and constrain the excitation temperature of this molecule.

\subsection{Cyclopropenylidyne: c-C$_3$H}
\label{sect.C3H}
The $3_{1,2}-2_{1,1}$ group of transitions at 174\,GHz is the only feature of c-C$_3$H identified in the survey.
Contamination by the feature by the $J=2-1$ transition of HN$^{13}$C makes the fitting uncertain. Thus, the 
velocity component were fixed to $V_{\rm LSR}=180$\, and\, $ 280\,\rm km\,s^{-1}$ as derived for most species and the width of the
lower velocity component to $\Delta v_{1/2}=41\rm \,km\,s^{-1}$ similar to that from the fitting to the high velocity one.
The $4_{1,4}-3_{1,3}$ transitions at 172.6\,GHz appear to be blended with a H$^{13}$CN line. 
At 132.9\,GHz, the $3_{1,3}-2_{1,2}$ group of transitions, which is expected to be similar in intensity to those at 174\,GHz, 
is likely present blended to a transition of CH$_2$NH and an also likely transition of CH$_3$OH, but the noise level at these frequencies do not 
allow us to separate these lines. 
Given the considered source size of $20''$ the 133\,GHz line would be $20\%$ fainter than that at 174\,GHz, so
beam dilution cannot account for the non-detection of these lines.
Therefore, its detection, similar to the HN$^{13}$C line to which it is blended (Sect.~\ref{sect.HNC}) is highly tentative.

\subsection{Cyanoacetylene: HC$_3$N}
\label{sect.HC3N}
Four out of the five transitions of cyanoacetylene detected in the observed frequency range 
(136.5, 145.6, 154.7, 163.8 and 172.8\,GHz) appear not to be blended with any other molecular line. 
Two velocity components have been fitted to each line.
The $J=16-15$ transition at 145.6\,GHz is blended with an H$_2$CO line. 
Since we have observed more HC$_3$N than H$_2$CO lines to accurately determine the physical parameters,
we have estimated the relative contribution of each of the velocity components of HC$_3$N.
For the HC$_3$N $J=18-17$ line, where the velocity components are not clearly differentiated, the velocity of each component has been 
fixed.
The $J=19-18$ transition seems to lie above the $T_{\rm rot}\sim20$\,K derived from the other observed transitions.
This may indicate that the lower energy transitions would not be optically thin as blending with any other molecular line 
seems unlikely.
Multiline observations by \citet{Mauers90} (towards a position $9''$ north of our position) also noticed an excess  of the
higher transition lines of HC$_3$N. Their comparison with model computations show that two molecular components are required to 
explain the observed line intensities, one with $\rm n(H_2)\sim 10^4 cm^{-3}$ and $T_{\rm kin}\sim 60$\,K, and a second with 
$\rm n(H_2)\sim 5\,10^5 cm^{-3}$ and $T_{\rm kin}\sim 150$\,K.
The line intensity of the $J=17-16$ transition measured in this work is about half that measured by \citet{Mauers90} which 
indicates the strong effect of a pointing offset in the total integrated intensity.
However, as far as the physical parameters are concerned, the use of the $J=19-18$ line, although causing a 
change in the determined rotation temperature of $\sim 50\%$, affects the total column density by less than a factor of 2.

\subsection{Methanimine: CH$_2$NH}
One transition of methanimine is clearly identified in the line scan.
The observed profile of the $2_{1,1}-1_{1,0}$ transition at 133\,GHz is heavily affected by non linear baselines and 
the fit shown in Table~\ref{tab.fits} is uncertain. 
The $1_{1,0}-1_{0,1}$ transition at 167\,GHz, expected to have an intensity twice as large as that at 133\,GHz, is not 
reliably detected due to the lack of sensitivity at this frequency.
Although the detection seems to be clear, further observations of transitions in other frequency bands are needed to 
confirm this detection.

\subsection{Cyclopropenylidene: c-C$_3$H$_2$}
\label{sect.cC3H2}
Three transitions of cyclopropenylidene (155.5 and 150.8\,GHz) are clearly detected.
However, only one transition, the $3_{2,2}-2_{1,1}$ line at 155.5\,GHz, appears not to be blended with any other line. 

The $4_{0,4}-3_{1,3}$ and $4_{1,4}-3_{0,3}$ lines, both at 150.8\,GHz, appear to be blended.
A double Gaussian profile was fitted by assuming equal width and velocity and a line intensity ratio determined by their 
spectroscopical parameters, assuming optically thin emission.

The intensities of the other blended transitions (145.1 and 150.4\,GHz) were estimated for a rotation temperature of 9\,K as derived 
from the rotation diagram (Fig.~\ref{fig.PopDiag}). 
Given the small dynamic range in energies of the unblended transitions used in the rotation diagram, 
the estimated $T_{\rm rot}$ is affected by a large uncertainty (see Table~\ref{tab.MolecDensity}) and, therefore, 
the estimates for the blended transitions are also quite uncertain.
The $5_{1,4}-5_{0,5}$ transition at 151.3\,GHz with an estimated intensity of $\sim 1$\,mK, well below the noise level of the spectra, 
has been calculated with the aim of estimating its contribution to the faint SO$_2$ transition at that frequency (Sect.~\ref{sect.SO2}).

\subsection{Cyanamide: NH$_2$CN}
Only one line of cyanamide is clearly detected in the survey, the $8_{1,7}-7_{1,6}$ transition at 161.0\,GHz which appears to be
partially blended with image sideband emission from H$_2$S. 
The emission of H$_2$S in the signal band has been corrected by the image sideband rejection and subtracted 
from the spectrum before fitting the NH$_2$CN line.
The other observed lines are close to the noise level and therefore all fitted parameters are strongly affected by the baseline.
The $8_{1,8}-7_{1,7}$ transition at 158.8\,GHz, seems to be blended with a SO (Sect.~\ref{sect.SO})
but the noise level does not permit a fit.
The high rotation temperature of 63\,K derived from the rotation diagram (Fig.~\ref{fig.PopDiag}) depends mainly on the fainter and 
therefore less reliably fitted transitions.
A much lower rotation temperature would lower the column density by a factor of about 2.

\subsection{Methyl cyanide: CH$_3$CN}
\label{sect.CH3CN}
Two spectral features are identified as methyl cyanide transitions (147.2 and 165.6\,GHz). 
Each of the $J-J'$ transitions of this symmetric top consist of a number of overlapped $K$-components ($K=0\ldots J-1$).
Single Gaussian profiles were fitted to the observed lines.
The contribution of the $K=0$ and 1 ladders, for the low rotation temperatures derived from other high dipole moment molecules,
account for $\sim 98\%$ of the line profile.
Therefore the $K>1$ contribution to the total line intensity has not been taken into account.
We derive a $T_{\rm rot}= 10$\,K from the rotation diagram in Fig.~\ref{fig.PopDiag} which has been obtained as explained in 
Sect.~\ref{Sect.BlendedPopDiag}.

\subsection{Methanol: CH$_3$OH}
\label{sect.CH3OH}
A total of nine transitions or groups of transitions of methanol is detected in the line survey
at 143.9, 145.1, 146.4, 157, 165 and 170\,GHz. 
The transitions at 145.1\,GHz are blended with the C$_3$H$_2$ $3_{1,2}-2_{2,1}$ line whose contribution has been estimated 
(Sect.~\ref{sect.cC3H2}) and subtracted from the observed feature.
In the same way, the lines at 165.0\,GHz are slightly blended with SO$_2$ transitions. 
The emission of these SO$_2$ lines, subtracted from the spectra (Sect.~\ref{sect.SO2}), mainly affects the fit of the 
$4_{1,3}-4_{0,4}$ transition.
The  $6_{-1}-5_0$ transition of CH$_3$OH expected to be detected at 132.8\,GHz, is heavily affected by baseline instabilities. 
The rotation diagram derived from the observed transitions has been plotted using both the single- and
multi-transition lines as described in Sect.~\ref{Sect.BlendedPopDiag}.
The derived rotation temperature is 12\,K.

\subsection{Methyl acetylene: CH$_3$CCH}
\label{sect.CH3CCH}
Three methyl acetylene features are observed (136.7, 153.8 and 170.9\,GHz).
As in the case of CH$_3$CN each of these corresponds to a $\Delta J=1$ transition consisting of a number of
unresolved $K$-components (with $K=0\ldots J-1$). Single Gaussian profiles have been fitted to the observed profiles.

The best fit to the rotation diagram as described in Sect.~\ref{Sect.BlendedPopDiag} results in a rotation temperature of $\sim 62$\,K. 
Given the high rotation temperature derived
only contribution of transitions with $K=0...4$, which have upper level energies $<160$\,K, have been taken into account in the fit.
For a temperature of 60\,K, transitions with $K=4$ contribute $\sim5\%$ to the observed profiles, while
higher $K$ lines represent less than $1\%$ of the integrated intensities.
Neglecting the contribution of these higher energy transitions (i.e. assuming the whole component is dominated by the $K=0$ transitions)
would result in an overestimation of the column densities by up to a factor of 4.

\subsection{Hydrogen recombination lines}
\label{sect.hrec}
Four H$\alpha$ recombination lines are present within the frequency range covered by our survey.
Only one of them, H34$\alpha$, is not blended and can be fitted.
The H33$\alpha$ line lies close to the $2-1$ line of C$_2$H at the upper end of the frequency range where the rms is similar to the 
expected intensity of the line.
The H35$\alpha$ and H36$\alpha$ lines are blended with CS\,$3-2$ and H$_2$CS\,$4_{1,4}-3_{1,3}$ lines, respectively.
If the relation of the line flux $S_{\rm L}\propto \nu_{\rm L}^2$ applies at these frequencies \citep{Puxley97}, the lines 
should have flux densities 20 and 30$\%$ below that measured for the H34$\alpha$ line. 
We have considered them to have a similar line profile as that of the measured H34$\alpha$ line 
given the uncertainty in the fitted intensity.

\begin{figure}[tb]
\centering
	\includegraphics[angle=-90,width=0.5\linewidth]{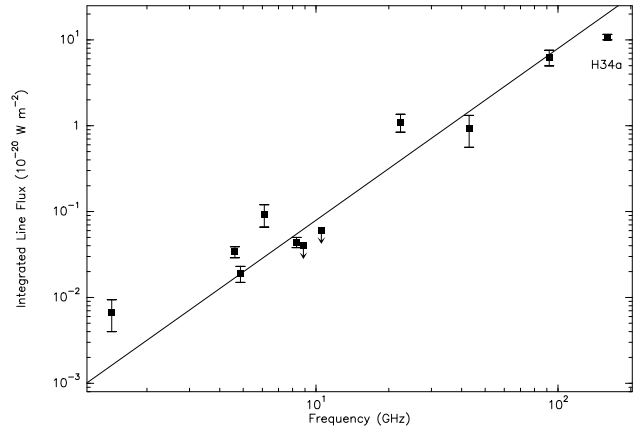}
	\caption{Recombination line fluxes of measured H$\alpha$ lines in the range from 1 to 160 \,GHz as a function of frequency.
	A best-fit to the line fluxes in the form $S_{\rm L}\propto \nu^2$ is shown as fitted by \citet{Puxley97}.}
	\label{fig.Halfa}
\end{figure}

To convert the measured H34$\alpha$ line temperature into flux densities we use the conversion factor for the 30\,m telescope
$S/T_{\rm MB}=4.95\, {\rm Jy/K}$. Following \citet{Puxley97}, a source to beam coupling factor
$(\theta_b^2+\theta_s^2)/\theta_b^2$ has been applied for a source size $\theta_s=7''$.
We calculate an integrated line flux of $(1.08\pm0.08)\,10^{-19}$ W\,m$^{-2}$.
As seen in Fig.~\ref{fig.Halfa}, this value closely follows the best fit to the relation between the integrated line flux versus
frequency derived by \citet{Puxley97}. In this figure the flux of the H34$\alpha$ line is plotted together with the recombination
lines observed in millimeter and centimeter wavelengths \citep[and references therein]{Puxley97}.
\clearpage
\clearpage

\end{appendix}



\end{document}